**Review**

# Prediction of biological radiation effects based on ionization clusters (nanodosimetry)

Hans Rabus

*Physikalisch-Technische Bundesanstalt (PTB), 10587 Berlin, Germany*

Dedicated to the memory of Bernd Grosswendt†

## Abstract

This article reviews approaches that link the formation of ionization clusters in nanometric volumes to radiobiological effectiveness. The corresponding models were developed as the field of nanodosimetry developed. Some address early biological radiation effects, such as DNA damage, while most aim to predict cell survival or inactivation. The models also differ in the nanodosimetric quantities considered, with many based on the probability distribution of ionization cluster formation in a single target. Some models account for the synergistic effects of pairs of ionization clusters formed in different targets. Several models feature macroscopic aggregation frameworks based on particle fluence, which are proposed for use in radiotherapy treatment planning, particularly in ion-beam radiotherapy. The models are presented here using harmonized terminology and notation for nanodosimetric quantities. An extension of the conceptual framework of nanodosimetry is also discussed. This extension transitions from a target-centered description to a track-centered description. It also introduces nanodosimetry-based analogs of dosimetric concepts, such as dose and linear energy transfer. This paper traces and summarizes the historical development of nanodosimetry-based biological effect models and discusses conceptual aspects of the models to reveal their underlying assumptions and the extent to which they are mechanistic or merely elucidate correlations. Eventually, an attempt is made to identify the key open questions in this field that still need to be addressed.

## 1 Introduction

Nanodosimetry is concerned with the quantitative description of the stochastic pattern of radiation interaction, known as the track structure, in terms of clusters of ionizations in targets with nanometric dimensions [1–4]. The development of this field was prompted by evidence that clustering of lesions within short segments of the deoxyribonucleic acid (DNA) molecule determines radiobiological effectiveness [5–7]. The first detectors for measuring particle track structure were developed 50 years ago [8]. The foundations for the present state-of-the-art nanodosimetric gas counters were established in the late 1990s [9–11] leading to parallel developments of nanodosimeters by the Italian National Institute for Nuclear Physics (INFN) in Legnaro, Italy [12], the Polish Nuclear Research Center (NCBJ) in Swierk, Poland [13], and a collaboration of Loma Linda University (LLU), USA and the Weizmann Institute of Science (WIS),

Israel [14]. These detectors simulate a single nanometric target volume in tissue or liquid water based on the principle of density and material scaling [15] and are the established reference instruments for nanodosimetry. However, the focus of development in this area has shifted towards the exploration of detectors with potential track imaging capabilities [16–19] or more compact detector designs [20–22].

The development of gas-counter nanodosimeters was accompanied by theoretical investigations into ionization cluster (IC) formation [23,24] and the development of a dedicated track-structure simulation code [25], which was later named PTra (PTB track structure code) [26]. In contrast to parallel developments of other track-structure codes such as KURBUC (Kyushu University and Radio-biology Unit Code) [27,28] and PARTRAC [29], PTra focused on numerical simulation of the nanodosimetric detectors and their performance. Later, it was also employed to study the significance of differences between cross sections of liquid water and DNA for predicting ICs in DNA [30]. The general-purpose Monte Carlo codes Geant4 and PHITS were extended to also enable track-structure simulations with their variants Geant4-DNA [31–34] and PHITS-ets [35], respectively.

This article reviews models proposed to predict the biological effectiveness of radiation from nanodosimetric features of the particle track structure. Initial attempts to achieve this objective were already made in parallel with the development of nanodosimetric gas-counter detectors [36]. Since then, a variety of approaches have been put forth, including the recently proposed cluster dose concept by Faddegon et al. for the use of nanodosimetric quantities in treatment planning for hadron therapy [37]. Table 1 provides a chronological overview of the evolution of nanodosimetry and these models, listing the first publications on each development as milestones.

All models are based on the hypothesis that nanodosimetry can provide radiation quality descriptors that capture biological effectiveness across different types of ionizing radiation. The proposed models aim at different applications, including radiation protection, radiation therapy, and cell radiobiology. They also pertain to different biological endpoints: some to DNA damage induction, and others to cell killing or inactivation. Most of the reviewed models are predicated on nanodosimetric outcomes from track structure simulations, while a few models are based on experimental nanodosimetric data.

Although they are built on the same foundations, the notations used for nanodosimetric quantities are inconsistent across articles proposing these models. Additionally, the cluster dose concept uses quantities related to, yet distinct from, those considered in the recent ICRU Report 98 [38]. Therefore, Section 2 first provides a description of the concepts of nanodosimetry and the dependencies of the nanodosimetric probability distributions. Extensions of these concepts beyond the framework covered by ICRU Report 98 [38] are also discussed in Section 2. Section 3 describes the biological-effect models, which are grouped by development strands and commented on where appropriate. Section 4 summarizes the different approaches, highlighting commonalities and differences and discussing open issues and future challenges identified by the author.

The objective of this review is to provide an overview of the proposed approaches and to present them alongside their nanodosimetric basis in a harmonized way. Where possible it follows the notation introduced by ICRU Report 98 [38]. For this reason, the notation used in this work deviates from that used in the articles reviewed in certain cases. In instances where this variant notation might lead to confusion, this is explicitly noted in footnotes, or a “translation” is included in figure captions of reproduced figures. Additionally, Supplementary Table 1 provides a comparison of the different notations used in literature with the notation used here and in ICRU Report 98.

Table 1: Selected milestones in the development of nanodosimetry and nanodosimetry-based models of biological radiation effects.

| Year | Nanodosimetry | Bio-effect modelling |
|---|---|---|
| 1975 | First Track Ion Counter [8] | |
| 1996 | Proposal Ion Counter nanodosimeter [10] | |
| 1998 | Particle track imaging detector OPAC [16] | |
| 1999 | First ionization cluster measurements [39] | |
| 2000 | Jet Counter nanodosimeter[13] | |
| 2001 | | Two compartment model [40] |
| 2002 | Startrack [12] and Ion Counter [41] nanodosimeters, Scaling principle [15] | |
| 2003 | | Biological weighting factor [42] |
| 2004 | | $F_2$ = bio effect cross section [43] |
| 2006 | Ion Counter with particle tracking [44] | Combinatorial model [45] |
| 2008 | | Nanodosimetry-based quality factors [46] |
| 2009 | Nanodosimetric detector array [47] | |
| 2010 | Verification of scaling procedure [48] | Upgraded combinatorial model [49] |
| 2014 | Track structure imaging detector [50] | |
| 2015 | Consistence Ion Counter – StarTrack [51]<br>Weighted nanodosimetry at cell level [52] | Track event theory (TET) [53]<br>Nanodosimetry in treatment planning [54] |
| 2016 | Consistence of 3 nanodosimeters [55] | |
| 2017 | | Inactivation cross-section model [56] |
| 2018 | Multi-target simulation [57]<br>Weighted nanodosimetry at voxel level [58]<br>Spatial correlation measurement [59] | Single-target survival model [60] |
| 2019 | Nanodosimetry without trigger detector [61] | TET with nanodosimetry [62], Treatment planning using nanodosimetry [63] |
| 2020 | Nanodosimetric cross-sections [64] | Logistic nanodosimetry model [65]<br>Nanodosimetry radiation action model [66] |
| 2021 | FIRE compact nanodosimeter [67]<br>Cluster frequency at cell level [68] | |
| 2023 | ICRU Report 98 [38] | Cluster dose concept [37] |
| 2024 | | Nanodosimetric-quantity weighted dose [69]<br>$R_2$-based nanodosimetric model [70] |
| 2025 | | Modified logistic nanodosimetry model [71]<br>Calibration of the cluster dose model [72] |
| 2026 | | Cluster-dose based treatment planning [73] |

## 2 Nanodosimetry

For the sake of clarity, it is worth to reiterate that the term “nanodosimetry,” utilized in this review, pertains to the study of the formation of ionization clusters in nanometric volumes and must be discerned from its application in different context. For instance, nanodosimetry should be distinguished both from microdosimetry in nanometric target volumes [74,75] and from the occasionally observed misconception that it refers to the measurement of absorbed dose in nanometric volumes.

### *2.1 Fundamental concepts of nanodosimetry*

Comprehensive reviews of the fundamental principles of nanodosimetry can be found in the literature [1,2], and only the key features are reiterated here. Conceptually, a nanodosimetric measurement constitutes a coincidence experiment as illustrated in Fig. 1: A primary particle moving along the black dashed line passes the target volume (labeled ‘site’) and is recorded by a primary particle detector. The black circles mark the points at which ionizing interactions by the primary particle or its secondary electrons occur. The number of ionizations produced in the target volume is called the ionization cluster size (ICS) $\nu$ of this primary particle event, which is defined by the detection of a primary particle in the primary particle detector.

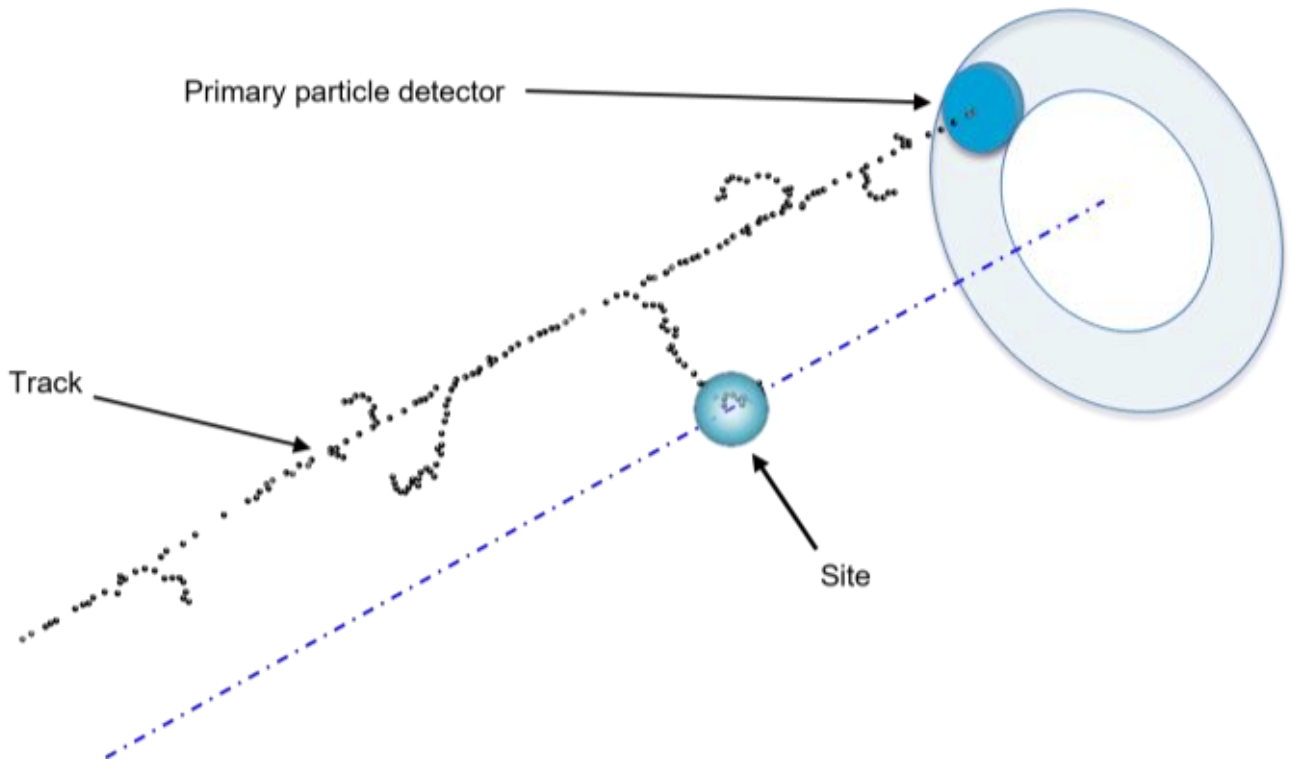

**Fig. 1:** Illustration of a nanodosimetric event in a spherical site passed by a track at a given impact parameter. The black dots represent points with ionizing interactions. The ionizations produced in the volume of this site are scored in coincidence with the detection of a passing primary particle. The annulus represents the area covered by primary particle detectors corresponding to the same impact parameter.[1]

The relative frequency $f_\nu$ of events producing an IC of size $\nu$ in an experiment or a simulation is an estimate of the probability $p(\nu)$ of the generation of an IC of size $\nu$ in the target. This probability pertains to a primary particle of this type and energy passing the site with a given impact parameter. The impact parameter is defined by the offset between the parallels to the particle trajectory passing the site centrally (dot-dashed line in Fig. 1) and the center of primary particle detector (dashed line in Fig. 1), respectively. Following the notation used in ICRU Report 98 [38], the probability distribution of IC size $\nu$ is denoted here by $p(\nu)$. However, different notations have been used in literature to designate this probability and quantities derived from it (Supplementary Table 1).

[1] Fig. 1 is reprinted from Radiation Physics and Chemistry 232, H. Rabus and L. Thomas, On a revised concept of an event that allows linking nanodosimetry and microdosimetry in nanometric sites with macroscopic dosimetry, 112640, Copyright (2025), with permission from Elsevier.

The probability distribution of ICs can also be described by the (cumulative) probabilities $F(k)$ of obtaining an IC up to a maximum size $k$ or the (complementary cumulative) probabilities $F^*(k)$ of an IC exceeding a size $k$. The complementary cumulative probabilities are obtained from the probabilities $p(\nu)$ of cluster size $\nu$ by using Eq. (1).

$$F^*(k) = \sum_{\nu>k} p(\nu)\,; \qquad F_k = \sum_{\nu\geq k} p(\nu) \tag{1}$$

The definition of $F^*(k)$ corresponds to the notation used in ICRU Report 98. The notation $F_k$, on the other hand, has been widely used in nanodosimetric literature (see Supplementary Table 1). Since the symbol $F_k$ appears in many of the figures reproduced in this review, both notations for complementary cumulative probabilities are used in parallel. The two quantities defined in Eq. (1) are connected by Eq. (2), where the symbol "≡" indicates the mathematical identity of the expressions of both sides of the equation.

$$F_k \equiv F^*(k-1) \tag{2}$$

The (cumulative) probabilities $F(k)$ of obtaining an IC up to a maximum size $k$ have not been used in nanodosimetric literature so far. However, some authors have considered the probability of an IC between a minimum size $a$ and a maximum size $b$ [36,54,69]. Inspired by the notation used by Yang et al. [69], these quantities are denoted here as $F_{a-b}$ (Eq. (3)).

$$F_{a-b} = \sum_{\nu=a}^{b} p(\nu) \tag{3}$$

With the coincidence setup shown in Fig. 1, the possible values of $\nu$ include 0, which means no ionization in the target. Cases of no ionization or only a single ionization evidently do not correspond to ionization clustering in a strict sense. Therefore, it makes sense to also consider conditional ICS distributions $p^c(\nu)$, which are defined for $\nu \geq 1$ by Eq. (4).

$$p^c(\nu) = \frac{p(\nu)}{F^*(0)} = \frac{p(\nu)}{1-p(0)} \tag{4}$$

The quantity $p^c(\nu)$ is the probability of obtaining an ICS of $\nu$, given that at least one ionization occurs in the target. Therefore, the conditional distributions defined by Eq. (4) pertain to an event concept similar to that employed in microdosimetry [76]. Replacing $p(\nu)$ in Eq. (1) with $p^c(\nu)$ gives the conditional complementary cumulative probabilities $F^{c*}(k-1) \equiv F_k^c$.

The conditional probabilities $p^c(\nu)$ can be calculated from the unconditional ICS distribution using Eq. (4). As was demonstrated by Pietrzak et al., conditional ICS distributions can also be determined directly from nanodosimetric measurements performed without a primary particle detector [61]. However, this faces challenges such as discriminating against multiple primary particle events that contribute to the same measurement.

Given that a single ionization does not constitute an ionization cluster in the strict sense, it is useful to also consider conditional distributions $p^{c_m}(\nu)$ for ICs having $\nu \geq m > 1$ ionizations defined for $\nu \geq m$ by Eq. (5):

$$p^{c_m}(\nu) = \frac{p(\nu)}{F^*(m-1)} \quad (5)$$

This second type of conditional ICS distribution focuses on targets receiving true ionization clusters (i.e., more than one ionization). Substituting $p^{c_m}(\nu)$ for $p(\nu)$ in Eq. (1) yields the conditional complementary cumulative probabilities $F^{c_m*}(k-1) \equiv F_k^{c_m}$, defined for $k \geq m$.

The probability distribution $p(\nu)$ of the different possible cluster sizes is characteristic of the radiation quality, which is specified by the particle type and energy and is often denoted by $Q$. This distribution is the result of a compound Poisson process and generally does not follow a Poisson distribution (see Supplementary Fig. 1).[2] However, one of the fundamental assumptions of nanodosimetry is that the properties of the ICS distribution are determined by the mean ionization cluster size $M_1$. The mean ICS is the first moment of the ICS distribution $p(\nu)$, where the $k$-th moments $M_k$ are defined by Eq. (6).

$$M_k = \sum_{\nu} \nu^k p(\nu) \quad (6)$$

The summation in Eq. (6) formally ranges from $\nu = 0$ to infinity. However, it can be truncated at the maximum possible ICS $\nu_{max}$, since $p(\nu) = 0$ for all $\nu > \nu_{max}$. The moments of the conditional probability distributions defined in Eqs. (4) and (5) are obtained by replacing $p(\nu)$ in Eq. (6) with $p^c(\nu)$ and $p^{c_m}(\nu)$, respectively. For $M_1^c$ and $M_1^{c_m}$, the summation starts at $\nu = 1$ and $\nu = m \geq 2$, respectively.

If only ICs between $a$ and $b$ are considered [54], the corresponding conditional moments $M_k^{a-b}$ are given by Eq. (7):

$$M_k^{a-b} = \frac{\sum_{\nu=a}^{b} \nu^k p(\nu)}{F_{a-b}} \quad (7)$$

Table 2 shows a comparison of the notation used in ICRU report 98 [38] and in this work, along with a condensed summary of the different notations found in the literature. For better legibility, only the dependence on the ICS is shown explicitly. A more detailed account listing individual articles can be found in Supplementary Table 1. Supplementary Table 1 also includes additional parameters such as radiation quality, target volume diameter, and impact parameter.

As Table 2 shows, the nanodosimetric literature uses very heterogenous notation. Two aspects are particularly cumbersome and may lead to misinterpretation: First, that the same symbols were used for conditional and unconditional distributions. Second, in the pioneering work of Grosswendt [43,77–80], the symbol $F_k$ was introduced to designate the probability of $k$ or more ionizations. Since the ICS and the minimum number of ionizations in an ICS are integers, they could be used in handier notation as indices of probability values instead of arguments of a probability mass function. However, to remain consistent with common practice in statistical literature, as ICRU report 98 [38] aspires, $F_k$ would have to encode $F(k)$, that is, the probability of obtaining a maximum of $k$ ionizations.

Therefore, this review takes a pragmatic approach: It uses the notation of probability mass functions from ICRU report 98 [38]. The probability of an ICS $\nu$ is denoted $p(\nu)$, and the probability if an ICS exceeding $k$ is denoted $F^*(k)$. Since the symbol $F_k$ is commonly used in nanodosimetric literature to designate the cumulative probability of at least $k$ ionizations, the definition contained in Eq. (2) is employed as a short-hand notation for facilitating comparison with the literature.

[2] It is therefore distinct from the $p(\nu)$ mentioned in the glossary of [38].

Table 2: Notation used in ICRU report 98 [38], the present work, and the nanodosimetric literature for ICS probabilities and the first moments of the ICS distribution. References are only given for symbols used in a single article. The details on which publications used which notation can be found in Supplementary Table 1. Work on experimental nanodosimetry that does not attempt to model biological consequences has been omitted from the overview.

| Quantity | ICRU Rep. 98 | This work | Literature |
|---|---|---|---|
| Ionization cluster size (ICS) | $\nu$ | $\nu$ | $\nu$, rarely: $n_{ion}$, $j$ [36] |
| Diameter of target volume | $d$ | $d$ | $D$, rarely: $d$ [81], $d_s$ [76] |
| Impact parameter | - | $b$ | $d$, rarely: $r$ |
| Probability of ICS $\nu$ | | | |
| unconditional | $p(\nu)$ | $p(\nu)$ | $P_\nu$, $P(\nu)$, rarely: $p(\nu)$, $f(\nu)$ |
| conditional on $\nu \geq 1$ | $p^c(\nu)$ | $p^c(\nu)$ | $P_\nu$, $P(\nu)$, rarely: $P^*(\nu)$ [54] |
| conditional on $\nu \geq 2$ | - | $p^{c2}(\nu)$ | $p_\nu$ [82] |
| Probability of ICS $\nu \leq k$ | $F(k)$ | $F(k)$ | - |
| Probability of ICS $\nu \geq k$ | | | |
| unconditional | $F^*(k-1)$ | $F^*(k-1)$, $F_k$ | $F_k$, rarely: $F(k)$ [83] |
| conditional on $\nu \geq 1$ | - | $F^{c*}(k-1)$, $F_k^c$ | $F_k$ |
| conditional on $\nu \geq 2$ | - | $F^{c_2*}(k-1)$, $F_k^{c_2}$ | $F_k^{c2}$, rarely: $f_k$ [82] |
| Probability of $a \leq \nu \leq b$ | | | |
| unconditional | - | $F_{a-b}$ | - |
| conditional on $\nu \geq 1$ | - | $F_{a-b}^c$ | $F_{a-b}$ [69] |
| Mean ICS | | | |
| unconditional | $M_1$ | $M_1$ | $M_1$ |
| conditional on $\nu \geq 1$ | - | $M_1^c$ | $M_1$, rarely: $M_1^*$ [54] |
| conditional on $\nu \geq 2$ | - | $M_1^{c2}$ | $M_1^{c2}$, rarely: $m_1$ [82] |

### *2.2 Parameters determining the ICS distribution*

The probability distribution $p(\nu)$ depends on radiation quality, the material, size and shape of the target, the material in the surrounding medium, and the impact parameter or range of impact parameters considered in simulations or measurements. Different ICS distributions are obtained with different target sizes and shapes, materials, and irradiation geometries (impact parameter ranges).

Therefore, strictly speaking, the unconditional ICS distribution $p(\nu)$ is still a conditional probability distribution, namely, contingent on the parameters considered in a simulation or pertinent to the measurement setup in an experiment. Many publications have accounted for this dependency by using notation from statistical literature for conditional probabilities such as, e.g., $p(\nu|\theta)$, where $\theta$ denotes one or more parameters such as target size or irradiation geometry (see Supplementary Table 1).

An important implication is that models may conceptually use the same nanodosimetric probabilities, which have very different values if they derive from substantially different probability distributions pertaining to different experimental or simulation conditions. This raises the question, under which conditions do we obtain (almost) the same ICS distribution for the same radiation quality in two different experiments or, e.g., an experiment in propane gas and a simulation in liquid water? The answer is that this occurs when target size and density satisfy the scaling relation illustrated in Section 2.2.1 and discussed in Appendix A.

The scaling relation requires that the ICS distributions obtained under different conditions have the same mean value $M_1$. This implies that the ICS distribution is uniquely determined once $M_1$ is known. This topic is discussed in Section 2.2.1, along with attempts to derive this functional dependence for the first few elements of the ICS distribution.

However, it should be noted that the theoretical foundations of both the scaling principle and the functional relations between the first elements of the ICS distribution and the mean ICS is based on an irradiation geometry in which the primary particle path intersects the target volume.

### *2.2.1 Dependence on target size*

The nanodosimeters developed so far determine the ICS by counting the number of ionizations produced in a dilute gas. Different approaches have been taken to define the actual target volume in the gas [2], and the equivalence of the ICS distributions from these macroscopic gas targets to targets in tissue or water of nanometric dimensions relies on a scaling relation. This scaling relation (discussed in more detail in Appendix A) accounts for density and material properties and sometimes also the detection efficiency [15,84,43,85,77]. The validity of the scaling between materials was experimentally confirmed for several operating gases in a nanodosimeter [48,86].

As an illustration of the scaling principle, Fig. 2 presents IC distributions for an alpha particle of 4.6 MeV energy traveling in water (boxes), nitrogen (circles), and propane (diamonds) traversing a cylindrical target of equal diameter and height centrally. The target in water has a diameter $d$ of 4 nm and a mass density $\rho$ of 1 g/cm$^3$, corresponding to $d\rho = 0.4$ μg /cm$^2$. Similar IC distributions are obtained with $d\rho$ values of 0.58 μg /cm$^2$ for nitrogen and 0.32 μg /cm$^2$ for propane. [3]

A closer look at Fig. 2 reveals that the mean values of the three distributions are different. In fact, the mean is about 7.7 for water, while it is about 8.6 for propane and 8.4 for nitrogen. The latter values are about 13 % and 9 % higher than the mean for water, respectively. This is because the scaling relation (see Appendix A) relies on the ratio of the total ionization cross sections of the two materials. For this ratio, 10 % may be assumed to be the order of magnitude of the associated uncertainty. In fact, Bug et al. [26] suggested a 16 % correction for the propane cross sections used in the PTra track structure code in subsequent work. This correction would reduce the area mass density used in the propane simulations and bring the propane results in closer agreement with the water data.

[3] Note that we adopt here the notation of ICRU Report 98 for the target diameter [38]. In nanodosimetry literature, the target diameter has often been designated by $D$ (Supplementary Table 1).

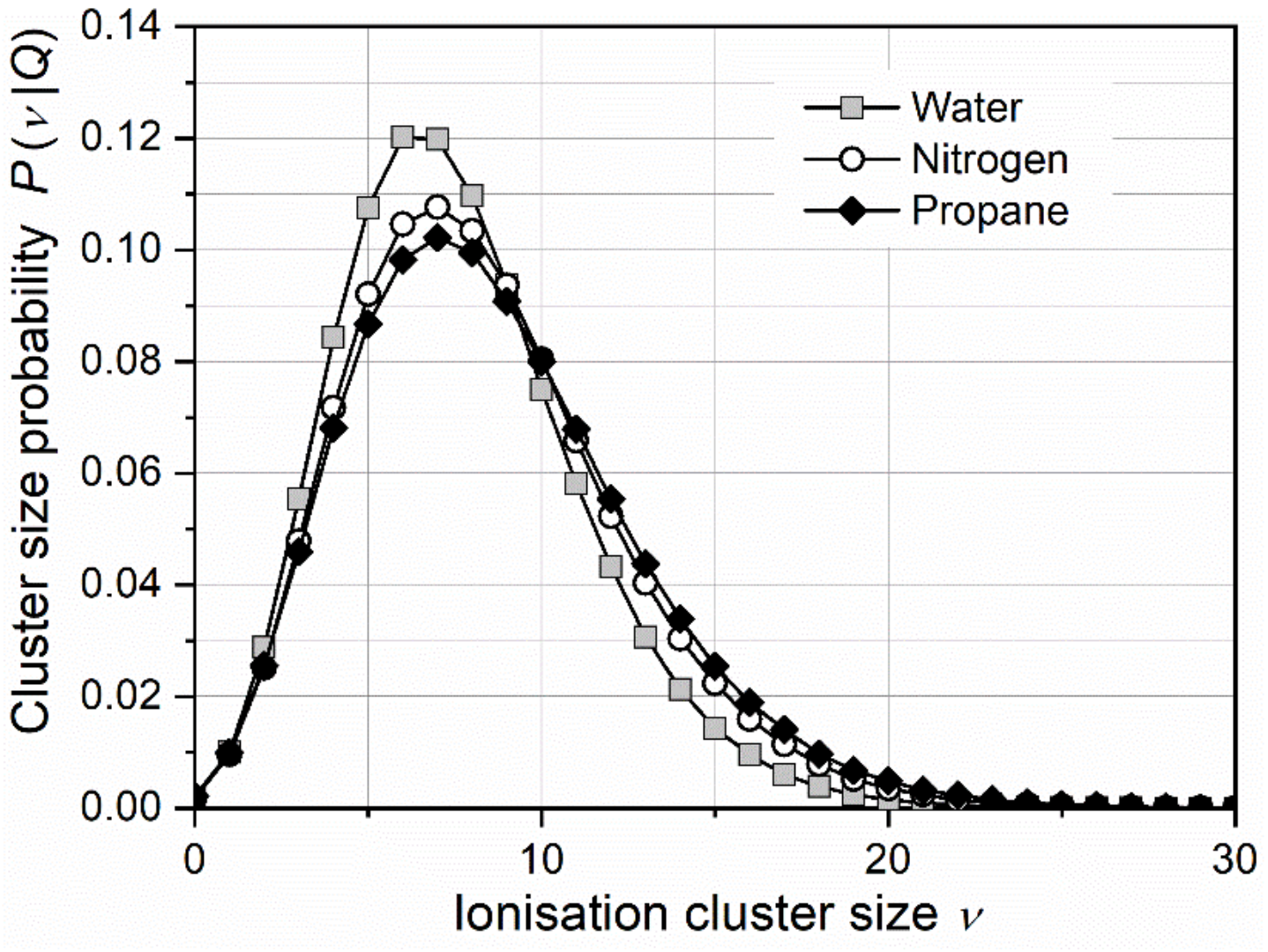

**Fig. 2:** Simulated ionization cluster size distributions for a 4.6 MeV alpha particle passing through the center of a cylindrical target in a direction perpendicular to the cylinder's axis. In the case of liquid water, the cylinder's height and diameter $d$ were set to 4 nm. To obtain the same mean ionization cluster size in the sensitive volume, the mass per unit area $d\rho$ of nitrogen and propane was 0.58 µg /cm$^2$ and 0.32 µg /cm$^2$, respectively, where $\rho$ denotes the mass density.[4]

Nevertheless, the primary purpose of the scaling relation is to convert the target diameter in a nanodosimeter to an equivalent target size in water [86]. The target volume in a nanodosimeter is not an ideal cylinder as was assumed in the simulations producing the data shown in Fig. 2. Rather, it is determined by the spatial distribution of collection efficiency of the produced ions. Different approaches have been taken to determine the size of the sensitive volume [86–88]. Therefore, a relative uncertainty of 10 % may be assumed for the equivalent target size in water.

The diameters of target volumes in liquid water simulated by existing nanodosimeters range between approximately 2 nm and 20 nm. The resulting measured ICS distributions obtained with primary particles intersecting the target volume are very different [56]. Their mean cluster sizes are essentially proportional to the site diameter and the instrumental detection efficiency in agreement with the scaling relation. However, as was demonstrated by Conte et al. [56], when the probabilities $F^*(k)$ measured by different nanodosimeters are plotted as a function of the corresponding mean IC size $M_1$, they fall on the same curves for the same $k$. This is illustrated by Supplementary Fig. 2, which shows the case of $F^*(1) \equiv F_2$. As discussed in Section 3.6, these curves can even be related to radiobiological cross-sections [60]. However, it must be noted that these universal relations apply to measurements in which the primary ions pass the target volume centrally.

The unique dependence of the complementary cumulative cluster probabilities $F^*(k)$ on $M_1$, as presented in [56,60], corresponds to data with different radiation qualities and different target sizes. To disentangle the effects of these two parameters, Conte et al. studied the dependence of the probabilities $F^*(k)$ on $M_1$ for given target sizes in later work [89]. They derived theoretical model functions based on a development of work by De Nardo et al. [90]. These model functions can be formally written as Eq. (8)[5]:

$$F^*(k|d) = 1 - \widetilde{\Pi}_k(M_1|d)e^{-C_1(d)M_1} \tag{8}$$

[4] Fig. 2 is reprinted from Radiation Measurements 46 (9), H. Nettelbeck and H. Rabus, Nanodosimetry: The missing link between radiobiology and radiation physics? 893-897, Copyright (2011), with permission from Elsevier.

Here, $d$ denotes the diameter of the target sphere, $\widetilde{\Pi}_k$ is a polynomial of degree $k$ in the independent variable $M_1$, and $C_1$ is one of the parameters of the function. The coefficient of the zeroth-order term of $\widetilde{\Pi}_k$ is unity for all $k$, and the other coefficients and $C_1$ depend on the target diameter but not on radiation quality [89]. However, the non-trivial coefficients of the polynomials depend on only two model parameters, $C_2(d)$ and $\tilde{C}_3(d)$ (see Table 3)[6].

The coefficients $C_1$, $C_2$, and $\tilde{C}_3$ were determined by fitting the model to the results of track structure simulations for protons, helium and carbon ions of different energies travelling in propane. Propane is the operating gas of the "StarTrack" nanodosimeter developed by their group. ICS distributions were obtained for the central passage of spherical target volumes of various sizes, equivalent to spheres in water with diameters ranging from 0.5 nm to 10 nm. As illustrated by Supplementary Fig. 3, the fit curves according to Eq. (8) reproduce the data well. Data from different ion types follow the same trend for the same value of $k$.

Table 3: Explicit expressions of the polynomials that appear in Eqs. (8) and (9). The coefficients $C_2$, $\tilde{C}_3$, $C_3$ and $C_4$ are model parameters and depend on the site diameter $d$.[6]

| Polynomials in [89] | Polynomials in [91] |
|---|---|
| $\widetilde{\Pi}_0(x) = 1$ | $\Pi_0(x) = 1$ |
| $\widetilde{\Pi}_1(x) = 1 + C_2x$ | $\Pi_1(x) = C_2x$ |
| $\widetilde{\Pi}_2(x) = 1 + \tilde{C}_3x + \frac{C_2^2}{2}x^2$ | $\Pi_2(x) = C_3x + \frac{C_2^2}{2!}x^2$ |
| - | $\Pi_3(x) = C_4x + C_2C_3x^2 + \frac{C_2^3}{3!}x^3$ |

By definition, the complementary cumulative frequencies $F_1$, $F_2$ and $F_3$ depend only on the probabilities of ICS values between 0 and 2. Therefore, further work [91] studied the dependence of the probabilities $p(\nu)$ on $M_1$ for $\nu \in \{0,1,2,3\}$, i.e. also including the parameter $F^*(3)$. The study used simulations of the tracks of protons, $^4$He, $^{12}$C, and $^{20}$Ne ions with energies per mass ranging from 1 MeV/u to 330 MeV/u in propane gas. The target size considered corresponds to a sphere with a diameter of 1 nm in liquid water. Based on the same theoretical assumptions as before, the expressions for the functional dependence of $p(\nu)$ on $M_1$ for $\nu \geq 1$ have the form given in Eq. (9)[7].

$$p(\nu|d) = \Pi_\nu(M_1)e^{-C_1(d)M_1} \tag{9}$$

[5] Note that Eq. (8) is an abstraction made by the author of this paper. Conte et al. [89] provided explicit expressions of Eq. (8) for each considered $k$ and used the notation $F_k$ for the complementary cumulative probability.

[6] The third model parameter is defined differently in [89] and [91]. Therefore, we use the notation $\tilde{C}_3$ here. The connection between the two definitions is $\tilde{C}_3 = C_2 + C_3$.

[7] Note that Eq. (9) is an abstraction made by the present author. Conte et al. [91] gave explicit expressions for Eq. (9) for each considered $\nu$.

Here, $\Pi_\nu$ is a polynomial of degree $\nu$ in the independent variable $M_1$, the coefficients of which depend on three parameters $C_2(d)$, $C_3(d)$, and $C_4(d)$ (see second column in Table 3). The polynomials $\Pi_\nu$ are related to the polynomials in Eq. (8) by $\widetilde{\Pi}_k(M_1) = \sum_{\nu=0}^{k} \Pi_\nu(M_1)$.

### *2.2.2 Dependence on the impact parameter*

In nanodosimetric experiments, the lateral extension of the primary particle detector is often comparable in size to the diameter of the target [2]. Under these conditions, the impact parameters of different primary particles hitting the detector are within a small interval around the mean value corresponding to a hit in the center of the detector. The resulting measured frequency distribution is an estimate of the probability distribution $p(\nu)$ for to this specific impact parameter.

The dependence of $p(\nu)$ on the impact parameter can be studied by conducting measurements with different lateral offset between primary particle detector and target volume [90,92]. Alternatively, on can use a significantly larger primary particle detector with position sensitive detection capabilities [44,93,94]. Using either approach, it was found that the conditional ICS probabilities are almost invariant in a certain range of impact parameters [92,94]. Using a larger primary particle detector also enables measuring in "broad beam" geometry, which determines the $p(\nu)$ corresponding to a range of impact parameters [94].

Fig. 3 illustrates this, showing ICS distributions produced in the sensitive volume of the "Ion Counter" nanodosimeter by 88 MeV carbon ions. The red symbols correspond to carbon ions that pass the target volume centrally. The blue symbols represent carbon ions that still intersect the target volume. The other colored symbols represent carbon ions that pass outside the target volume, producing ICs only by secondary electrons. It can be seen that, for this radiation quality, the ICS distribution shows a maximum at nine and five ionizations for the red and blue datapoints, respectively. When the ion misses the target volume, the maximum IC is found at $\nu = 0$. Note that the legend shows water-equivalent values of the product of impact parameter (denoted by $d$) and the mass density $\rho$. Thus 1 μg/cm$^2$ corresponds to 10 nm.

The large black symbols correspond to ICs measured in coincidence with the detection of a carbon ion anywhere across the primary particle detector. This detector had a sensitive surface of 30 mm×3 mm, corresponding to impact parameters up to 17 nm in liquid water. For comparison, the equivalent cylindrical target in water had a diameter of 2.3 nm and a height of 6.1 nm [94].

A prominent feature of the broad beam data is the high peak at $\nu = 0$, which occurs because the probability of ionizations in a target decreases with increasing impact parameter values. Thus, the probability of no ionization increases, and ICS distributions have an increasing probability of $\nu = 0$ as the range of impact parameters increases [95]. Conversely, conditional ICS distributions were found to be less dependent on range of impact parameters and to converge quickly with increasing impact parameter (Fig. 18 in [94]).

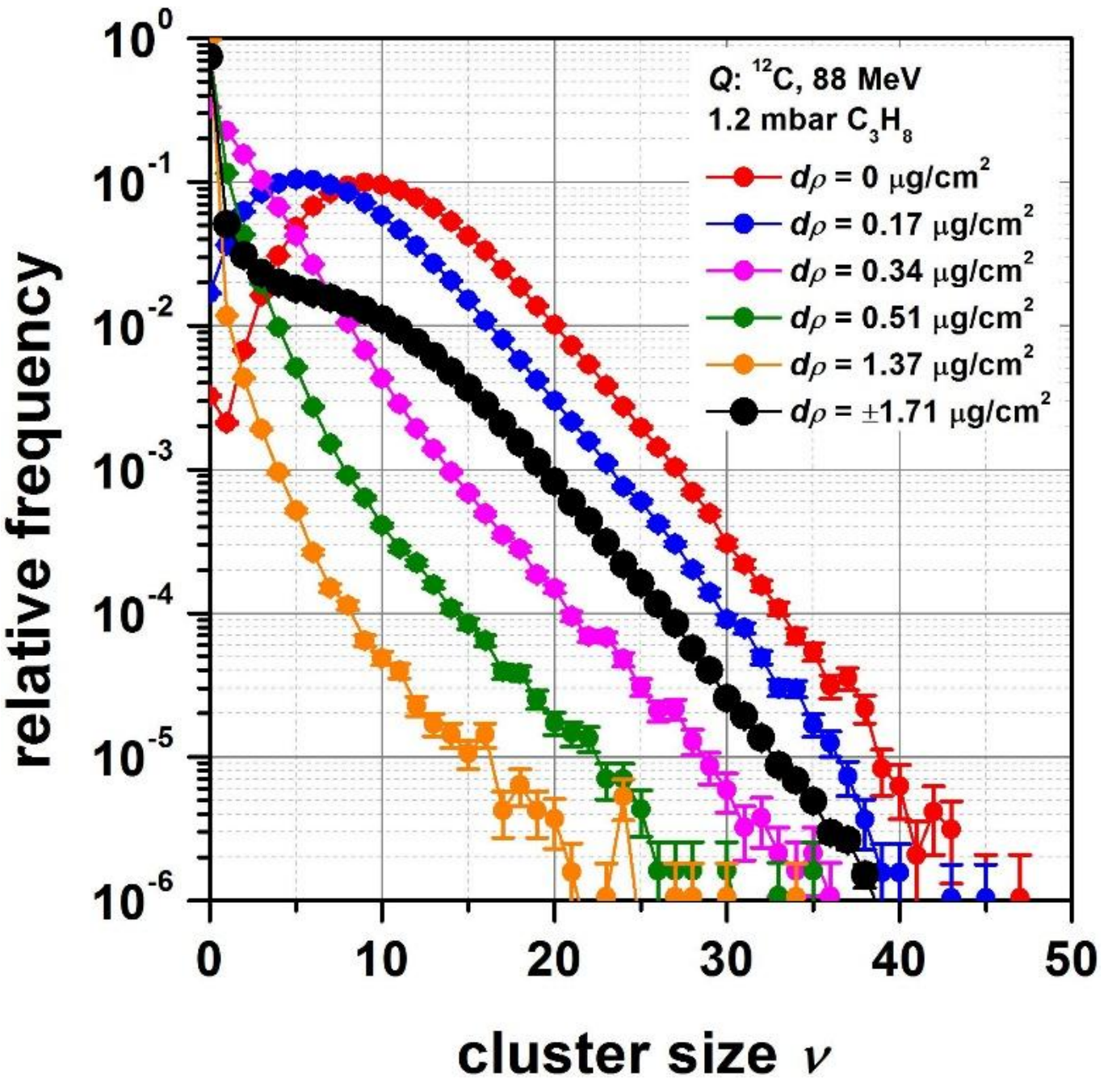

**Fig. 3**: Measured ionization cluster size distributions produced by carbon ions with an energy of 88 MeV in the Ion Counter nanodosimeter operated with propane gas at a pressure of 1.2 mbar. The primary particles were detected with an extended position-sensitive detector (for details see [94]). The colored symbols correspond to measurements at different values of impact parameter, indicated in the legend as $d\rho$, where $d$ denotes the impact parameter and $\rho$ denotes the mass density. The black symbols correspond to the results for the whole range of impact parameters covered by the primary particle detector. [8]

The radiation effect models presented later all consider ICS distributions (determined by simulations) in liquid water targets with comparable diameters of a few nanometers or in gas-phase targets that produce equivalent ICS distributions (in experiments or simulations) to such nanometric targets in water. The main difference between the data used in the models is the range of impact parameters considered when determining the ICS distributions.

The range of impact parameters that can be studied experimentally is generally limited by constraints of the experimental setup, such as the dimensions of the vacuum chamber or the size of the available detectors. Track structure simulations are not faced with these practical limitations. However, many simulation studies have used configurations analogous to those of real or potential experimental setups: a beam of particles directed at a single target. In such setups, increasing the range of impact parameters can significantly reduce counting statistics.

Simulating a single target with a centered beam is essentially equivalent to simulating a pencil beam with targets in a cylindrical region whose diameter is equal to the intended beam diameter [57]. However, if the targets are implemented explicitly in the simulation, as in [57,96], no efficiency is gained. Substantial variance reduction can be achieved by performing simulations that consider only (virtual) targets located in the vicinity of energy transfer points with ionizations. “Virtual” means that the target volumes are not predefined in the simulation geometry but are rather derived when tallying the energy transfer points in the simulated tracks. In contrast to approaches that utilize both an extended beam and a spatial distribution of targets [54,58], track-centered approaches allow investigating the influence of different impact parameters on the final ICS distribution.

[8] Fig. 3 is reprinted from Physics in Medicine and Biology 62, G. Hilgers et al., Measurement of track structure parameters of low and medium energy helium and carbon ions in nanometric volumes, 7569 (2017) under the CC BY-3.0 license (http://creativecommons.org/licenses/by/3.0).

Track-centered scoring was employed by Braunroth et al. [64,97–99] to study the variation of ICS distributions with impact parameter over the full range of impact parameters generating ICs in the target. (Earlier work by Rabus et al. [100,101] employed a nearly identical sampling approach but only considered impact parameters up to 100 nm.) In this study, they used a multitude of cylindrical target volumes for impact parameters up to about 25 nm (track core). At larger impact parameters, sectors of concentric cylinder shells were used that had the same volume as the cylinders used in the track core. In later work, Ngcezu et al. used polyhedrons representing the unit cell of a face-centered cubic Bravais lattice as approximations of spherical targets and spherical symmetry [68]. More recent studies employed the associated volume clustering approach with spherical targets [102,103,76].

Results from the study of Braunroth et al. [64,97–99] are reproduced in Supplementary Fig. 4. This figure illustrates the variation of the probabilities of ICs with at least 1, 2, 3, and 4 ionizations as a function of the impact parameter for protons with energies of 1 MeV and 30 MeV. The presentation of data is analogous to the plotting style used in microdosimetry, i.e., the area under the plotted curve is proportional to the contribution to the integration over all radial distances. It was considered that the number of targets at a given radial distance from the ion trajectory grows linearly with the impact parameter. Or, in the target-centered view presented in Fig. 1, that the probability distribution of impact parameter is proportional to the impact parameter.

It can be seen from Supplementary Fig. 4 that the radial distances at which a primary particle may contribute to the probability of forming an IC in a target extent over several 100 nm and even into the µm range (depending on primary particle energy). This implies that realizing broad beam conditions may require scoring at larger impact parameters than those used in experiments or some simulation studies, in which the beam cross-section matched the target cross-sectional area [70,77] or the beam diameter was about double [104] or three to five times [94,105,106] the target's diameter.

### 2.2.3 *Dependence on radiation quality*

Nanodosimetric probability distributions are studied because they relate to radiation quality. For a given ion type, the radiation quality varies with particle energy. Several studies have analyzed the energy dependence of nanodosimetric parameters [58,91,96]. Fig. 4 shows a comparison of results for protons and carbon ions, with the data presented on a linear *y*-axis in the left panels and a logarithmic *y*-axis in the right panels.

The squares and blue line pertain to the results reported by Alexander et al. [96] for protons with energies between 1 MeV and 100 MeV and carbon ions with reduced energies between 10 MeV/u and 400 MeV/u. The particles travelled in a water cube of 20 µm side starting from the center. $10^6$ cylindrical target volumes, each with a diameter of 2.3 nm and a height of 3.4 nm, were randomly placed in the scoring volume. The scoring volume was a cuboid with lateral dimensions slightly larger than the largest distance of an ionization from the primary particle trajectory. This means that the whole range of relevant impact parameters was covered. To avoid bias from build-up effects, the scoring volume began 1 µm downstream from the starting point of the primary particles. Only targets containing at least one ionization were tallied, so that the conditional ICS distribution was determined.

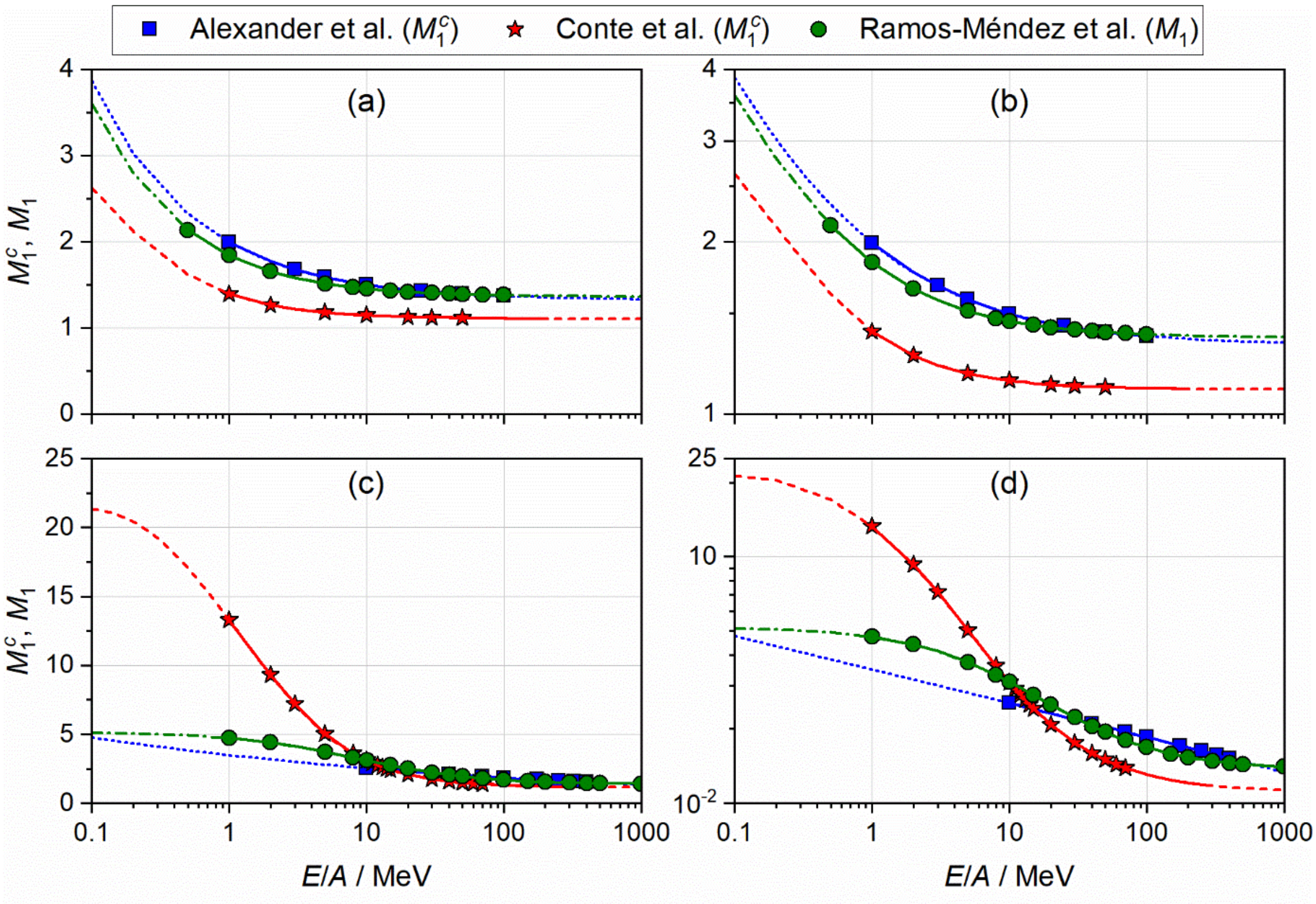

**Fig. 4**: Comparison of the best-fit interpolation curves for the mean ICS as a function of the reduced projectile energy of (a) and (b) protons and (c) and (d) carbon ions reported by different authors [58,91,96]. The solid lines correspond to the energy range of the data used to determine the interpolation functions; dashed lines indicate extrapolations to lower and higher energies. The simulations producing the data to which the model curves were fitted (symbols) have been obtained for different impact parameter distributions (see text).

The energy dependence of the mean ICS was fitted with a modified power function (Eq. (10)):

$$M_1^c(E) = a_p \left(\frac{E/A}{\mathrm{MeV}}\right)^{b_p} + c_p \qquad (10)$$

The fit parameters $a_p$, $b_p$, and $c_p$ depend on the particle type; $E$ is the projectile energy, and $A$ is the number of nucleons.[9]

The asterisks and red lines in Fig. 4 correspond to the results that Conte et al. [91] obtained for spherical targets with a diameter of 1 nm and for protons, helium, carbon, and neon ions with reduced energies ranging from 1 MeV/u to 200 MeV/u (300 MeV/u for carbon). The primary particles traversed the target volume centrally. The unconditional ICS distribution and its mean value $M_1$ were determined in the simulations. The energy dependence of $M_1$ was fitted by the empirical model function given in Eq. (11).

[9] Note that the formula given in [96] does not contain the division by $A$. However, when the formula given in [96] is evaluated using the values of the best-fit parameters stated therein, the calculated $M_1$ values are smaller than those of the data points shown in Fig. 11 of [96] by a factor of about 4. The function values calculated with Eq. (10) agree with the data points.

$$M_1(Q) = a\, Z_{\mathrm{eff}}^2 \left(\frac{E/A}{\mathrm{MeV}}\right)^{b+c\log\frac{E}{A}} \tag{11}$$

Here, the argument $Q$ indicates the radiation quality, i.e. particle type and energy. The fit parameters $a$, $b$, and $c$ are independent of particle type. $E$ is the projectile energy, $A$ is its nucleon number, and $Z_{\mathrm{eff}}$ is the effective projectile charge according to the empirical Barkas formula [107], where $Z$ is the number of protons in the projectile's nucleus and $\beta$ is its relativistic speed:

$$Z_{\mathrm{eff}} = Z \times \left(1 - e^{-125\beta Z^{-\frac{2}{3}}}\right) \tag{12}$$

Conte et al. also determined the value of the parameter $C_1$ of the model functions given in Eq. (8). This value was used here to calculate complementary cumulative probability $F_0^*$ and to obtain $M_1^c$ by dividing the model curves by $F_0^*$ according to Eq. (11). The data marked in Fig. 4 by asterisks and red lines are the resulting conditional mean ICS values.

The circles and green lines in Fig. 4 indicate the results from Ramos-Méndez et al. [58], who used target volumes contained within a cylinder that represented a segment of a chromatin fiber. The nanometric targets were cylinders with a diameter of 2.3 nm and a height of 3.4 nm, corresponding to a DNA segment of 10 base pairs. The scoring volume had a diameter of 30.4 nm and a length of 161 nm and contained 1800 randomly oriented and distributed target cylinders. Simulations were performed for protons with energies between 0.5 MeV and 100 MeV, $^{4}$He ions with energies per nucleon between 1 MeV and 100 MeV, and $^{7}$Li, $^{9}$Be, $^{11}$B, $^{12}$C, $^{14}$N, and $^{16}$O ions with energies per nucleon between 1 MeV and 1000 MeV.

Three different irradiation geometries were considered. One was a source located on the surface of the scoring volume emitting particles inward, with isotropic distribution in 2π. The second source was located on a spherical surface, which fully contained the scoring volume and emitted particles towards the scoring volume. The third source was a pencil beam directed along the rotation axis of the scoring region cylinder. For the third source, the range of impact parameters is between 0 nm and 15.2 nm. For the other two geometries, impact parameters up to about 163 nm are possible. However, as illustrated in Supplementary Fig. 5, the impact parameter distribution is far from uniform. This figure shows a comparison of the estimated impact parameter distributions for the three geometries and that used by Faddegon et al. [37].

Each curve in Supplementary Fig. 5(a) shows the probability distribution of impact parameter for the considered irradiation geometry. In the track-centered view, the proportion of the area under the curve in an interval on the *x*-axis is the probability of finding a nanometric target at this radial distance from the ion trajectory. In the target-centered view, this proportion is the probability of an ion beam passing the target with an impact parameter in this interval. These probabilities can be translated into local fluence profiles, which are shown in Supplementary Fig. 5(b). The relevant quantities for calculating the radial integral of the impact parameter dependence shown in Supplementary Fig. 4 are these fluence profiles.

For geometry 3, the probability increases linearly with the impact parameter up to the maximum possible value of 15.2 nm. For the other geometries, almost identical probabilities are obtained. The results for $M_1$ obtained with the first two geometries are also almost identical [58] and were fitted by the phenomenological function reproduced here in Eq. (11)[10]:

$$M_1(Q) = \frac{P_1}{1 + \exp\left(\frac{\ln(P_3) - \ln(E/A/\mathrm{MeV})}{P_4}\right)} + P_2 \tag{13}$$

Fig. 4 shows an agreement within about 10 % between the simulation results of Ramos-Méndez et al. and Alexander et al., even though the data obtained by Ramos-Méndez et al. (circles and green lines) are for unconditional ICS distributions. This suggests that $F^*(0)$ is close to one for the impact parameter range and radiation qualities considered in [58]. Larger discrepancies can be seen in the energy ranges where at least one of the two models is an extrapolation. This is presumably due to the phenomenological nature of the fit functions used.

The results of Conte et al. for protons are systematically lower by about 25 % than the result of the other papers (Fig. 4(a) and (b)). The likely explanation is that larger clusters occurring at the track ends of secondary electrons are included when an average over a range of impact parameters is performed, while these are absent for a pencil beam of protons which passes the target centrally.

The comparison for carbon ions shows also smaller values for a pencil beam at larger particle energies, which amount to about 20 % at 50 MeV energy per nucleon. Below about 10 MeV per nucleon, larger $M_1^c$ values are found for the pencil beam. At 1 MeV per nucleon, this discrepancy is by a factor of about 4.5. The reason for this is that at these energies, the interactions of the primary particle produce larger clusters in targets they intersect than secondary electrons do in targets in the penumbra [108]. Therefore, averaging over a larger impact parameter range reduces the mean ICS in targets that receive ionizations.

### *2.3 Average nanodosimetric quantities in mixed radiation fields*

The majority of nanodosimetric experiments were conducted under conditions of a well-defined ion beam quality, i.e., ion type, charge state, and kinetic energy. In a recent proof-of-principle investigation, a nanodosimeter was operated behind an absorber hit by a clinical carbon ion beam [109], where data analysis becomes more involved owing to the complex mixed radiation field [110]. However, such mixed fields are typical of practical applications of ion beams in the domain of radiotherapy.

The setup in aforementioned experiment was similar to that used in earlier work [94] to produce the data shown in Fig. 3. That is, an extended primary particle detector was used with a single target volume to cover a range of impact parameters up to about 17 nm. The measured ICS distribution corresponded to a superposition of those produced by primary $^{12}$C ions of different energies and those produced by secondary nuclear fragments [110].

However, in ion beam irradiation, the ICs produced relate to the entire range of impact parameters, particle types and energies. The results discussed in Section 2.2.3 show that there is a complex interplay between impact parameter distribution and radiation quality. Therefore, averaging nanodosimetric distributions over energy and impact parameter ranges is needed.

---

[10] Note that the formula given in [58] does not contain the division by *A* and the unit MeV. However, when the formula given in [58] is evaluated using the values of the best-fit parameters state therein, the calculated $M_1$ values for $^{12}$C are 40 % smaller than those of the data points shown in Fig. 3 of [58]. The function values calculated with Eq. (11) agree with the data points.

Two approaches have been used to obtain average nanodosimetric quantities in a mixed radiation field. Casiraghi and Schulte [54] performed simulations to determine these quantities in voxels of 5 mm side in spread-out Bragg peaks (SOBPs). The SOBPs were formed using 10 initially monoenergetic beams of either protons or carbon ions. The proton beams were degraded by range shifters, and the carbon ion beams were degraded using ripple filters. The beams had a two-dimensional Gaussian profile with a standard deviation of 3 nm. ICs were tallied in five adjacent voxels along the beam direction, and scoring regions were placed at three different positions in each voxel. Each scoring region consisted of a regular two-dimensional array of $10^4 \times 10^4$ cylinders, each with a height and diameter of 500 nm. Each cylinder contained $10^4$ cylindrical target volumes, which were randomly placed and oriented and had a diameter of 2 nm and a height of 16 nm. IC formation was tallied in these nanometric cylinders.

Averaged nanodosimetric IC distributions were obtained for each voxel and beam. From these distributions, the following parameters were determined: mean values $M_1$ and $M_1^c$, probabilities $F_{2-3}$ and $F_{4-10}$ of small and large clusters, and modified conditional mean ICS $M_1^{2-10}$ according to Eq. (7).[11] The latter was named the biologically effective mean ICS.

The second approach constructs the macroscopic average starting from nanodosimetric ICS distributions of monoenergetic particle tracks. In the context of the BioQuaRT (Biologically weighted quantities for radiotherapy) project, synergistic effects of several ions traversing volumes of micrometric dimensions corresponding to biological cells or cell nuclei on averages of ICS distributions were studied [52]. This development built on the parameterizations of the energy dependence of nanometric probabilities for protons and carbon ions from the work of Alexander et al. [96] discussed in Section 2.2.3. Ramos-Méndez et al. have generalized this approach to arbitrary volumes, such as voxels used for treatment planning, and created a comprehensive database for carbon ions and their most abundant nuclear fragments [58]. They also proposed a scheme for weighted superposition in mixed radiation fields. Dai et al. [65] built on the database and definitions of average quantities of Ramos-Méndez et al. [58], though they used slightly different notation.

In essence, for a nanodosimetric quantity $Q$ (named ionization detail in [58]), the fluence-weighted[12] and dose-weighted averages, $\bar{Q}_\Phi$ and $\bar{Q}_D$, are conceptually defined by Eqs. (14) and (15), respectively.

$$\bar{Q}_\Phi = \frac{\sum_p \int Q_p(E) \Phi_E^p(E) dE}{\sum_p \int \Phi_E^p(E) dE} \tag{14}$$

$$\bar{Q}_D = \frac{\sum_p \int Q_p(E) \Phi_E^p(E) S_p(E) dE}{\sum_p \int \Phi_E^p(E) S_p(E) dE} \tag{15}$$

In Eqs. (14) and (15), $p$ denotes the type of (charged) particle and $Q_p(E)$ is the value of the nanodosimetric quantity produced in a considered target by a particle of type $p$ and energy $E$, $\Phi_E^p$ is the spectral particle fluence of particles of type $p$ in the volume considered for averaging, and $S_p$ is the stopping power of the material in the voxel for a particle of type $p$ and energy $E$.

[11] Note that Casiraghi and Schulte used the notation $M_1^*$ for the conditional mean ICS and $Y_{SC}$ for small and $Y_{LC}$ for small and large clusters, respectively, which they called yield instead of probability.
[12] Alexander et al. [52] used in their Eq. (7) the arithmetic mean of tracks for the purpose of fluence averaging. However, this is not exactly the same as fluence averaging.

The notation used by Ramos-Méndez et al. [58] and Dai et al. [65] differs from the one used here in that they replaced the integrals by sums over the bins of fluence or dose histograms as would be obtained from simulations. Additionally, Ramos-Mendez et al. [58] used track length and imparted energy (termed "deposited energy") instead of fluence and dose. Since fluence is the ratio of track length to volume and dose is the ratio of energy imparted to mass, their expressions are equivalent to those used here (and in [65]). Ramos-Méndez et al. omitted the summation over particle type in the denominator in their Eqs. (1) and (2) and did not explicitly show the dependence of the contributions to accumulated track length and energy imparted on the particle type in their formulae. However, their notion that the increments of these quantities relate to successive interactions implies that they depend on particle type and should be summed over particles as well.

The practical evaluation of Eqs. (14) and (15) follows a three-step procedure. In the first step, the nanodosimetric quantities, $Q_p(E)$, are determined by performing track structure simulations for a set of relevant particle types and energies. Subsequently, interpolating functions of the energy dependence of $Q_p(E)$ are determined for each distinct particle type. In the final step, volume-averaged fluences of pertinent particles and energies are obtained through condensed-history simulations. Subsequently, the integrals are estimated through the summation over energy bins. This evaluation is performed as averages over a given volume.

The volume averaging was done by Alexander et al. [52] in micrometric volumes representing a cell nucleus. Ramos-Méndez et al. [58] implicitly considered cuboid-shaped voxels as used in treatment planning. In contrast, Dai et al. considered cylinder shells with outer dimensions corresponding to a decrease of the absorbed dose to 50 %, 10 %, 1 %, and 0.1 % of that at the beam axis of a circular beam with a diameter of 4 mm [65].

It is worth noting that the nanodosimetric quantities considered in the aforementioned publications [58,65,96] are cumulative probabilities and mean ICS, which depend linearly on the ICS probabilities. Therefore, fluence and dose averaging can also be performed on the ICS probabilities $p(\nu)$ to obtain average ICS distributions $\bar{p}_\Phi(\nu)$ and $\bar{p}_D(\nu)$ for each voxel. All other nanodosimetric quantities of interest can then be calculated using these average ICS distributions by replacing $p(\nu)$ with the voxel-averaged ICS probabilities in the definitions in Eqs. (1) to (7).

### *2.4 Extension of the conceptual framework of nanodosimetry*

The nanodosimetric concepts presented in Section 2.1 pertain to a target-centered view in which the quantity of interest is the probability distribution of ICS in a single target volume. These ICS probabilities are conditional on radiation quality. Most of the models presented in Section 3 remain within this conceptual framework. Section 3.8 describes the concept of cluster dose, which implies a paradigm shift in nanodosimetry by transitioning from the target-centered view to a track-centered view.

In the track-centered view, the quantity of interest is the number of ICs produced along the track. That is, instead of one stochastic variable (the number of ionizations in a target), we have a set of (statistically correlated) stochastic variables, the number of ICs per path of the primary particle. Superposition of the components of a mixed radiation field then gives a distribution of the number of ICs in a volume of matter. The expectations of the frequency distributions of the number of ICs per path correspond to the macroscopic dosimetric concept of linear energy transfer (LET).

### 2.4.1 From IC probabilities to yields of clusters

In nanodosimetry, radiation quality is characterized by the probability distribution of ICS or derived quantities. In macroscopic dosimetry, radiation quality of charged particles is characterized by LET, the mean energy imparted by the particle per path traveled. The ratio of LET to the mean energy imparted per ionization, $W_i$, yields the mean number of ionizations per particle path, $d\overline{N}_i/dx$.

Dividing $d\overline{N}_i/dx$ by the conditional mean ICS $M_1^c$ in nanometric targets of a given size yields the mean number of such targets per primary particle path that receive at least one ionization. This quantity is addressed here as the linear nanodosimetric event frequency, $d\overline{N}_0^*/dx$. If we interpret the conditional probabilities $p^c(\nu)$ as the proportions of sites with an IC of $\nu$, then the quantity $d\overline{N}_\nu/dx$ defined by Eq. (16) is the expected number of ICs of size $\nu$ per primary particle path.

$$\frac{d\overline{N}_\nu}{dx} = \frac{d\overline{N}_0^*}{dx} \times p^c(\nu) = \frac{d\overline{N}_i}{dx} \times \frac{p^c(\nu)}{M_1^c} = \frac{LET}{W_i} \times \frac{p^c(\nu)}{M_1^c} \tag{16}$$

The quantities $d\overline{N}_\nu/dx$ and $d\overline{N}_k^*/dx$, where the latter are obtained by replacing $p^c(\nu)$ in Eq. (16) by $F^{c*}(k)$, may be seen as nanodosimetry-based analogs to LET or a nanodosimetry-based extension of the LET concept. Here, they are referred to as linear yield of ICs of size $\nu$. These quantities are the expectation values of stochastic quantities $dN_\nu/dx$ and $dN_k^*/dx$, i.e., the number of ICs produced per path length of a particular particle track. In analogy to microdosimetric notation, these stochastic quantities could be called lineal yield of ICs.

Averaging the quantities $d\overline{N}_\nu/dx$ by fluence or dose weighting according to Eqs. (14) and (15) gives quantities that are nanodosimetry-based extensions of frequency-mean or dose-mean LET. However, calculating the weighted integral of spectral particle fluence using the linear IC yield gives the mean number density $d\overline{N}_\nu/dV$ of ICs produced.

$$\frac{d\overline{N}_\nu}{dV} = \sum_p \int \frac{d\overline{N}_\nu}{dx}(E,p)\Phi_E^p(E)dE \tag{17}$$

Here, $d\overline{N}_\nu/dx(E,p)$ and $\Phi_E^p$ are the linear IC yield and spectral particle fluence of particles of type $p$ with energy $E$. The integral of $d\overline{N}_\nu/dV$ over a volume is the expected number of IC with ICS $\nu$ in that volume for the irradiation represented by $\Phi_E^p(E)$.

### 2.4.2 Determination of linear yields of ICs

Measuring the quantities $dN_\nu/dx$ required experimental setups that detect ICs along a stretch of the particle track. First steps towards this goal were taken with detectors for the spatial imaging of particle track segments [17,18,50]. In principle, these approaches should allow for the measurement of the yield of ionization clusters per path length. However, substantial further development will be required before this goal can be achieved. A major drawback of track-imaging devices based on secondary electron detection in optical scintillators [17,18] is their limited spatial resolution due to the large lateral drift of electrons. The device based on ion detection in multi-channel dielectric gas-electron multipliers had low detection efficiency and large dead times [50].

The expected number of ICs per path of the primary particle, $d\bar{N}_\nu/dx$, can be related to the concept of the effective track cross section (ETCS) for formation of ICs introduced by Braunroth et al. [64]. The ETCS is defined as the two-dimensional integral of the probability of a nanometric target to receive an ICS of given properties, e.g., at least $k$ ionizations in the cluster, in a plane perpendicular to the primary particle trajectory. That is, it corresponds to an integral over the impact parameter range with IC probabilities as weights. In the author's view, this is analogous to how interaction cross sections are defined.

Multiplying the ETCS by the number density of target volumes gives the number of ICs produced per projectile path length. This is analogous to how macroscopic interaction cross sections are obtained by multiplying atomic or molecular interaction cross-sections by the number density of scatterers. Since $d\bar{N}_\nu/dx$ relates to all ICs formed per primary particle path, the corresponding target density is one target per volume of a target as is obtained in close-packed targets arrangements [64,68,100]. Therefore, the expression relating $d\bar{N}_\nu/dx$ with ICS probabilities is given by Eq. (18), where $b$ denotes the impact parameter, $V_t$ is the target volume, and the angle bracket indicates the isotropic average over all target orientations, if applicable.

$$\frac{d\bar{N}_\nu}{dx} = \frac{1}{V_t} \times \int_0^\infty \langle p(\nu|b) \rangle 2\pi b db \tag{18}$$

The determination of $d\bar{N}_\nu/dx$ from simulations can also be achieved by directly scoring the absolute numbers of ICs in the simulations [111]. The yield of ICs per particle path could also have been determined in the simulation studies of Alexander et al. [96] and Selva et al. [57], in which the scoring was done in a track-centered view. That is, a pencil beam of primary particles was simulated, and a multitude of targets was placed around it.

As previously mentioned, Alexander et al. [96] simulated protons and carbon ion tracks in a water cube of 20 µm side. The particles started from the center of this cube. Targets containing at least one ionization were scored, and the conditional ICS distribution was determined. If instead the absolute mean number of targets with ionization clusters per particle track were divided by the length of the primary particle track segment, a quantity related to $d\bar{N}_\nu/dx$ could have been obtained. To obtain$d\bar{N}_\nu/dx$, this quantity must be divided by the ratio of the total volume of targets to the volume of the scoring region.

Selva et al. [57] placed spheres with diameter of 3 nm and 1 nm in a cylindrical scoring volume with diameter and height of 20 nm. The primary particles moved along the axis of the scoring volume, such that an impact parameter range between 0 nm and less than 10 nm was covered. Selva et al. determined the mean hit frequency per primary track, where a hit was defined by the occurrence of an IC of at least 2. The yield of ICs with at least two ionizations, $d\bar{N}_1^*/dx$, would have been obtained by dividing the absolute number of hits per track by the length of the primary particle path segment and the ratio of the total volume of all targets to the volume of the scoring region.

The different impact parameter ranges used in these two studies correspond to different analogs to LET: $d\bar{N}_\nu/dx$ and $d\bar{N}_k^*/dx$, when determined over the entire impact parameter range, correspond to the unrestricted LET. In contrast, these quantities, when determined over a restricted range of impact parameters, correspond to the restricted LET, which includes only energy transfers up to a given maximum value. However, there is probably no one-to-one correspondence with restricted LET.

Of the simulation geometries used by Ramos-Méndez et al. [58], only the third had a constant primary particle path length and a uniformly covered impact parameter range. Similar to the work of Selva et al, a pencil beam passed along the center of the cylindrical scoring region. For the other two geometries, the distribution of impact parameters resulted in a non-uniform fluence profile (see Supplementary Fig. 5). Additionally, the pertinent length of the primary particle is not easily obtained. This is because secondary electrons are not only produced along the chord of the primary particle trajectory inside the scoring region, but also outside of it.

In the subsequent work of Faddegon et al. [37], the radiation beam was perpendicular to the cylinder axis of the scoring volume and had a rectangular cross-section matching that of the scoring region. The corresponding impact parameter distribution is also shown in Supplementary Fig. 5 and is comparable to the ones from irradiation geometries 1 and 2 in [58]. The yield of ICs was normalized to the mean chord length of the primary particles in the scoring region, which may underestimate the actual relevant path length of the primary particle.

Therefore, Ortiz et al. [112] employed a more complex simulation setup. In this setup, a cubic scoring region with 100 nm side was filled with closely packed target cylinders measuring 2.3 nm in diameter and 3.4 nm in height. Primary particles were emitted isotropically from randomly distributed source points within a spherical void centered within the scoring region and having a diameter of 34 nm. Secondary particles were subject to periodic boundary conditions, which caused them to be reinjected at the opposite face of the cube after leaving it. Although a retrospective estimation of the impact parameter distributions does not seem possible in this case, the periodic boundary conditions ensure that all ICs produced by secondary electrons are scored, except those that would be formed in the void containing the source. Therefore, the results should correspond to $d\bar{N}_\nu/dx$ if the relevant particle path is that used in [112], namely the mean chord length of the scoring cube.

### *2.5 Correlated ionization clusters*

A particle track is a collection of statistically correlated energy transfer points according to the event concept in microdosimetry [113]. Therefore, ICs formed in a particle track are statistically correlated. As posited in a generic model proposed by the BioQuaRT project [114] and subsequent independent studies [66], there is the possibility that correlated ICs play a role for the effectiveness of ionizing radiation.

The aforementioned detectors for the spatial imaging of particle track segments [17,18] could eventually provide comprehensive information on correlated cluster formation. Experimental studies have already been conducted on the correlations of ionization clusters in two targets in proximity using existing nanodosimeters with different drift time windows, [59,105,106]. Pietrzak et al. [59] studied the correlation between the ICS distributions in a pair of sensitive volumes $V_1$ and $V_2$. $V_1$ is traversed by the primary particle. They found that the mean cluster size in $V_2$ varies linearly with the ICS in $V_1$. The result corroborates the contribution of ionizations in $V_2$ resulting from secondary electrons produced in the other target volume.

Hilgers and Rabus [105] performed measurements using a primary particle detector that was larger than the cross-section of the sensitive volume, with position-sensitive detection along and perpendicular to the axis of the sensitive volume. This allowed studying various geometrical arrangements of the two target volumes. In addition to a narrow beam traversing one of the volumes (geometry A), they also studied a beam passing between the volumes (geometry B), and a beam passing outside both of them (geometry D). The case of a broad-beam irradiation (geometry C) was also investigated. In geometry B, the volumes had the same impact parameter. In geometry D, both volumes were aligned in the same radial direction, perpendicular to the beam, but at different impact parameters. To study the dependence on the target size, the experiments were performed with three different operating gases. The simulated target sizes in water were approximately 0.88 nm, 2.9 nm, and 4.6 nm [86].

They studied the occurrence of correlations by considering the ratio of the relative frequency of observations with ICs in both volumes to the product of the relative frequencies of ICS in each volume separately. If the simultaneous occurrence of ICs is due to coincidence, this ratio should be approximately one. If there is correlation, the ratio should be greater than one; if there is anti-correlation, the ratio should be less than one.

For geometry A, this frequency ratio was 20 % greater than one for the smaller target sizes and about 10 % less than one for the largest one. For geometry B, the ratio was generally less than one and decreased with increasing target size, reaching about 0.5 for the largest targets. With broad-beam geometry C, the ratio was generally below about 0.1. Only geometry D produced large values of this ratio of relative frequencies; as high as three for the smallest target size and ICS exceeding 2. As the target size increased, this effect decreased. For the largest targets, the probability ratio was only 1.2 for an ICS exceeding 2.

The interpretation of these results is that, in geometry A, electrons produced in the target volume traversed by the primary particle contribute to the IC in the second volume, provided that the two volumes are not too far apart. In geometry B, the anti-correlation is due to the fact that secondary electrons produced between the two target volumes will only be emitted towards one of them. In broad-beam geometry C, electrons from tracks with larger impact parameters only interact with one of the volumes. The large correlation observed in geometry D is due to the fact that electrons reaching the target volume at larger impact parameters have a high probability of having traversed and interacted with the closer volume.

The correlated occurrence of ICs was also studied using simulations [68] in which ICs in proton tracks we examined using a virtual scoring volume approach based on Wigner-Seitz cells of a face-centered cubic Bravais lattice. These Wigner-Seitz cells are polyhedrons that completely fill space and are approximately spherical [68]. The lattice spacing was chosen such that their volume was equal to that of a sphere with diameter of 2.0 nm, 2.5 nm, or 3.0 nm. The latter sphere diameter corresponds to a volume equal to that of a cylinder with a diameter of 2.3 nm and a height of 3.4 nm.

When tallying ICs, the number of ionizations in each target cell was counted and the center of gravity of each IC was determined. This information was then used in a second round of tallying with a larger-spaced Bravais lattice to tally clusters of ICs. In this round, the targets (referred to as cluster volumes) corresponded to spheres with diameters of either 7.5 nm, 12.0 nm or 18.0 nm. The results of this study show that the frequency of clusters of ICs increases with increasing size of the cluster volume. However, the results also indicate that predicting the frequency of occurrence of clusters of ICs in a segment of the particle track based on the IC formation probability leads to an underestimation by orders of magnitude.

None of these reported studies has considered a formalization of the effect such as a linear yield of clusters of ICs within a given spatial regime. This endeavor is also beyond the scope of this review but mentioned here as a potential route for future development of nanodosimetric concepts.

# 3 Biological effect models based on nanodosimetry

## 3.1 *The two compartment model of cell survival*

Chronologically, the first attempt to link nanodosimetry with biological effects was the two-compartment model proposed by Schulte et al. [36] in 2001, at a time when most of the state-of-the-art gas-counter nanodosimeters were still under development.

The two-compartment model considers cylindrical nanodosimetric sites with a diameter of 2 nm and a height of 16 nm (corresponding to a sequence of approximately 50 base pairs in the DNA molecule). The choice of target volume appears to have been primarily motivated by the availability of published data accessible at that time regarding energy deposition by protons and helium ions with energy per mass between 0.3 MeV and 5 MeV. Furthermore, the dimensions were similar to those of the sensitive volume of the ion counter nanodosimeter when operated without a time window for ion collection.

It is assumed that ICs between 2 and 5 in such a target result in repairable DNA damage, whereas ICs between 6 and 10 create irreparable damage. Charge and radical recombination are assumed to lead to a reduced biological effectiveness for ICS exceeding 10, similar to, albeit not analogous to, the known decrease in RBE at high LET values. It is further assumed that reparable DNA damage may be misrepaired and that lethal damage resulting from repairable lesions follows a linear-quadratic dose dependence. In this model, the quadratic term corresponds to the implicit assumption that two repairable lesions may interact to form a lethal lesion. Irreparable damage is assumed to occur proportionally to the dose.

The number of lethal damages is assumed to follow a Poisson distribution. Thus, the resulting surviving probability $S$ of cells as a function of absorbed dose $D$ is given by Eq. (19).

$$S = e^{-(\alpha_1 q_1 + \alpha_2 q_2)D - \beta {q_1}^2 D^2} \tag{19}$$

The parameters $\alpha_1$ and $\alpha_2$ are LET-independent. Parameter $\alpha_2$ corresponds to the induction of direct lethal DNA damage, and $\alpha_1$ to lethal damage resulting from misrepair of repairable damage. The parameter $\beta$ corresponds to the conventional quadratic term of the dose dependence, and the weighting factors $q_1$ and $q_2$ are defined by Eq. (20) for a given radiation quality $Q$ and reference radiation quality $Q_{ref}$.

$$q_1 = \frac{\sum_{\nu=2}^{5} p(\nu|Q)}{\sum_{\nu=2}^{5} p\left(\nu|Q_{ref}\right)} \ ; \quad q_2 = \frac{\sum_{\nu=6}^{10} p(\nu|Q)}{\sum_{\nu=6}^{10} p\left(\nu|Q_{ref}\right)} \tag{20}$$

In other words, the radiation quality factors are the ratios of the probabilities of small and large clusters of radiation quality $Q$ to the probabilities of reference radiation quality $Q_{ref}$.

The values of the radiation quality parameters were obtained from published data for energy deposition in the considered target volume in water. From this data, IC distributions were estimated according to the law of total probability. The conditional probabilities of an ICS for a given energy imparted, $\varepsilon$, was assumed to be uniquely determined by the mean ICS $M_1$, estimated as $M_1 = \frac{\varepsilon}{W}$, where $W$ is the macroscopic mean energy per ion pair. The ICS distribution was assumed to be a modified Poisson distribution with $M_1$ as expectation and a variance reduced by a chosen Fano factor of 0.3. The inverse of the Fano factor is the modifying parameter $m$ in Eq. (21).

$$P_\nu(M_1) = m \frac{(mM_1)^{m\nu}}{\Gamma(m\nu + 1)} e^{-mM_1} \tag{21}$$

The denominator of Eq. (21) is the gamma function, which satisfies the identity $\Gamma(n+1) = n!$ for integer values of $n$.

The resulting model function was tested on literature data of V79 cell survival [115] and it reasonably reproduced the corresponding survival curves.

### *3.2 The approach of Grosswendt to link nanodosimetry and DNA damage*

The fundamental model assumptions of the approach developed by Grosswendt in several publications [77–79] are as follows: 1. The yield of DNA single-strand breaks (SSBs) produced by a specific type of radiation is proportional to the probability $p(1)$ of an ICS of $\nu = 1$ in a nanometric cylindrical target volume. 2. The yield of DNA double-strand breaks is proportional to the complementary cumulative probability $F^*(1)$ ($F_2$) of generating an ICS with $\nu \geq 2$ in the same target volume. The dimensions of this volume are implicit model parameters, and different choices were made at different stages of the model development.

In the initial study, the cylindrical volume had a diameter and height of 2 nm and was irradiated by a circular monoenergetic electron beam of 2 nm in diameter [78]. The initial energies ranged from 12 eV to 100 keV. In addition to the ICS distribution, the mean energy imparted in the cylinder, $\Delta E$, was also determined in the simulations. (The aforementioned paper does not provide these details, but it can be reasonably inferred that a planar electron source in contact with the target cylinder was used and that the scored energy encompassed energy deposits from ionizations, electronic excitations, and stopping electrons.) The yields of SSBs and DSBs, $G_{SSB}$ and $G_{DSB}$, were then estimated by utilizing Eq. (22), where $c_1$ and $c_2$ represent dimensionless parameters.

$$G_{SSB} = c_1 \frac{p(1)}{\Delta E} \; ; \quad G_{DSB} = c_2 \frac{F^*(1)}{\Delta E} \tag{22}$$

This was motivated by the fact that yields of DNA strand breaks are generally quantified as the ratio of the number of breaks to the product of absorbed dose and mass of irradiated DNA. If the dose is uniform in the part of the irradiated region containing DNA, the denominator of this ratio is the expectation of the energy imparted to DNA.

Although not explicitly determined, the parameters $c_1$ and $c_2$ were implicitly determined by comparing the predictions according to Eq. (22) with results obtained by the PARTRAC code [116] requiring a match at the highest electron energy. For SSBs this scaling produced agreement at energies above 8 keV, whereas for DSBs up to roughly 10 % discrepancies were found in this energy range. Overall, the data from Eq. (22) gave a comparable qualitative energy dependence as the PARTRAC data, while the absolute values differed by up to a factor of almost 2 for SSBs and almost 3 for DSBs.

A comparison of the ratio of the yields for DSBs to that of SSBs revealed up to a factor of 4 higher values than obtained with PARTRAC, with experimental data [117] (available only at a few energy points) falling in between the two predictions. In addition, a comparison was presented between the energy dependence of $F_2$ and the bio-effect cross-section data of Simmons and Watt [118], where a scaling factor was once again applied to match the data at an electron energy of 150 eV. This comparison showed agreement with respect to the trend; the absolute deviations remained below several 10 % in many cases, but there were also instances of up to a factor of 3 discrepancy.

In further development of the approach [77,79], the size of the cylindrical target was set to a diameter of 2.3 nm and a height of 3.4 nm, corresponding to a segment of 10 base pairs in DNA in its B-type configuration [119]. This choice implicitly introduced a mechanistic model assumption, namely that nanodosimetric quantities relate to direct radiation effects in the DNA, thus neglecting indirect damage by radiolytic species. For DSB induction, this can be expected to provide a good approximation, since the radical species undergo Brownian motion so that concurrent arrival of different species at the same short DNA segment and reactions forming strand breaks may be expected to be of low probability.

A comparison of the SSB and DSB yields from electrons with results from PARTRAC again yielded the same qualitative energy dependence, with agreement for energies of 8 keV or higher for both yields [77]. The discrepancies observed at smaller electron energies were still by up to a factor of two for SSBs. Furthermore, the ratio of the maximum yields predicted according to Eq. (22) and those from PARTRAC was reduced to a factor of about 1.7. When the scaling factors $c_1$ and $c_2$ for the nanodosimetric predictions according to Eq. (22) were determined by comparison with the yields obtained with PARTRAC for 220 kV x-rays [120], about 14 % higher values than with PARTRAC were obtained for the SSB yield at high electron energies and almost a factor of two difference to data reported by Nikjoo et al. [121]. For the ratio of DSB to SSB yields, a factor of four higher maximal value was obtained compared with PARTRAC [79].

For monoenergetic protons and alpha particles in linear energy transfer (LET) ranges between 0.4 keV/μm and 30 keV/μm and between about 9 keV/μm and 170 keV/μm, respectively, a peak in the LET dependence was found at about 50 keV/μm for the estimate from Eq. (22) [79]. The findings from PARTRAC [120,122] and Nikjoo et al. [123] indicated this peak to appear at about 150 keV/μm, at which also a peak in the RBE for cell killing is observed in experimental studies [124]. A comparison with the corresponding data from PARTRAC and Nikjoo et al. for the ratio $R_{ds}$ of DSB to SSB yields showed that a comparable LET dependence was obtained with the nanodosimetric estimator given by Eq. (23).

$$R_{ds} = c_3 \frac{M_2}{M_1} \tag{23}$$

The value of the proportionality constant $c_3$ was obtained by requiring a match with the yield ratio obtained by PARTRAC for protons of 20 MeV [79].

The model postulates that the probabilities $F_2$ of generation of an IC of size 2 or higher and of DSB induction in a short DNA segment are proportional to each other. The validity of this assumption was scrutinized for DSB formation in plasmid DNA by different ion types and energies [125–127]. Fig. 5 shows a comparison between the cross-section for DSB induction in viral DNA [128] as a function of LET and the corresponding probability $F_2$, obtained by simulations in water for a target cylinder with a diameter of 2.3 nm and a height of 3.4 nm. In the simulations, an incident rectangular beam with a width of 4.6 nm and a height of 6.8 nm was used. The two y-axes in Fig. 5 are proportional to each other, both spanning 2.5 orders of magnitude. Thus, Fig. 5 suggests that the assumed proportionality is approximately valid (with some experimental data points deviating from the curve defined by the simulations by a factor of up to two.

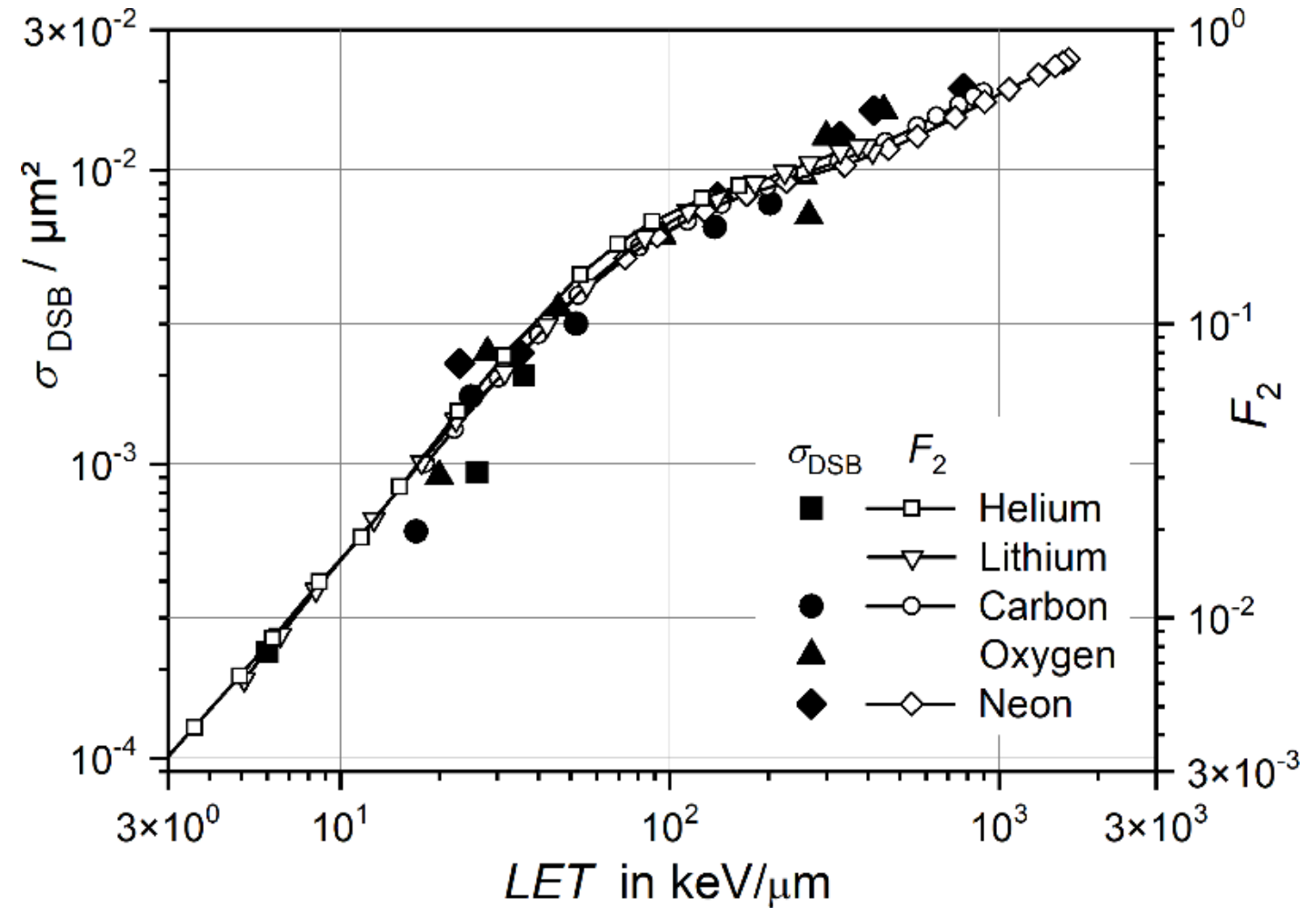

**Fig. 5**: Dependence of measured cross-section data for double-strand break (DSB) formation, $\sigma_{DSB}$, in SV40 viral DNA [128] on LET (filled symbols, left y-axis) and the cumulative probability, $F_2$, of obtaining an ionization cluster size of two or more (open symbols, right y-axis). These $F_2$ values were derived from Monte Carlo simulations of helium, lithium, carbon and neon ions (of different energies), using a water cylinder with a diameter of 2.3 nm and a height of 3.4 nm as the target volume [126,127]. The primary particle beam had a rectangular cross section with dimensions twice the magnitude of the target's dimensions. The direction was perpendicular to the cylinder's axis.[13]

### *3.3 The combinatorial model of Garty et al. for DNA damage*

The approach of Garty et al. [45,49] is based on ICS distributions measured with the Ion Counter nanodosimeter [14,41]. These distributions are assumed to be representative of the ICS distribution in an equivalent target in biological matter. The determination of the size of the equivalent volume relies on the scaling relation proposed by Grosswendt (see Appendix A). The variant for produced ionizations [43] was used, which implicitly assumes a 100 % detection efficiency for the nanodosimeter. Using the same scaling factor for lateral dimensions as well, the beam size used in the experiments corresponds to a beam size in biological matter of 56 nm.

The first version of the model [45] had only one parameter $p_s$, which represented the constant probability of an ionization to result in a DNA strand break, regardless where the ionization occurred in the target volume. The nanometric target volume was assumed to be a cylinder of 4.5 nm in diameter and 8 nm in height, representing a DNA segment of 16 base pairs (2.3 nm in diameter and 5.4 nm in height) surrounded by a hydration layer with a thickness of 1.3 nm. Thus, the model includes indirect effects from water radiolysis and the subsequent reactions of produced radical species with DNA, which result in a strand break. The model parameter $p_s$ corresponds to the average probability resulting from the combined direct and indirect effects.

A later version of the model [49] includes a second parameter, $p_b$, which represents the constant probability that base damage will occur when an ionization takes place anywhere in the target cylinder. In this study, the simulated nanometric target volume was estimated to be a cylinder of 4.3 nm in diameter and a height of 6 nm, interpreted as a segment of DNA surrounded by a 1 nm-thick water shell. Assuming the presence of a water shell on the top and bottom of the target cylinder as well, the DNA content corresponds to approximately 12 base pairs. (The height of the cylinder corresponds to 18 DNA base pairs.)

---

[13] Fig. 5 is reprinted from Radiation Measurements 46 (12), H. Rabus and H. Nettelbeck, Nanodosimetry: Bridging the gap to radiation biophysics, 1522-1528, Copyright (2011), with permission from Elsevier.

The joint probability of having $n_s$ strand breaks and $n_b$ base damages is obtained by considering the multinomial probability of obtaining these values through random sampling from $\nu$ ionizations in the target cylinder (Eq. (24)).

$$p(n_s, n_b) = \sum_{\nu \geq n_s + n_b} f_\nu \frac{\nu!}{n_s!\, n_b!\, (\nu - n_s - n_b)!}\, p_s^{n_s} {p_b}^{n_b} (1 - p_s - p_b)^{\nu - n_s - n_b} \qquad (24)$$

Note that in Eqs. (24) and (25), $f_\nu$ is used instead of $p(\nu)$ to emphasize that the model relies on measured ICS distributions (or ICS distributions from simulations of the measurement). The probability of inducing a DSB, $p_{DSB}$, is obtained by multiplying Eq. (24) by the probability of at least one SB per strand (Eq. (27)) and summing over $n_b$. The result is given in Eq. (25) [49].

$$p_{DSB} = \sum_{\nu \geq 2} f_\nu \left[1 - 2\left(1 - \frac{p_s}{2}\right)^\nu + (1 - p_s)^\nu\right] \qquad (25)$$

Additionally, double-strand lesions (DSL) were considered in which each strand contains at least one break or base damage. The corresponding probability $p_{DSL}$ is obtained by replacing $p_s$ in Eq. (25) by $p_t = p_s + p_b$. The probability of a non-DSB lesion affecting both strands is obtained by subtracting $p_{DSB}$ from $p_{DSL}$ [49].

The model parameters were obtained by fitting predicted yields of DSBs and DSLs to the experimental data from plasmid DNA assays [129]. Conversion from probabilities to yields was performed by estimating the average energy imparted in the target as $\Delta E = M_1 \times W_i$, where $M_1$ is the mean ICS and $W_i$ is the (macroscopic) mean energy per ionization. The absorbed dose $D$ was then obtained by dividing the energy imparted, $\Delta E$, by the mass of the nanometric target volume, $m_{SV}$. Using Eq.(26), the yields of different DNA damage are obtained, where the subscript *d* indicates the type of damage (SSB, DSB, DSL) and $m_{DNA}$ represents the mass content of DNA in the target volume.

$$G_d = \frac{p_d}{D \times m_{DNA}} = \frac{m_{SV}}{M_1 \times W_i \times m_{DNA}} \times p_d \qquad (26)$$

It is worth noting that in both papers of Garty et al. [45,49] the expressions for the yields were written in a dimension-free style by which the mass of the DNA content did not appear explicitly. In [49] it was mentioned explicitly that the yield was obtained by dividing "by the mass of the equivalent DNA segment 6500 Da". The numerical factor appearing in Eq. (3) of [45] is $9.6\times10^{-10}$, and the target volume is expressed in $nm^3$. In Eq. (5a) of [49], the numerical factor is $9.6\times10^{11}$, and the volume is in $cm^3$. Considering the 21 orders of magnitude between $nm^3$ and $cm^3$, this is the same factor. Hence, also in [49] a value of 6500 Da was used for $m_{DNA}$. This choice of $m_{DNA}$ seems inconsequential given the actual lengths of the DNA segments assumed, which is different from 10 base pairs to which the value of 6500 Da applies.

Furthermore, both papers wrote the mass of the target volume as the product of the mass density $\rho$ and the volume of the target volume $V_{SV}$, where the latter is stated to be scaled to unit density based on the nanodosimetric scaling relation [45,49]. The second paper [49] states that $\rho$ and $V_{SV}$ are operational constants of the nanodosimeter, so that all parameters entering the yield, except $W_i$ , were obtained experimentally. Taken literally, this statement appears inconsistent with the scaling relation according to which the diameter of the target volume (and not the target volume itself) is inversely proportional to the mass density (Appendix A). Thus, the masses of the macroscopic and nanometric volumes are different. (In fact, they differ by twelve orders of magnitude.) However, in this author's interpretation, this is presumably an issue of wording. What seems to be meant is that the product of the (macroscopic) sensitive volume of the nanodosimeter and the mass density of the gas in it can be converted into the mass per area of a nanometric target in water or tissue via the scaling relation.

The yields predicted by the model were benchmarked against PARTRAC simulations of protons over a wide energy range [120], where good agreement of better than 10 % was generally found. The predictions were also compared with experimental data from various studies (see Section 3.4 in [49]). The model reproduced the LET dependence of the yield of DSBs in irradiated V79 cells in the LET range up to 100 keV/µm. However, the trend of the LET dependence of non-DSB lesion clusters was not correctly predicted. Radiochemical reactions within large ionization clusters, which are not considered in the model, were discussed as one possible explanation.

### *3.3.1 Consequences for Grosswendt's approach*

The results presented in Fig. 5 suggest that the probabilities $F_2$ of generation of an IC of size 2 or higher and of DSB induction in a short DNA segment are proportional to each other, as was hypothesized by Grosswendt. However, from the heuristics in the work of Garty et al. [45,49], one may question whether of such proportionality is expected. If there are two or more ionizations in the DNA segment for which no further details are available regarding their spatial locations, then the probability $p_{2s}$ of damage on both strands of DNA depends on the ICS, $\nu$, and is given by Eq. (27) [45] under the assumption that each ionization results in a strand break:

$$p_{2s} = 1 - \left(\frac{1}{2}\right)^{\nu-1} \tag{27}$$

That is, this probability is 50% for $\nu = 2$, 75% for $\nu = 3$, and 87.5% for $\nu = 4$. Considering this and taking into account that an ionization may have a probability of less than 1 for resulting in a strand break, the predicted probability of DSB induction shows non-linear dependence on $F_2$ as illustrated in Fig. 6. These data correspond to results obtained using the track structure codes PTra [26] and Geant4-DNA [31–34] employing the probability of DSB induction from [49]. The data show the expected saturation-type behavior when $F_2$ approaches its upper limit of unity. This implies that proportionality between $F_2$ and DSB induction is only expected for small values of $F_2$, i.e., when the probability $p(2)$ for two ionizations dominates.

Bug et al. [83] later confirmed this in their work using a Bayesian approach to determine the likelihoods that an SSB results from an IC of size $\nu$ and that a DSB results from an IC of at least $k$ ionizations in a DNA target of 10 base pairs. Among the four radiation qualities considered, a DSB was most likely to result from an IC with $k = 2$ for the sparsely ionizing 5 MeV proton and 20 MeV alpha particle. For these two radiation qualities, an SSB was also most likely to result from an IC of $\nu = 1$. However, for more densely ionizing radiation, this was no longer the case. For example, for a 60 MeV $^{12}$C ion, a DSB had an IC with $k = 6$ as its most likely origin. One caveat of these findings is that they used the value for the probability of an ionization resulting in a strand break as determined in the work of Garty et al. [45]. However, the probability derived by Garty et al. conceptually did not pertain only to direct strand breaks.

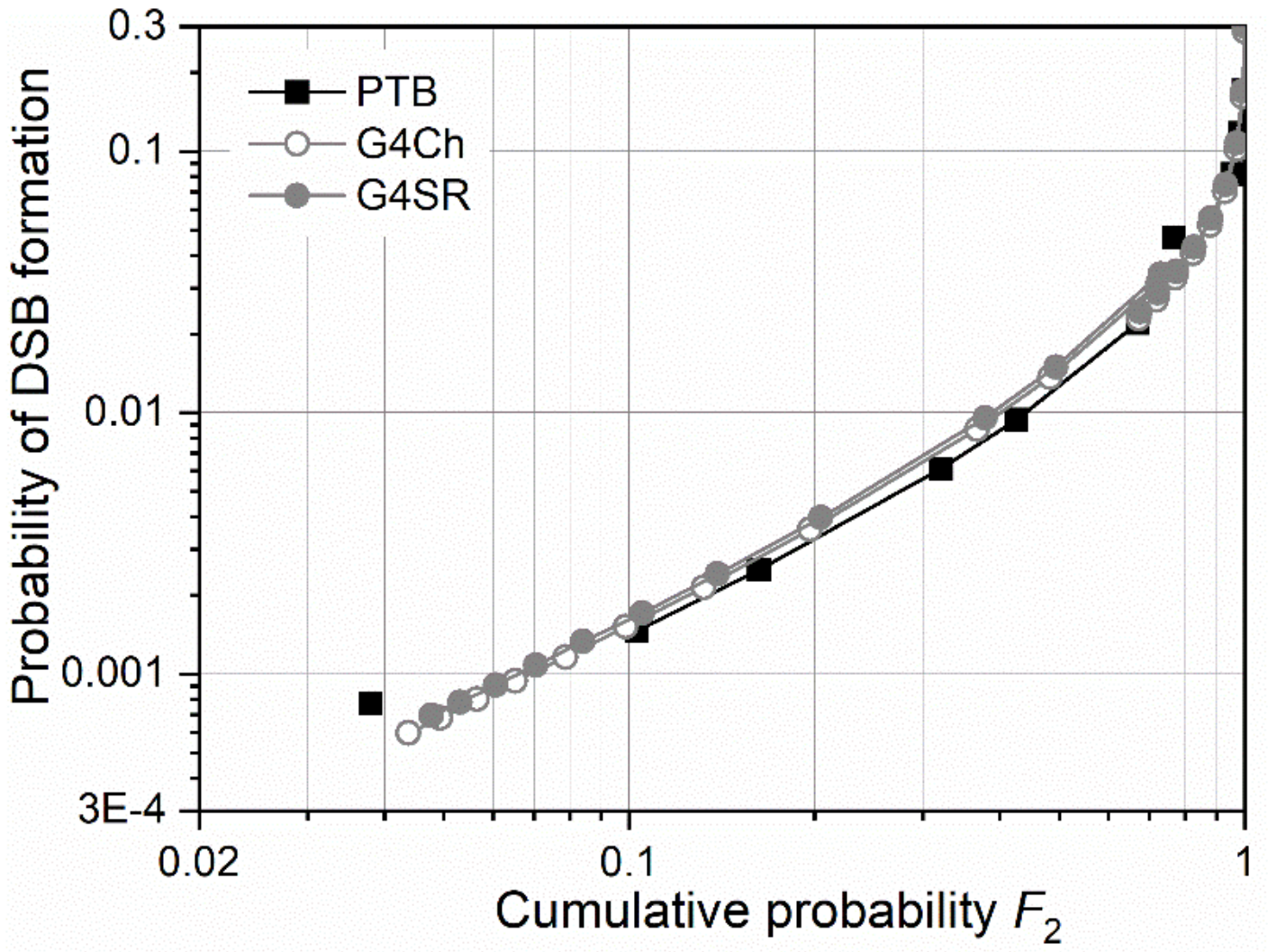

**Fig. 6:** Correlation between the probability of double-strand break (DSB) induction, as predicted by the combinatorial model (Section 3.3), and the complementary cumulative probability $F_2$ for an ionisation cluster size of at least two by particles with an LET between 5 keV/µm and 220 keV/µm. These values were derived from the ionization cluster size distributions simulated with the PTra code (PTB) and Geant4-DNA, where G4Ch and G4SR correspond to the respective Screened Rutherford and Champion models used for elastic scattering of electrons.[14]

## 3.4 *Nanodosimetry based radiation quality factors for radiation protection*

Schulte et al. used a variant of the first version of the model of Garty et al. [45] to derive nanodosimetry-based predictions of radiation quality factors for high-LET radiations for applications in radiation protection dosimetry in space [46]. They simulated measurements with the Ion Counter nanodosimeter using a dedicated track structure code [130] for electrons with an energy of 100 keV, protons with energies ranging from 0.5 MeV to 250 MeV, $^{4}$He ions with energies ranging from 1 MeV to 30 MeV, and $^{12}$C ions with energies ranging from 10 MeV to 250 MeV. These combinations of particle type and energy covered the LET range between 0.4 keV/µm and 800 keV/µm. The model parameter $p_s$ was assumed to be independent of LET and to have a value of 0.15 ± 0.05. Three different heights of the sensitive volume were considered, 7 nm, 16 nm, and 150 nm, which correspond to about 20, 50, and 500 base pairs, respectively. The quantity derived from the ICS distributions was the probability $p_{cDSB}$ of forming a complex DSB. This probability is given by Eq. (28), where the symbol $\binom{\nu}{2}$ represents a binomial coefficient.

$$p_{cDSB} = p_{DSB} - \sum_{\nu \geq 2} f_\nu \binom{\nu}{2} \frac{{p_s}^2}{2} (1 - p_s)^{\nu - 2} \tag{28}$$

The resulting yield of complex DSBs $G_{cDSB}$ is given by Eq. (29), where $W_i$ is again the mean energy per ionization:

$$G_{cDSB} = \frac{p_{cDSB}}{M_1 \times W_i} \tag{29}$$

[14] Fig. 6 is reprinted from Radiation Measurements 46 (12), H. Rabus and H. Nettelbeck, Nanodosimetry: Bridging the gap to radiation biophysics, 1522-1528, Copyright (2011), with permission from Elsevier.

As can be seen by comparing with Eq. (26), it was implicitly assumed that the sensitive volume is identical to the volume occupied by DNA.

Nanodosimetric quality factors were obtained by calculating the ratio of the yield for complex DSBs, according to Eq. (29), from the ICS distributions produced by particles of given type and energy, divided by the yield obtained for a reference radiation, which was 100 keV electrons. Fig. 7 shows a comparison of the quality factors obtained for protons (diamonds), helium ions (triangles) and carbon ions (squares) with the quality factors defined in ICRP publication 60 (thick line). It can be seen that the nanodosimetric quality factors reproduce the increasing trend of the quality factors at LET values up to 80 keV/µm but fail to capture the decrease at higher LET values. There is also a clear difference between protons and heavier ions. As shown in [46], this difference among particle types and general trends are essentially insensitive to the choice of model parameter within the considered range between 0.1 and 0.2. Comparing results for different $p_s$ values and heights indicated that Q-factors decrease with LET for higher LET values when larger $p_s$ values and heights of the sensitive volume are used.

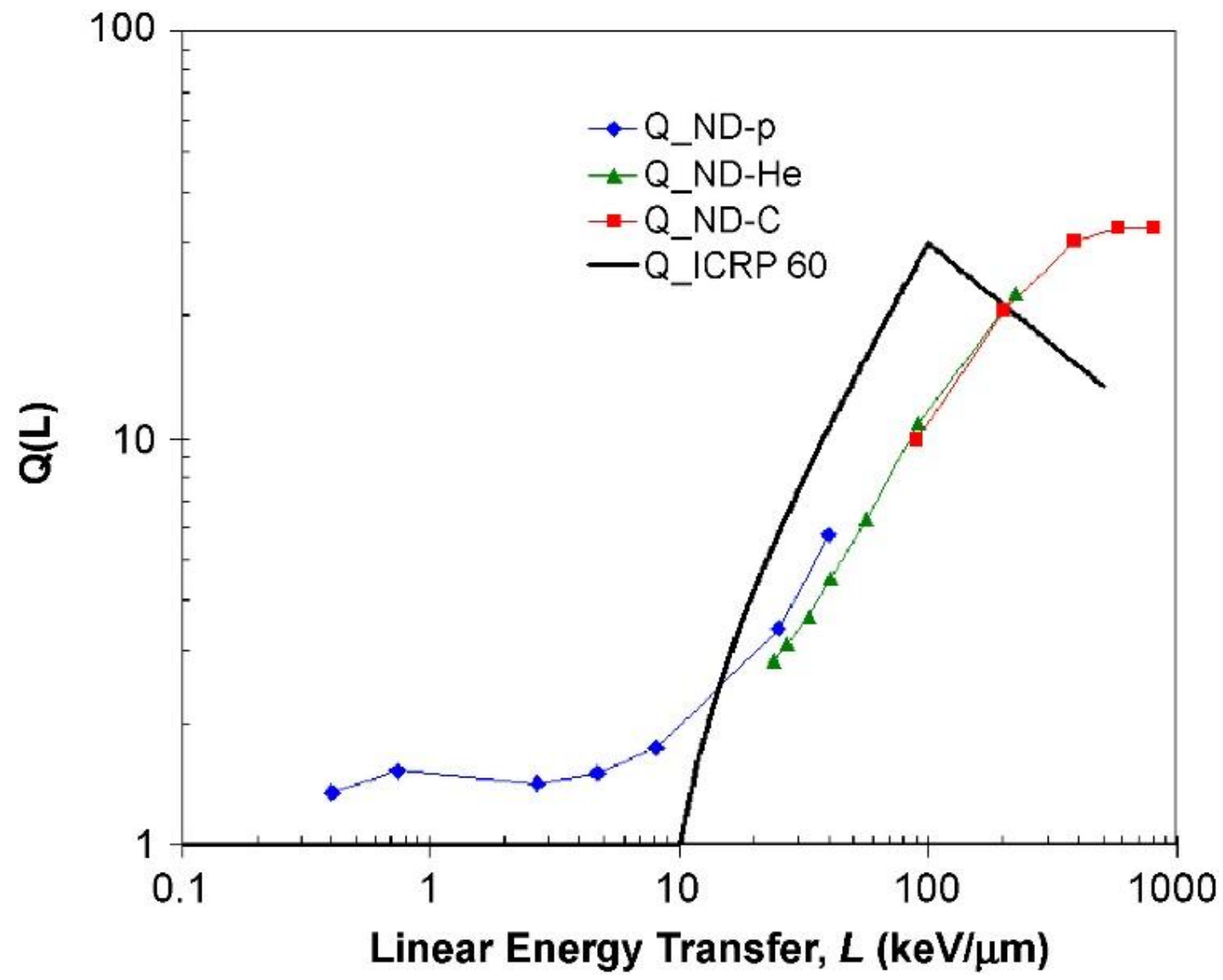

**Fig. 7:** Comparison of the radiation quality factor Q as a function of LET from ICRP Publication 60 (thick line) and nanodosimetry-based quality factors derived by Schulte et al. from ICS distributions obtained for a cylindrical sensitive volume with a diameter of 2 nm and a height of 16 nm.[15]

## 3.5 *The track event theory of cell survival*

### 3.5.1 *The original track event theory*

The track event theory (TET) was formulated by Besserer and Schneider [53,131] as a model of biological radiation effects with a potential relation to nanodosimetry, that was only elaborated on in later work [132,62]. Note that the term "event" used in the TET has the same meaning as in stochastics: a potential outcome of a random process, such as a probabilistic biological outcome. This differs from the use of this term in the terminology of microdosimetry, where an event refers to the occurrence of a physical radiation interaction, such as a track passage, energy deposition, or ionization-cluster formation.

[15] Fig. 7 is reprinted from Zeitschrift für Medizinische Physik 18 (4), R. Schulte et al., Nanodosimetry-based quality factors for radiation protection in space, 286-296, Copyright (2008), with permission from Elsevier.

The original TET [53] assumes that all DNA SSBs and the occurrence of a single (direct) DSB in a cell nucleus are non-lethal. In contrast, two (direct) DSBs formed on the same or two different chromosomes are always assumed to be lethal. The model distinguishes between one-track events (OTEs) and two-track events (TTEs). An OTE corresponds to the induction of two DSBs in the nucleus by a single track; a TTE corresponds to the induction of one DSB by a track. TTEs induced by two different tracks result in a lethal event. OTEs and TTEs are assumed to be statistically independent events "in the terminology of nanodosimetry" [53], without further explanation what this means. In this author's interpretation, the intended message is that OTEs and TTEs are assumed to follow independent Poisson processes.

It is further assumed that the number of tracks interacting with a cell, the occurrence of OTEs, and the occurrence of TTEs are all Poisson distributed. The frequencies of occurrence of OTEs and TTEs are assumed to be proportional to the absorbed dose with different proportionality factors, $p$ and $q$, respectively. Under these assumptions, a cell survives if it experiences no OTE and no more than one TTE. Thus, the probability of survival, $S$, as a function of absorbed dose, $D$, is given by Eq. (30).

$$S(D) = (1 + qD)e^{-(p+q)D} \tag{30}$$

When $qD \ll 1$, Eq. (30) can be transformed into the expression of the LQ model with $\alpha = p$ and $\beta = \frac{q^2}{2}$ [53]. In the asymptotic limit of $qD \gg 1$, the model approaches an exponential dose dependence (linear in the conventional semi-logarithmic plots). The model was used to fit a set of experimental survival data. The quality of fit was found to be comparable to that of the (three-parameter) model of Hug and Kellerer [133], while the purely phenomenological LQ-model generally showed better performance than both [53].

Ngcezu et al. [68] pointed out that the model assumptions of exactly two DSBs per OTE and the interaction of exactly two TTEs to form a lethal lesion should have been phrased as at least two DSBs per OTE and at least two TTEs. Assuming then a Poisson distribution for the number of tracks interacting with a cell and equal probabilities of all tracks to produce an OTE or a TTE, the frequencies of OTEs and TTEs are Poisson distributed and the frequency distributions are statistically independent. This also leads to Eq. (30) but requires fewer model assumptions.

### *3.5.2 Track event theory with repair*

The second version of the TET also accounted for repair by introducing a third parameter, $R$, which represents the probability of repairing two DSBs produced either by an OTE or two TTEs. This results in a more complex dose dependence, wherein the factor preceding the exponential is modified such that the parameter $q$ is replaced by a second-order polynomial $\Pi_2$ in the absorbed dose [131].

$$S(D) = (1 + \Pi_2(D|p, q, R) \times D)e^{-(p+q)D} \tag{31}$$

The coefficients of this polynomial depend on the parameters $p$, $q$, and $R$.

To reduce the number of free parameters to two again, it was argued that the ratio of the parameters $p$ and $q$ should depend only on the chromatin structure, so that cells having the same basic chromatin structure would have the same ratio, $\varepsilon = \frac{p}{q}$, when irradiated with the same radiation quality.

The second-version model was fitted to a large set of radiobiological survival data comprising 42 different cell lines irradiated at comparable dose rates with either $^{60}$Co, $^{137}$Cs, or X-rays of at least 220 kV peak voltage to ensure comparability of radiation quality across all experiments. Initially fitting all model parameters $p$, $q$ and $R$ allowed deriving an estimate for $\varepsilon$. In a second pass, this value of $\varepsilon$ was kept fixed, and only $q$ and $R$ were allowed to vary. With the second-version TET using only two free parameters (i.e., keeping $\varepsilon$ fixed), statistically significant fits were obtained for 36 of the datasets, whereas with the conventional LQ model this was only the case for 32 datasets.

In their analysis of the TET, Ngcezu et al. [68] noted that the model assumptions regarding repair cover only a fraction of the cases possible. Since the model considered only the repair of a pair of DSBs (produced by an OTE or two TTEs), it does not address how cases involving more than two lesions produced by a single track, more than two tracks producing one lesion each, or combinations of OTEs and TTEs would be affected by repair. Additionally, they pointed out that the derivation of Eq. (31) seemed to implicitly assume that lesions produced in the same cell from an OTE and by two TTEs are repaired independently, i.e., as if they occurred in two different cells. The illustration of the model in [131], according to Ngcezu et al. [68], suggested that four or more DNA lesions in a cell would always be lethal; however, the consequences of more than two DNA lesions were not explicitly addressed. Therefore, Ngcezu et al. [68] proposed that the model required an additional parameter to distinguish between always-fatal and potentially-fatal events. They also suggested that the outcome of repair depends on the number of lesions induced, not on how they were induced. Considering cells in which more than two DSBs are produced and assuming these to be always lethal yields a simpler model equation than Eq. (31). This equation has no mixed terms between OTEs and TTEs, but it requires a fourth model parameter [68].

### *3.5.3 Track event theory and nanodosimetry*

Although nanodosimetry was mentioned in the first work on the TET [53], the relationship between the parameters of the TET and nanodosimetry was first discussed in later work of Schneider et al. [134]. They proposed that the ratio $\varepsilon = \frac{p}{q}$ could be factorized into two parts: one depending only on radiation quality and the other one depending only on chromatin organization. This geometrical part, $\varepsilon_{geo}$, was determined by considering voxelized models of tetranucleosomes with cubic voxels with a side of 2 nm. Two different approaches were explored to determine whether voxels were hit. One approach assumed uniformly distributed ionizations, while the second used ray-tracing simulations. In the latter approach, two tetranucleosomes were placed in close proximity at the center of a sphere with a diameter of 32 nm, and chords through the sphere were determined by randomly choosing entrance and exit positions. Intersections of a voxel by a chord were counted as a DSB.

To determine the part $\varepsilon_{ion}$, which depends on radiation quality, Grosswendt's assumptions [77–79] were used: an ICS of at least two in a voxel of the considered tetranucleosome model corresponds to a DSB and an ICS of one corresponds to an SSB. Using the corresponding probabilities, $F_2$ and $p(1)$, the parameter $\varepsilon_{ion}$ is determined according to Eq. (32):

$$\varepsilon_{ion} = \left(\frac{F_2}{p(1)^2}\right)^2 \tag{32}$$

However, it should be noted that this implies an implicit modification of the model assumptions. In the original TET as well as in the text of [134], a TTE was defined as the induction of one DSB. Now a TTE is assumed to correspond to two SSBs induced in different voxels. The lethal combination of two TTEs now requires that both tracks produce SSBs in the same voxels. In addition, the production of two DSBs in different voxels by the same track is assumed to occur statistically independently.

In later work by Schneider et al. [132], a different connection between the TET and nanodosimetry was discussed. They modified two of the original model assumptions such that an OTE is now equivalent to the induction of one DSB and a TTE is equivalent to an SSB. The lethal combination of two TTEs corresponds to the occurrence of SSBs in the same volume by different tracks. Using Grosswendt's assumptions, the parameter $\varepsilon$ is now determined by using Eq. (33).

$$\varepsilon_{ion} = \frac{p_1^e \times F_2}{f_2^e \times p(1)^2} \tag{33}$$

Here, $p_1^e$ and $f_2^e$ represent the probability of finding one or at least two electron tracks traversing the sensitive volume. The frequency of electron tracks interacting with the volume is assumed to be Poisson distributed again. The nanodosimetric probabilities $p(1)$ and $F_2$ are probabilities for one or at least two ionizations in the target volume by an electron hitting it, respectively.

The values of $p_1^e$ and $f_2^e$ were determined for electrons with energies of 100 keV, 500 keV, and 1000 keV using simulations based on the NOREC code [135] in a voxelized geometry with cubic voxels of 2 nm side. The union of all voxels traversed by (straight-line) steps of primary or secondary electrons between interactions was defined as the track volume $V_{Tra}$. The sum of all[16] electron steps between interactions was defined as the track length $l_{Tra}$. Three different sensitive biological targets were considered: a single DNA segment, two adjacent chromatin fibers, and chromatin loops. Each target is characterized by a typical dimension $l_{Sen}$ (diameter or length).

The number of sensitive targets affected by an electron track, $n_{Sen}$, was estimated as $n_{Sen} = \frac{l_{Tra}}{l_{Sen}}$. The number of sensitive targets, $n_{Rad}$, in the irradiated volume, $V_{Rad}$, was estimated as $n_{Rad} = \frac{V_{Rad}}{V_{Sen}}$, where $V_{Sen}$ denotes the volume of the sensitive structure. It appears that the irradiated volume is assumed to be completely filled with sensitive structures. The expected number of electron tracks traversing a sensitive volume is estimated by the proportion of sensitive volumes that are traversed by a track.

### *3.5.4 RBE model based on track event theory and nanodosimetry*

Another version of the connection between the track event theory (without repair) and nanodosimetry was presented by Schneider et al. [62]. They considered two relevant target volumes of different sizes. First, a basic interaction volume (BIV) in the form of a sphere with a diameter of 2 nm represents a DNA segment of 5 to 10 base pairs and constitutes the target for formation of DSBs. On the other hand, a larger sensitive volume containing several BIVs is named the lethal interaction volume (LIV). An OTE is defined as the induction of two or more DSBs in the LIV by a single track, while a TTE is defined as two DSBs (in two different BIVs) produced by two tracks, each producing one DSB.

---

[16] In fact, Schneider et al. [132] give separate results for the track length of primary and all secondary electrons. Since this distinction is not mentioned with the introduction of the symbol for track length, it appears logical that the total track length is meant.

The probability of production of a DSB in a BIV is assumed to be equal to the probability $F_2$ of forming an IC of two or more. The formation of ICs in different BIVs within the LIV is assumed to occur statistically independently. This allows us to derive expressions for the TET parameters $p$ and $q$ in terms of $F_2$. Solving Eq. (30) for dose $D$ then provides an expression for RBE under the assumption that the parameter $q$ is independent of radiation quality. The resulting RBE is a function of the ratio of the LIV and BIV diameters and the probability $F^*(1)$, which depends on particle type and energy. The energy dependence of RBE was obtained by performing track structure simulations for photons, protons, helium ions, and carbon ions. Comparing the results to experimental data allowed to identify the best-fitting LIV diameter. The result was that the optimal LIV diameter depends on the particle type, ranging from 35 nm for photons to 6 nm for carbon ions [62].

In their analysis of the TET, Ngcezu et al. [68] noted that the model parameters $p$ and $q$ in Eq. (30) have a dimension of reciprocal dose. In contrast, the expressions for $p$ and $q$ derived in [62] are probabilities, that is, dimensionless quantities. Since the parameters are derived from geometry and the single-track quantity $F^*(1)$, it is unclear how they relate to absorbed dose. It is not explained in [62] how the link to absorbed dose would be established, presumably because the focus was on determining the ratio $\varepsilon$ of the parameters $p$ and $q$. Any normalization to dose would cancel out in this ratio. Since the expression for RBE only depends on $\varepsilon$, Normalization to dose also does not affected it. However, to compare the cell survival model with experimental data, it would be necessary to specify how the connection to dose was established. Ngcezu et al. also noted that the model considers only one nanometric LIV, that is, a small sub-cellular region, while the original TET model assumed OTEs and TTEs to occur at the chromosome level. They estimated that there are at least in the order of $10^3$ LIVs in a cell and showed that the resulting modifications to the theory presented in [62] results in a purely exponential dose dependence [68]. However, this objection is only valid if a cell contains more than one LIV.

### 3.5.5 Summary of the TET development

The development of the TET and its link with nanodosimetry was accompanied by several changes in the interpretation of the different ingredients. Key aspects of the different stages of the track-event theory are summarized in Table 4.

The first row in Table 4 shows the considered target region to which the model is applied. The second row lists the target, in which lethal or sub-lethal damage is produced. The following three rows show the definitions of OTE and TTE and whether repair is considered.

The free parameters listed are adjusted when fitting the survival model to data from cell assays. Those in the “nanodosimetry” line are determined from of ICS distribution parameters within a binomial probability framework assuming only coincident simultaneous occurrence of ICs. The “confounders” line shows whether additional information is considered when determining the model parameters based on nanodosimetry. This occurs only in one case, where the confounder variables are the probability of electron tracks to interact with the specific target volume.

Table 4: Comparison of key aspects of the TET across the publications discussed in the previous sections. For details see text.

| Ingredient | [53] | [131] | [134] | [132] | [62] |
|---|---|---|---|---|---|
| Target region | cell nucleus | | tetra-nucleosome | ? | lethal interaction volume (LIV) |
| Specific target for radiation damage | chromosome | | voxel | several | basic interaction volume (BIV) |
| OTE | 2 DSBs | 2 DSBs | 2 DSBs | 1 DSB | 1 DSB |
| TTE | 1 DSB | 1 DSB | 1 DSB / 2 SSBs | 1 SSB | 1 SSB |
| Repair | no | yes | yes | no | no |
| Free parameters | $p, q$ | $\varepsilon = p/q, R$ | $R$ | - | - |
| Nanodosimetry | - | - | $\varepsilon_{ion}$ | $\varepsilon_{ion}$ | $\varepsilon = p/q$ |
| $p_{OTE}$ | - | - | $F_2$ | $F_2$ | $P(>1\vert n, F_2)$ |
| $p_{TTE}$ | - | - | $p(1)$ | $p(1)$ | $P(1\vert n, F_2)$ |
| confounders | - | - | - | yes | - |

### *3.5.6 The model of radiation action based on nanodosimetry*

The model of radiation action based on nanodosimetry, as proposed by Schneider et al. [66], was essentially a reformulation of the TET with some developments. The fundamental model equation for the survival rate is given by Eq. (30), where the terms $qD$ and $pD$ are replaced by the mean numbers of unrepaired[17] sublethal lesions (SLs) and clustered lesions (CLs), denoted by $\bar{n}_{SL}$ and $\bar{n}_{CL}$, respectively. A lesion was assumed to be a DNA DSB and assumed to be equivalent to an IC of two or more in a spherical BIV with a diameter of 2.5 nm. A CL is assumed to be two or more DSBs produced by a single track in a sphere with a diameter of between 3 nm and 10 nm, called the cluster-lesion volume (CLV). The CLV corresponds to the LIV used in section 3.5.4. An SL is one DSB in a CLV. A pair of SLs formed by one or more tracks in different CLVs is called a distant lesion, where the two SLs potentially interact.

The mean number of unrepaired CLs, $\bar{n}_{CL}$, is related to particle fluence $\Phi$ by Eq. (34).

$$\bar{n}_{CL} = \Phi \times \sigma^* \times p_{CL} \quad (34)$$

The free parameter $\sigma^* = \sigma \times (1 - R_{CL})$ of the model is called “intersection-cross-section” and is defined such that $\Phi \times \sigma$ is the probability of a particle track to intersect any of the CLVs in a cell nucleus[18]. The term $R_{CL}$ is the probability of repairing a CL[17], and $p_{CL}$ is the probability of a track passing through the CLV to produce a CL. The mean number of SLs, $\bar{n}_{SL}$, is assumed to be given by Eq. (35).

[17] This is not stated explicitly in [66]. However, including ”cell specific parameters ... [such as] phase in cell cycle, radioresistance, repopulation and repair capability” in parameter $\sigma^*$ (written as $\sigma$ in [66]) suggests that this was intended. We believe the model is clearer when cell-specific factors are explicitly included in the repair probability $R_{CL}$.

[18] Schneider et al. write “any BIV in the cell nucleus” [66], but this appears to be a typo given that CLs and SLs refer to the CLV.

$$\bar{n}_{SL} = \Phi \times \sigma^* \times \frac{1}{8} \times (1 - R_{SL}) \times p_{SL} \tag{35}$$

The factor 1/8 was empirically determined by comparing the ratio of persistent simple and complex DSBs in radiobiological data (Appendix A of [66]), assuming two DSBs per CL. $R_{SL}$ is the dose-rate dependent (fast) repair probability of SLs and is assumed to converge to zero at high dose rates.

The relationship between particle fluence, $\Phi$, and dose, $D$, is given by Eq. (36), where $d_{BIV}$ is the diameter of the BIV, $\rho$ is the mass density, $M_1$ is the mean ICS and $W_i$ is the mean energy imparted per ionization.

$$\Phi = \frac{2}{3} \times d_{BIV} \times \frac{D\rho}{M_1 W_i} \tag{36}$$

The probabilities of CLs and SLs were related to nanodosimetry under the assumption that the probabilities of BIVs receiving an ICS of two or more are statistically independent, and that the mean number of BIVs traversed by a track, $n$, is equal to the ratio of the mean chord length of the CLV to the diameter of the BIV. When the diameter of the CLV is chosen so that $n$ is an integer, these two probabilities can be calculated from the binomial distribution, yielding Eqs. (37) and (38).

$$p_{CL} = 1 - (1 - F^*(1))^n - p_{SL} \tag{37}$$

$$p_{SL} = nF^*(1)(1 - F^*(1))^{n-1} \tag{38}$$

Since low-energy photons were considered as the primary radiation in [66], the track-averages of the nanodosimetric probabilities $F^*(1)$ were used in Eqs. (37) and (38). These averages were determined in a two-step procedure. First, the energy dependence of $F^*(1)$ was determined for electrons with initial energies between 10 eV and 1 MeV. In the second step, the spectral fluence of electrons was determined from track-length distributions obtained by transport simulations of initially monoenergetic electrons with energies equal to the mean energy of secondary electrons produced by the interaction of monoenergetic photons with energies between 280 eV and 8 keV or by 250 kV x-rays or $^{137}$Cs or $^{60}$Co gamma radiation.

The doses for survival levels of 0.1 and 0.9 were determined by inverting the model function, and RBE predictions were obtained using $^{60}$Co as the reference radiation. A comparison of the prediction for CLV diameters corresponding to $n = 1, \dots, 4$ BIV diameters per mean chord length with experimental data showed best agreement when $n$ was 2.

The model with $n = 2$ and $R_{SL} = 0$ was fitted to six radiobiological datasets of cell survival as a function of dose. This yielded values for the parameter $\sigma^*$ that correspond to circles with diameters between 5 µm and 8 µm, depending on the cell type and phase of the cell cycle.

In their analysis of the TET and derived models, Ngcezu et al. [68] pointed out that Eqs. (37) and (38) resulted from the model assumptions of statistical independence of IC formation in neighboring targets. However, this assumption contradicts the conceptual definition of tracks as collections of statistically correlated energy transfer points [38]. For proton tracks with projectile energies between 1 MeV and 100 MeV, Ngcezu et al. showed that only about 25 % of the ICs in spherical BIVs with a diameter of 3 nm are produced by primary particles traversing the BIV. Additionally, they demonstrated that for a proton fluence corresponding to a dose of 2 Gy, the probability of an IC in a BIV is in the order of magnitude $10^{-6}$-$10^{-5}$, making the probability $p_{CL}$ according to Eq. (37) negligibly small.

Ngcezu et al. [68] suggested that a model in the spirit of the TET must consider IC formation in tracks on the micrometer scale. They demonstrated this by first scoring ICs in the polyhedral unit cells of a face-centered cubic Bravais lattice with the same volume as the BIV. Then they scored clusters of BIVs with ICs in unit cells with the volume of the CV. The resulting frequency distributions of CVs containing a single BIV with ICS of at least two and multiple BIVs with ICS of at least two, respectively, were not Poisson distributed. The estimated number of CVs with a single IC or multiple ICs at a fluence corresponding to a dose of 2 Gy in a cell nucleus of 6 μm diameter was estimated to be thousands of times greater than the number of DSBs typically observed in radiobiological experiments.

This discrepancy can be explained by the fact that only a small proportion of the matter in a cell nucleus is DNA. Assuming a uniform distribution of true DNA targets in the cell nucleus covering 1 % of the nucleus volume and using this value for the success probability of a binomial distribution, the frequency distribution of single and multiple DSBs in spherical cluster volumes was obtained. Examples are shown in Fig. 8(a) and (b) for a dose of 2 Gy produced by proton irradiation at an energy of 3 MeV and 50 MeV, respectively. The results indicate that Poisson distribution can be applicable to less densely ionizing radiation but not to more densely ionizing radiation.

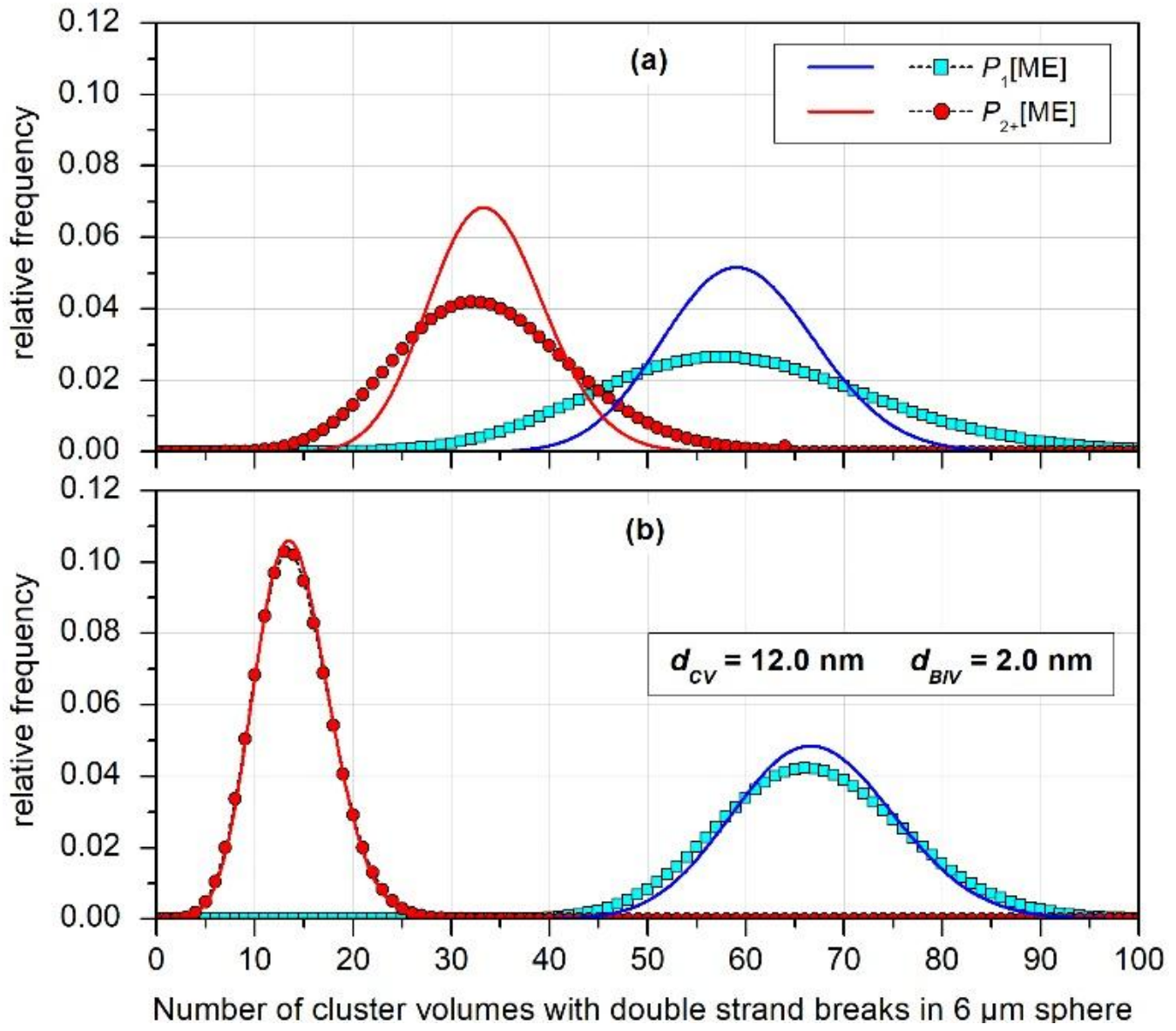

**Fig. 8:** Multi-event (ME) distributions of cluster volumes (diameter of 12 nm) inside a spherical region with a radius of 6 μm that receive a single DSB (squares) or two or more DSBs (circles) from protons of (a) 3 MeV and (b) 50 MeV energy. The data applies to a particle fluence corresponding to an absorbed dose of 2 Gy and a constant probability of 0.01 that an IC is converted to a DSB. The solid lines are Poisson distributions with the same expectation as the data shown by symbols. [19]

[19] Fig. 8 reproduced under the CC BY 4.0 license (https://creativecommons.org/licenses/by/4.0/) from Radiation and Environmental Biophysics 60, S. A. Ngcezu and H. Rabus, Investigation into the foundations of the track-event theory of cell survival and the radiation action model based on nanodosimetry, 559–578. Copyright Ngcezu and Rabus (2021).

### 3.6 The model of Conte et al. linking nanodosimetry and Inactivation cross-sections

The approach of Conte et al. [56,60,81,89,91] was based on findings within the European joint research project BioQuaRT (Biologically weighted quantities in radiation therapy) [114,136] and the Italian project MITRA (Microdosimetry and track structure). As part of the BioQuaRT project, pairwise comparisons of the three nanodosimeters existing at that time were performed by measuring the same ion beams. The Ion Counter and Startrack nanodosimeters, simulating target diameters of 2 nm and 20 nm, respectively, were compared by performing measurements at the accelerator facilities of the Italian Nuclear Research Center in Legnaro, Italy. Protons and $^{12}$C-ion beams with energies per mass between 3.5 MeV/u and 20 MeV/u were used. The comparison of the Ion Counter and the Jet Counter was performed at the Heavy Ion Accelerator Laboratory of Warsaw University in Poland. The simulated target sizes at unit density in nitrogen were 1.3 nm for the Ion Counter and 1.6 nm and 3.2 nm for the Jet Counter. Measurements were taken for 45 MeV and 76 MeV carbon ions.

The results of these measurements, along with data measured previously with the nanodosimeters for protons, $^{4}$He, and $^{12}$C ions [92,94,137–139], suggested the existence of a universal relationship between the measured mean ICS and the first three cumulative probabilities for ionization clusters of minimum sizes of one, two, and three [56,60]. This universal relation showed qualitative similarity with the LET dependence of cell inactivation cross section data in the particle irradiation data ensemble (PIDE) [124].

Cell inactivation cross sections, $\sigma_i$, are derived from the relative slopes of survival curves according to the linear-quadratic (LQ) model of cell survival, $S$, as a function of absorbed dose, $D$ [140]. It is given by Eq. (39), where $\Phi$ denotes the particle fluence.

$$\sigma_i(D) = -\frac{1}{S(D)}\frac{dS(D)}{dD} \times \frac{D}{\Phi} = -\frac{1}{S(\Phi)}\frac{dS(\Phi)}{d\Phi} \tag{39}$$

As indicated by the second identity in Eq. (39), $\sigma_i(D)$ can be interpreted in a Poisson lethal-lesion model as the expected number of additional lesions leading to inactivation caused by an additional ion hitting the cell. Obviously, $\sigma_i(D)$ depends on the dose value. In the work of Conte et al., two cases were considered: the dose corresponding to an expected survival rate of 5 % and the limit of very small doses. The corresponding values of $\sigma_i$ were denoted as $\sigma_{5\%}$ and $\sigma_\alpha$, respectively.

Since the $M_1$ and $F_k$ parameters of ICS distributions are dimensionless quantities, a linear model was assumed to fit the universal curves from nanodosimetric measurements to the experimental radiobiological data from the PIDE database. This model used proportionality factors between $M_1$ and LET, as well as between $F_k$ and $\sigma_{5\%}$ and $\sigma_\alpha$, respectively [56].

The analysis was performed on three different cell lines, namely V79, HSG and CHO cells, for which corresponding data were available in the PIDE database. For $\sigma_{5\%}$, the best match was found for the cumulative probability $F_2$ as a function of $M_1$, as illustrated in the left panel of Fig. 9 for the case of V79 cells and in the left panel of Supplementary Fig. 6 for the other cell types. In contrast, the best-fitting curve for $\sigma_\alpha$ was that of $F_3$ as a function of $M_1$, as shown in the right panels of Fig. 9 and Supplementary Fig. 6. The scaling factors for the $F_k$ were in the range between 50 μm$^2$ and 80 μm$^2$, i.e., comparable to the geometric cross sections of the cell nuclei. These values depended on cell type and varied up to 20 % between $\sigma_{5\%}$ and $\sigma_\alpha$. The scaling factors between $M_1$ and LET were established by track structure simulations. The best fits between nanodosimetric and radiobiologic data were found when the ICS distribution was determined in a target equivalent to a cylindrical water target with a diameter and height 1 nm for the $\sigma_{5\%}$ data and to a water cylinder with a diameter and a height of 1.5 nm for the $\sigma_\alpha$ data [56].

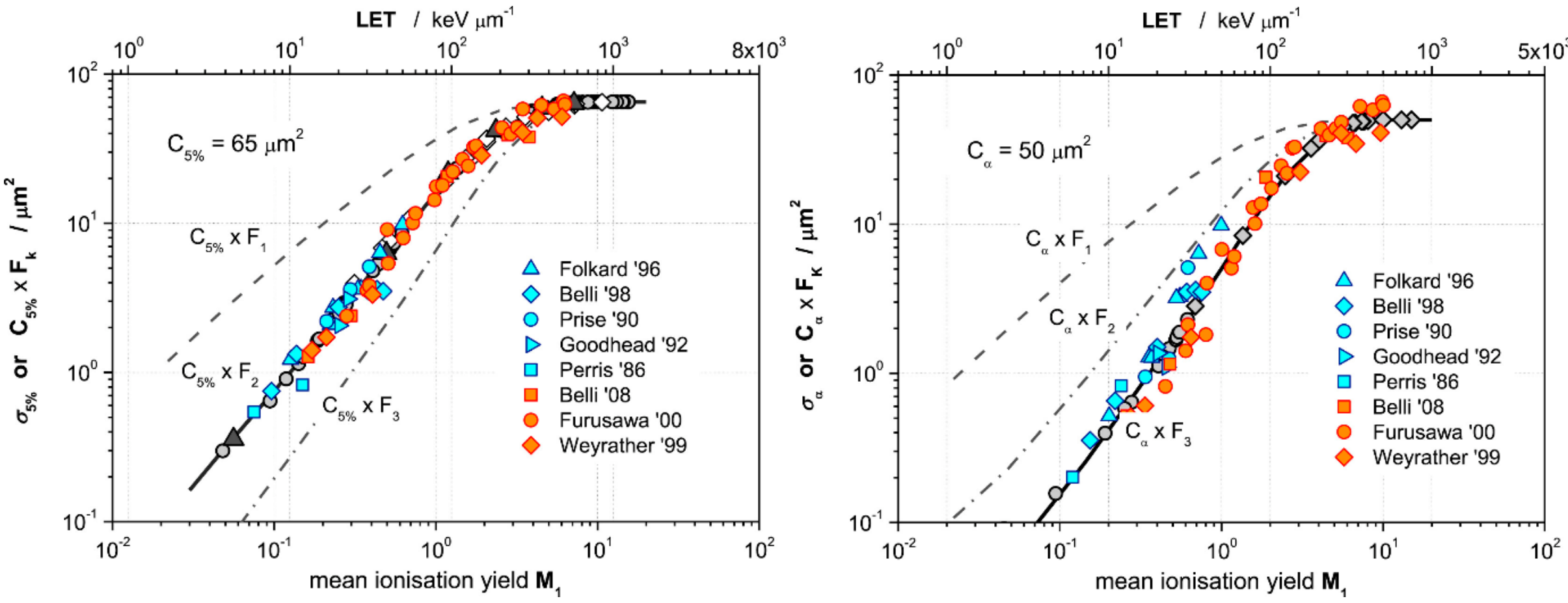

**Fig. 9**: Results of the linear fit of the measured dependence of nanodosimetric parameters $F_k \equiv F^*(k-1)$ as a function of the mean ICS $M_1$ to radiobiological data for the inactivation cross section at 5 % survival rate (left) and in the limit of small dose values (right). The symbols represent data from the PIDE database [124] plotted as a function of LET (upper x-axis). The lines represent the best fit to the nanodosimetric data scaled bi-linearly such as to match the radiobiological data.[20]

The same analysis was also performed for repair-deficient XRS5 cells, for which the best fit was obtained using the parameter $F_1 \equiv F^*(0)$ as a function of $M_1$. The scaling factor for $F_1$ was found to be 190 µm², i.e., comparatively large. The scaling factor for $M_1$ corresponded to a water cylinder with a diameter and height of 0.3 nm.

In follow-up work, the nanodosimetric dataset was extended by including nanodosimetric data taken from the literature, including electrons as primary particles [60]. A sensitivity analysis was performed for the case of $\sigma_{5\%}$ by also considering water cylinders of equal diameters and heights of 0.5 nm and 1.5 nm, demonstrating the optimality of the value 1 nm found earlier. In addition, expressions were obtained for the parameters $\alpha$ and $\beta$ of the LQ model in terms of the nanodosimetric parameters $F_2 \equiv F^*(1)$ and $F_3 \equiv F^*(2)$, determined in targets of sizes of 1.0 nm and 1.5 nm, respectively. These expressions had the form given in Eq. (40), where $S$ is a survival rate of 5 % and $c_\alpha$ and $c_S$ are proportional to the ratio of fluence $\Phi$ to absorbed dose $D$.

$$\alpha = c_\alpha F_3 \;\; ; \;\; \beta = \frac{\alpha^2 - (c_S F_2)^2}{4 \ln S} \tag{40}$$

The corresponding proportionality factors are assumed to be independent of radiation quality, so that they can be determined by “calibration” with the known values of the parameters $\alpha$ and $\beta$ for a reference radiation quality. This enabled the parameters $c_\alpha$ and $c_S$ to be calculated for any radiation quality and was shown to lead to reasonable fits of the survival curves of V79 for a selection of LET values [60,81].

Using the model functions $F^*(k|d\,)$ according to Eq. (8) and their dependence on the target size $d$ , the optimality of a target diameter of 1 nm was reconfirmed by comparison with radiobiological data. Additionally, the known functional dependence enabled the investigation of whether using linear combinations of $F^*(1)$ and $F^*(2)$ improves the fits between nanodosimetric parameters and radiobiological cross sections for inactivation. The outcome showed that, when both $F^*(1)$ and $F^*(2)$ are determined for a spherical target with a diameter of 1 nm, $\sigma_{5\%}$ and $\sigma_\alpha$ are best reproduced as proportional to $0.8F^*(1) + 0.2F^*(2)$ and $0.2F^*(1) + 0.8F^*(2)$, respectively. Notably, these combinations can also be written as $0.8p(2) + F^*(2)$ and $0.2p(2) + F^*(2)$. That is, the main difference between high and low-dose inactivation cross-sections appears to be the contribution of clusters of size 2.

---

[20] Fig. 9 is a merger of two figures reprinted from Radiation Measurements 106 (1), V. Conte et al., Track structure characterization and its link to radiobiology, 506-511, Copyright (2017), with permission from Elsevier.

The resulting functional expressions were then fitted to radiobiological data from the PIDE database [124,141], distinguishing between experiments conducted under aerobic and hypoxic conditions [91]. The results showed the best fit between $\sigma_{5\%}$ and $F^*(2)$ for hypoxic V79 and HSG cells (cf. Supplementary Fig. 7a and c). In contrast, for aerobic cells a linear combination of $F^*(2)$ and $p(2)$ had to be used, where the weight of $p(2)$ depended on the cell type (cf. Supplementary Fig. 8a and c). For V79 cells, $\sigma_\alpha$ correlated with $F^*(2)$ for aerobic conditions and with $F^*(3)$ for hypoxic conditions (cf. Supplementary Fig. 7b and d). Additional components of $p(2)$ and $p(3)$ had to be considered for aerobic and hypoxic HSG cells, respectively (cf. Supplementary Fig. 8b and d).

### 3.7 *The logistic nanodosimetry model of Dai et al. for RBE of carbon ions*

Dai et al. [65] developed the logistic nanodosimetry model (LNDM) building on the database of Ramos-Méndez et al. [58]. This model assumes that the number of lethal lesions produced by direct radiation effects is proportional to the number of targets with an ICS of two or more. For a cell nucleus in which $N$ ionizations occur, this number is estimated as $F^{c*}(1) \times \frac{N}{M_1^c}$, where both $M_1^c$ and $F^{c*}(1)$ are defined as fluence averages defined according to Eq. (14).[21] The proportionality factor, $p_L$, is the probability that a true IC (ICS at least two) will result in a lethal lesion. It is assumed to be a function of a dose-averaged conditional mean cluster size $M_1^{c2}$, where the corresponding value for a particle of given type and energy is calculated using Eq. (6) with the conditional probabilities $p^{c_2}(\nu)$ according to Eq. (5).[22] The dependence is assumed to be a logistic function with three free parameters ($k$, $r$, $m_0$) as shown in Eq. (41).

$$p_L(M_1^{c2}) = \frac{k}{1 + e^{-r(M_1^{c2} - m_0)}} \tag{41}$$

Additionally, it is assumed that all ICs (including those of ICS of 1) that do not result in a lethal lesion constitute sub-lethal lesions and that there is a constant probability of any pair of these sub-lethal lesions combining to produce a lethal lesion. This probability, $p_{sl}$, is a fourth model parameter.[23]

The model also assumes that the number of ionizations in a cell nucleus is Poisson distributed and can be related to the specific energy using a constant mean energy per ionization, $W_i$. Using these model assumptions, the parameters $\alpha$ and $\beta$ of the LQ model can be expressed as given in Eqs. (42) and (43):[23,24]

[21] Note that Dai et al. [65] used the notation $P(\nu)$, $M_1$ and $F_2$ for conditional probabilities and moments.
[22] In Eq. (9) of Dai et al. [65], the average is calculated by weighting with the product of fluence and deposited energy. Since they built on the work of Ramos-Méndez et al. [58], we assume that they used dose weighting according to Eq. (15) instead of the physically meaningless weighting by the product of fluence and energy imparted.
[23] Note that $p_{sl}$ is twice the parameter $P_{s\to l}$ used by Dai et al., who omitted the necessary division by two to obtain the number of pairs in their Eq. (14) [142].

$$\alpha = \left[F_2^c p_L(M_1^{c2}) - \frac{p_{sl}}{2}\left(1 - F_2^c p_L(M_1^{c2})\right) + \frac{p_{sl}}{2M_1^c}\left(1 - F_2^c p_L(M_1^{c2})\right)^2\right] \times {d_c}^{-1} \tag{42}$$

$$\beta = \frac{p_{sl}}{2}\left(1 - F_2^c p_L(M_1^{c2})\right)^2 \times {d_c}^{-2} \tag{43}$$

The quantity $d_c$ is the average contribution to the absorbed dose from the energy imparted to a nanometric target volume that receives at least one ionization. It is given by Eq. (44), where $W_i$ is the mean energy imparted per ionization, $M_1^c$ is the conditional mean ICS[21], and $\rho$ and $V$ are the mass density and volume of the cell nucleus, respectively.

$$d_c = \frac{W_i M_1^c}{\rho V} \tag{44}$$

It is worth emphasizing that the LNDM assumes that the fluence-averaged conditional ICS distribution is uniquely determined by the dose-weighted conditional mean ICS $M_1^{c_2}$for clusters of size $\nu \geq 2$. The LNDM also assumes that all ICs, including those of size 1, produce lesions that may result in a lethal outcome and that each pair of sub-lethal lesions can combine to form a lethal lesion. These appear as rather strong phenomenological assumptions.

The model parameters were determined by fitting Eqs. (42) and (43) to experimental data of Furusawa et al. [143]. It is not stated in [65] how the $\beta$ values were obtained, which are not reported explicitly in [143]. However, they are encoded in the $D_{10}$ values shown in [143]. The best-fit parameters for human salivary gland (HSG) cancer cells are given as $k = 9.3 \times 10^{-5}$, $r = 3.602$, $m_0 = 3.296$, and $p_{sl} = 4.0 \times 10^{-11}$.[23] Since the denominator of Eq. (41) is greater than 1, $k$ is an upper bound for $p_L$ and, as $F_2^c$ is less or equal to one, $F_2^c p_L$ is also less than $k < 10^{-4}$. Therefore, $F_2^c p_L$ can be neglected in the expressions in parentheses in Eqs. (42) and (43). This means that $\beta$ depends only on $p_{sl}$, and the expression in square brackets in Eqs. (42) deviates from $F_2^c p_L$ by a term in the order of $p_{sl}$, which is more than six orders of magnitude less than the upper bound of $F_2^c p_L$. Therefore, unless $F_2^c$ is very small, the additional terms in the square bracket may be neglected compared to the first term.

A later development of the logistic nanodosimetry model [71] used the approach of Garty et al. [49] for connecting ionizations and DNA strand breaks (Eqs. (24) and (25)), employing the value $p_s = 0.15$ from [46] and an upper limit of 10 for the summation in Eq. (24), as in the work of Casiraghi et al. [54]. Additionally, it is assumed that a DSB does not necessarily lead to a lethal event.

The probability $\hat{p}_L$ of a DSB resulting in a lethal lesion is assumed to be a function of the mean ICS, $M_1^c$, calculated with the conditional probabilities of Eq. (4). This function is given by Eq. (41) after substitution of $\hat{p}_L$ for $p_L$ and $M_1^c$ for $M_2^c$. The same substitutions also apply to the formulas for $\alpha$ and $\beta$ (Eqs. (42) and (43)). However, Yang et al. [71] only keep the first term in the brackets of Eq. (42), which corresponds to DNA lesions due to direct effects only. This simplification aligns with the order of magnitude estimate given above and results from a twofold reason. First, unlike the work of Dai et al. [65], it is not assumed that the number of ionizations follows a Poisson distribution. Second, the number of indirect lesions (formed by combinations of two sub-lethal lesions) is assumed to be proportional to the square of the number of sublethal lesions (see Eq. (11) in [71]). This second assumption implies that a sub-lethal lesion could "interact" with itself to form a lethal lesion, which appears paradoxical. This inconsistency must be considered as a caveat of the modified logistic nanodosimetric model.

[24] The corresponding formulas given in [65] are obtained by substituting $d_c$ from Eq. (44).

The order of magnitude of the parameters $k$ and $p_{sl}$ reported in [71] are the same as those obtained for HSG cells in [65]. Therefore, the argument given above that $\beta$ depends only on $p_{sl}$ still holds. Thus, both the original and modified LNDM essentially model the parameters $\alpha$ and $\beta$ independently. In the work of Yang et al. [71], data for two further cells lines were studied, in addition to the HSG cells considered in [65]. The best-fit values of the parameters $r$ and $m_0$ were the same for all three cell lines. This suggests that the relative shape of the logistic expression (Eq. (41) with $M_1^{C_2}$ replaced by $M_1^c$) as a function of $M_1^c$ may be cell-line independent. The LET-dependence of the parameters $\alpha$ and $\beta$ was fitted to both the original and the modified LNDM. No differences were found for $\beta$, which is expected since the model functions for the two LQ-model parameters are essentially uncorrelated. However, the LET-dependence of $\alpha$ obtained by the two models differs slightly. From the qualitative comparison shown in Fig. 3 of [71], a potential improvement cannot be judged. A quantitative comparison of the quality of fit is not given in [71].

### 3.8 *The cluster-dose concept of Faddegon et al.*

The concept of cluster dose was introduced by Faddegon et al. [37] and builds on the concept of ionization detail introduced by Ramos-Méndez et al. [58] and its proposed use in treatment planning for hadron therapy [63]. In the original work on the cluster dose concept and follow-up articles [73,112,144], the computational realization was presented as part of the concept. For clarity, the explanation of the concept and its realization are treated separately below.

Table 5: Comparison of terminology and notation used in the present work and in literature related to the cluster dose concept.

| Quantity | **This work** | **Literature** |
|---|---|---|
| Type and energy of particle | $Q = \{E, p\}$ | Particle class $c$ |
| Type and energy interval of particle | $\{E_i, \Delta E_i, p\}$ | Particle class $c$ |
| mean number of ICs of size $\nu \geq 1$ per path of a primary particle | linear event frequency, $\frac{d\bar{N}_0^*}{dx}$ | - |
| mean number of ICs of size $\nu$ per path of a primary particle | linear yield of ICS $\nu$, $\frac{d\bar{N}_\nu}{dx}(E,p)$ | frequency distribution of $\nu$, $f^c(\nu)$ [37], $f(\nu)$ [72,112] |
| mean number of ICs of size $\nu \geq k$ per path of a primary particle | linear yield of ICS $\nu \geq k$, $\frac{d\bar{N}_{k-1}^*}{dx}(E,p)$ | sum of ICS frequencies with $\nu \geq k$, $F_k^c$ [37], $F_k$ [112,144], $C_k$ [72] |

The concept proposed by Faddegon et al. [37] differs from the work Ramos-Méndez et al. [58] in two aspects. First, instead of considering nanodosimetric ICS probabilities $p(\nu)$, the absolute frequency distribution of ICs per unit path of a particle is considered and denoted by $f^c(\nu)$. The superscript "$c$" denotes the particle class defined by particle type and energy. In this author's assessment, $f^c(\nu)$ is identical to the linear yield of clusters of size $\nu$, $d\overline{N}_\nu/dx(E,p)$, introduced in Section 2.4, where the parameters defining the particle class (particle energy *E* and type *p*) are shown explicitly. Table 5 shows a comparison of the terminology and notation used in this work with that used in literature [37,72,73,112,144].

The second difference is the introduction of ionization detail (ID) parameters $I_p$, which are defined as a mapping of the frequency distributions $f^c(\nu)$ onto a single value by applying an operator defined on the space of all possible frequency distributions. The second difference is the introduction of ionization detail (ID) parameters $I_p$, which are defined as a mapping of the frequency distributions $f^c(\nu)$ onto a single value by applying an operator defined on the space of all possible frequency distributions. This appears conceptually equivalent to the definition of a metric on this space. Explicitly determining this metric or operator and, thus, the ultimate ID parameter are "left for future research" [37]. It is assumed that these ID parameters scale with material density.

Examples for ID parameters listed in [37] are the sums of ICS frequencies with $\nu \geq k$, denoted by $F_k$, and the sum of cluster size frequencies for $\nu \geq k$ weighted by cluster size $\nu$, denoted by $N_k$. According to our notation, the former would be denoted by $dN^*_{k-1}/dx$, and the latter can be written as $M_1^{Ck} \times dN^*_{k-1}/dx$, where $M_1^{Ck}$ is the mean ICS conditional on at least $k$ ionizations in the cluster. In later work [72,73,112,144], only $F_k$ was used.

In all these cases, the ID parameter $I_p^c$ is a linear function of the $f^c(\nu) \equiv dN_\nu/dx(E,p)$ , which can be expressed by (45).

$$I_p^c = \sum_\nu w_\nu \frac{dN_\nu}{dx}(E,p) \tag{45}$$

Here, $w_\nu$ is the weight associated with an IC of size $\nu$. One possible weighting scheme assigns a weight of unity to ionization clusters that contain at least a minimum number of ionizations and a weight of zero to clusters with fewer ionizations than this minimum value. This special case has been employed in different studies [72,73,103,144,111], albeit with different minimum values. However, the concept also permits cases in which ionization clusters of different complexity have different weights to account for their varying relevance in inducing biological damage. For example, for the quantity $N_k$ (in the notation of [37]), the weights are zero for $\nu < k$ and identical to $\nu$ for $\nu \geq k$.

However, then cluster dose can be defined as the expectation of the weighted sum of ionization clusters per mass. Similar to absorbed dose, which under conditions of charged particle equilibrium can be expressed as the integral of the fluence of charged particles, weighted by the stopping power (unrestricted linear energy transfer), over energy and particle types, cluster dose $g^{(w)}$ can be expressed by Eq. (46).

$$g^{(w)} = \frac{1}{\rho}\sum_p \int \Phi_{p,E}(E) \sum_\nu w_\nu \frac{d\overline{N}_\nu}{dx}(E,p;\rho)\, dE \tag{46}$$

Herein, $\rho$ is the mass density, $\Phi_{p,E}(E)$ is the spectral fluence of particle type *p* (number per area and energy interval), $\nu$ is the ionization cluster size, $w_\nu$ is the weight associated with an ionization cluster of size $\nu$, and $\frac{dN_\nu}{dx}$ is the expected number of ionization clusters of size $\nu$ produced per pathlength by a particle of type $p$ and energy $E$. Note that the term particle type includes the state of charge, so that atoms and their ions as well as ions of different charge are considered as distinct particles.

Analogous to absorbed dose, the definition of cluster dose given here in Eq. (46) is for a point quantity. However, for practical purposes, averages over voxels are considered (Eq. (47)).

$$\bar{g}_j^{(w)} = \frac{1}{\rho_j} \sum_p \int \bar{\Phi}_{p,E}(E) \sum_\nu w_\nu \frac{d\bar{N}_\nu}{dx}\left(E, p; \rho_j\right) dE \tag{47}$$

Here, $g_j^{(w)}$ denotes the mean cluster dose in voxel $j$, $\rho_j$ its mass density, and $\bar{\Phi}_{p,E}$ the volume average of the spectral fluence in the voxel. Additionally, the particle fluence is typically obtained by simulations (or measurements) as an average fluence in energy intervals. Therefore, the integral over the energy is approximated by a summation over the energy intervals, and the voxel-average ionization cluster dose can be expressed by Eq. (48).

$$\bar{g}_j^{(w)} = \frac{1}{m_j} \sum_p \sum_i \bar{\Phi}_{p,E}(E_i) \Delta E_i V_j \times \sum_\nu w_\nu \frac{d\bar{N}_\nu}{dx}\left(E_i, p; \rho_j\right) \tag{48}$$

Here, $V_j$ denotes the volume of the voxel, $m_j$ denotes its mass, $\rho_j$ denotes the resulting mass density, and $E_i$ is the center of the i$^{th}$ energy bin.

The definition in Eq. (48) is the same as that given in Eq. (5) of [37] which is reproduced here as Eq. (49):

$$g_j^{(I_p)} = \frac{1}{m_j} \sum_c \frac{\rho_j}{\rho_0} t_j^c I_p^c = \frac{1}{m_j} \sum_c t_j^c \times \left(\frac{\rho_j}{\rho_0} I_p^c\right) \tag{49}$$

Here, $m_j$ is the mass of voxel $j$, $\rho_0$ is the mass density of the material in which the ID parameter $I_p^c$ is determined (e.g., water), and $t_j^c$ is the track length of particles in class $c$ in voxel $j$.

To see that Eqs. (48) and (49) are equivalent, we remember that particle class is defined by particle type and energy interval. Therefore, summing over particle classes means summing over both particle type $p$ and energy bins. We further consider that the ratio of track length in a volume to the content of the volume is the fluence [145,146]. Therefore, the track length of particles in class $c$ is equal to the fluence in the energy interval of width $\Delta E_i$ around $E_i$ multiplied by the volume $V_j$ , that is, $\bar{\Phi}_{p,E}(E_i)\Delta E_i V_j$ . According to the scaling assumption, the last term in parentheses is the ID parameter in the medium with density $\rho$. Under the assumption that the ID parameter is a linear function of the linear yield of clusters, the term in parentheses in Eq. (49) is therefore equal to the summation over $\nu$ in Eq. (48), so that Eqs. (48) and (49) are equivalent.

In practical applications, particle fluence in voxels is obtained from condensed history simulations of particle transport that consider the average material composition of voxels within a given geometry. However, since track structure codes generally only allow simulations in water (or a limited combination of water and DNA constituents, as in Geant4-DNA), the determination of the number of clusters per pathlength generally relates to a different material than that in a voxel. Therefore, Schwarze et al. [147] proposed a density and material scaling approach for the quantities $\frac{d\bar{N}_\nu}{dx}(E_i, p)$ and used it to determine cluster dose in carbon ion therapy [111].

It is important to emphasize that, although the quantities $f^c(\nu) = \frac{d\bar{N}_\nu}{dx}(E_i, p)$ were determined by simulations using randomly distributed target cylinders in a larger cylinder of dimensions corresponding to a chromatin fiber, the quantity determined is the number of ionization clusters per particle pathlength [37]. Therefore, even though the corresponding figures in the articles may suggest that spatial filtering to DNA regions was applied, this is not the case. Therefore, the quantity cluster dose is defined in full analogy to the absorbed dose as described above. Thus, scaling of the number of clusters produced per pathlength in different materials, analogous to the one introduced to nanodosimetry by Grosswendt [25], seems appropriate. Otherwise the cluster dose would be independent of the material in the voxel [147].

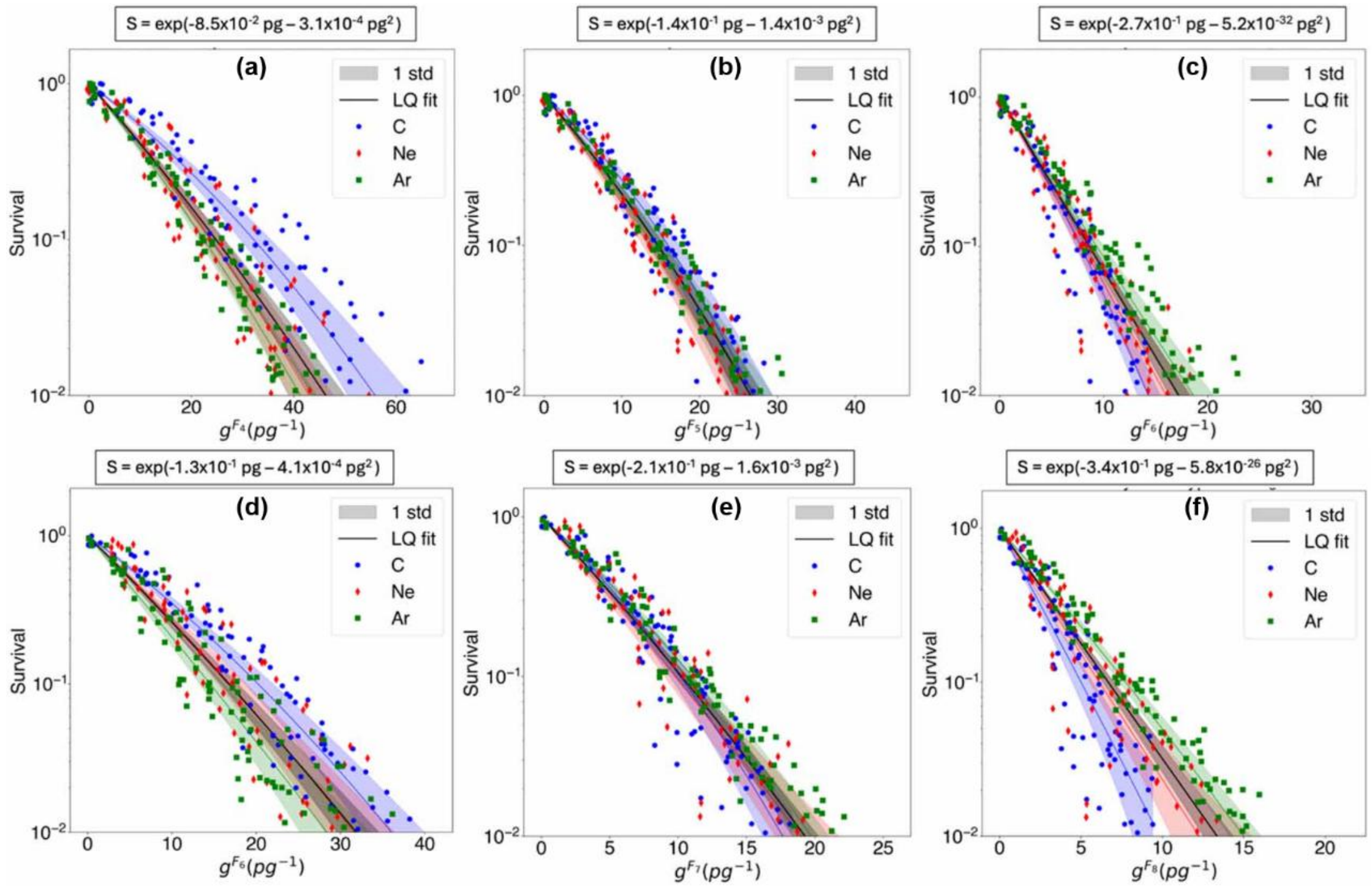

**Fig. 10**[148]bottom[25]

Ortiz et al. [72] investigated the degree of association between the cluster dose concept and cell survival considering cluster dose from ICs of a given minimum number of ionizations and three datasets from cell survival studies. The underlying nanodosimetric database was constructed from simulations using cylindrical targets with diameters of 2.3 nm and heights of 3.4 nm that were closely packed in cubes of 100 nm side to which periodic boundary conditions were applied for electron tracks [112]. An optimized energy binning was used to produce the database, ensuring that the relative variation of the parameters $F_k$ between adjacent energy points did not exceed 5 %.

Fig. 10 shows survival data of human kidney cells under heavy ion irradiation under aerobic (top row) and hypoxic conditions (bottom row). The data were taken from [148], and the reported dose values were converted to cluster doses for different minimum ICS for each ion type and energy. The middle column shows the minimum ICS for which the data for the different ions and the corresponding fit curves agree best. The left and right panels show how the picture changes when the minimum ICS is 1 smaller or larger, respectively. For the aerobic cell data shown in the top row, the best agreement occurs for a minimum ICS of 5. In contrast, the best agreement for the hypoxic cell data in the bottom row occurs for a minimum ICS of 7.

The degree to which cell survival is associated with cluster dose was assessed in [72] using three different statistical metrics, including the associated uncertainties. The first metric was a sliding window approach, which essentially averaged survival data and fit values across dose intervals. The second metric was the mean ratio of absolute fit error to the value of the fit curve at each dose. The third method was based on a Bayesian information criterion from pattern recognition using neural networks based on the likelihood concept. All metrics were determined using the combined dataset of all ion types employed to irradiate the cells, which was fitted to an LQ-model of cell survival as a function of cluster dose. Additionally, the metrics were also evaluated separately for each ion type. The variation of these three metrics with the minimum ICS used for calculation of the cluster dose can be seen in Supplementary Fig. 9. The rows correspond to the different metrics, and the columns correspond to the datasets of different cell types and oxygenation states. The minimum, indicating the best association, is consistently found across all metrics at a cluster dose with ICs of at least 5 for aerobic cells and at a cluster dose with ICs of at least 7 for hypoxic cells (Supplementary Fig. 9).

[25] Fig. 10 reproduced under the CC BY 4.0 license (https://creativecommons.org/licenses/by/4.0/) from Physics in Medicine and Biology 70, R. Ortiz, J. Ramos-Méndez, J.-H. Mao, R. Schulte, B. Faddegon, Evaluation of nanodosimetric quantities for ion radiotherapy treatment planning based on the degree of association of survival with cluster dose, 195004, merged into a single figure, adding the labels (a) to (f). Copyright Ortiz et al. (2025).

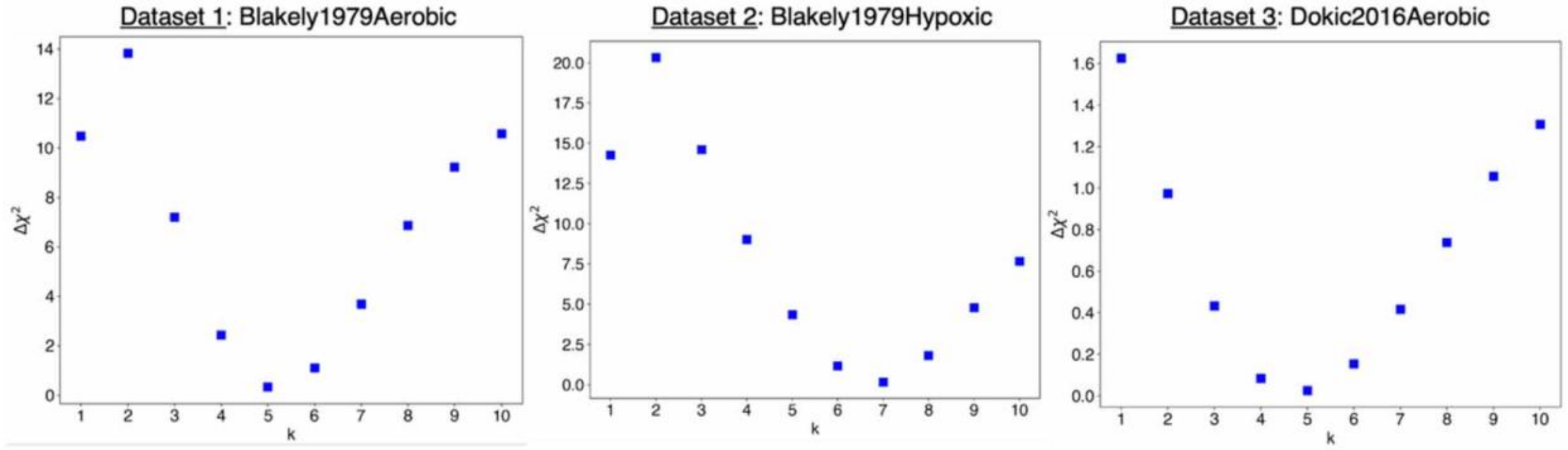

**Fig. 11**: Variation of the difference in the $\chi^2$ value between a fit of the data from all ion types to the LQ model and the sum of the $\chi^2$ values from fits per ion type. The *x*-axis of the plots is the minimum number of ions in an IC used in the calculation of the cluster dose. The left and middle are results for survival of human kidney cells under heavy ion irradiation under aerobic and hypoxic conditions, respectively [148]. The right panel is for data of aerobic A549 cells irradiated with protons, helium, carbon and oxygen ions [149].[26]

In addition, $\chi^2$ values were determined for fits of the LQ-model for the combined dataset of all ion types, as well as for separate fits to the data of each ion type (shown as lines in Fig. 10). Fig. 11 shows the difference, $\Delta\chi^2$, between the LQ-fit to the combined datasets and the sum of the $\chi^2$ values for the separate fits per ion species. This figure confirms the visual impression from Fig. 10 that, for aerobic cells, the smallest $\Delta\chi^2$ indicating the best association for all ion types is obtained with a cluster dose corresponding to a minimum ICS of 5, even when different cell types are considered. For hypoxic kidney cells and heavy ion irradiation, Fig. 11 shows the smallest $\Delta\chi^2$ at a minimum ICS of 7.

### 3.9 *The binomial weighting model for cell inactivation by Mietelska et al.*

Unlike the correlation approach of Conte et al., which was completely agnostic to the mechanism connecting ionization clusters and radiobiological effects, the approach of Mietelska et al. [70] includes a mechanistic component in the form of a weighting function for the probabilities of different ICs. Similar to the combinatorial approach of Garty et al. [45,49], a binomial distribution is assumed for the conditional probability of an IC of size $\nu$ to result in $k$ DNA lesions. The success probability, $p_s$, of the binomial process is assumed to be independent of the ICS.

[26] Fig. 11 reproduced under the CC BY 4.0 license (https://creativecommons.org/licenses/by/4.0/) from Medical Physics 51, R. Ortiz, B. Faddegon, Creating uniform cluster dose spread-out Bragg peaks for proton and carbon beams. 4482–4488.Copyright Ortiz et al. (2024).

The probability of a given number of DNA lesions, $m$, is then obtained using the derived nanodosimetric parameters, $R_m$, as defined by Eq. (50).

$$R_m(p_s) = \sum_{\nu=m}^{\infty} \sum_{k=m}^{\nu} p(\nu) B(k|\nu, p_s) \tag{50}$$

Given the normalization of the conditional probabilities and the definitions of $F_m$ and the binomial distribution, Eq. (50) can be rewritten as Eq. (51).

$$R_m(p_s) = F_m - \sum_{\nu=m}^{\infty} p(\nu) \sum_{k=0}^{m-1} \binom{\nu}{k} p_s^k (1-p_s)^{\nu-k} \tag{51}$$

Mietelska et al. only considered the case $m = 2$, which results in Eq. (52).

$$R_2(p_s) = F_2 - \sum_{\nu=2}^{\infty} p(\nu)(1-p_s)^{\nu} - p_s \sum_{\nu=2}^{\infty} \nu p(\nu)(1-p_s)^{\nu-1} \tag{52}$$

The parameter $R_2$ is assumed to represent the probability of a DNA double-strand break. Therefore, while the approach of Mietelska et al. [70] contains elements of the approach of Garty et al. [45,49], it considers two DNA lesions sufficient for induction of a DSB without taking into account the location of these lesions on the two strands. In this respect, the approach of Mietelska et al. appears to align with the approach of Grosswendt. Additionally, it is assumed that a single DSB is sufficient to inactivate a cell.

Mietelska et al. [70] performed simulations using Geant4-DNA for pencil beams of $^{1}$H, $^{4}$He, $^{7}$Li, $^{11}$B, $^{12}$C, $^{14}$N, $^{16}$O, and $^{28}$Si ions in water. The simulations covered a range of energies between 200 keV and 1 GeV and a range of LET between 6 keV/μm and 650 keV/μm. The target volumes were cylinders with equal diameter and height, ranging from 0.8 nm to 2.5 nm. They also considered cylinders with a height of 1.48 times the diameter and diameter values between 1.0 nm and 2.3 nm.

For a cylindrical target with a diameter of 2.3 nm and a height of 3.4 nm, the ratio $\frac{R_2}{M_1}$ plotted versus LET showed clear differences between hydrogen ions, helium ions, and all heavier ions. However, the data of the different ions coincided when plotted against $M_1$, except for mean ICS values below about 1.5. At these lower $M_1$ values, the heavier ions showed much lower values than hydrogen and helium ions [70]. These findings presumably pertain to a value $p_s = 0.35$, i.e., the best fitting parameter for this target size. Unfortunately, the value is not stated in [70], neither in the figure caption nor the text related to the figure.

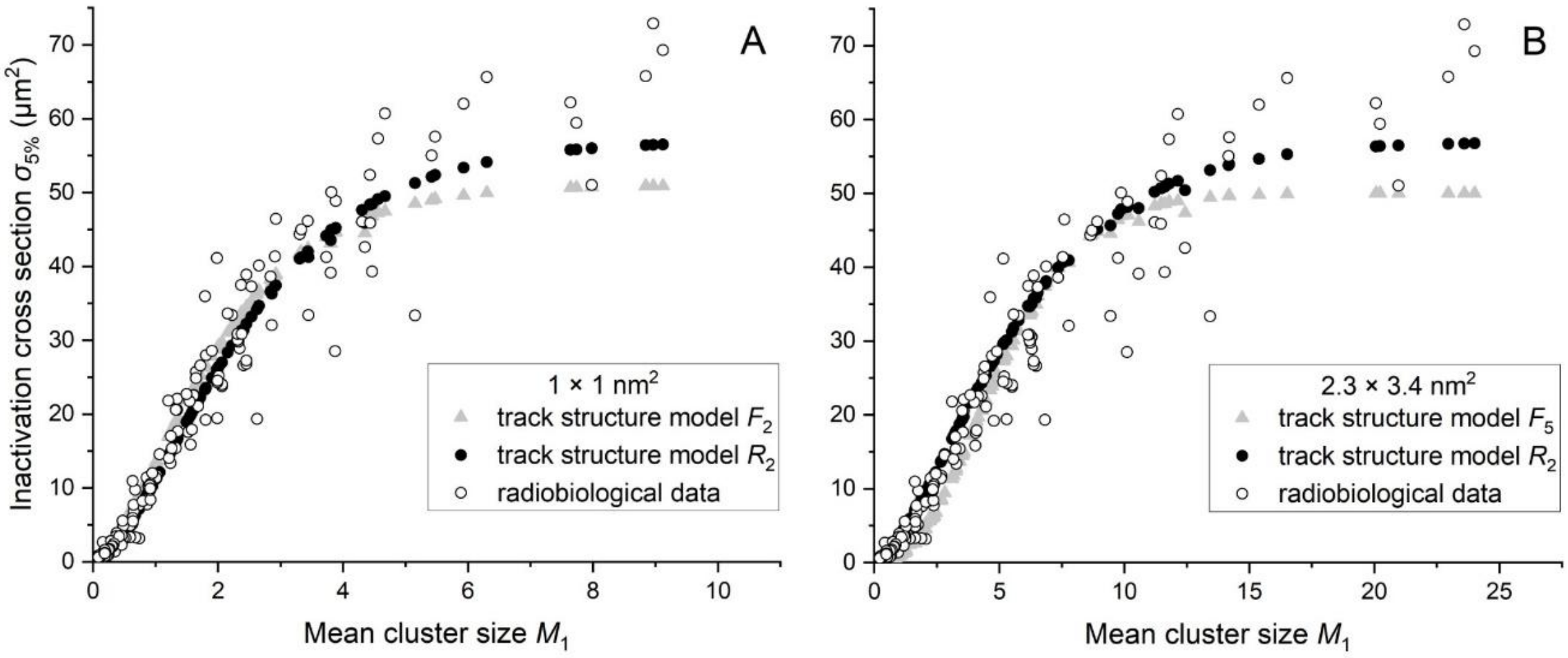

**Fig. 12:** Comparison of $R_2$ and $F_k$ curves yielding the best fit with radiobiological data for cylindrical targets of (A) both diameter and height of 1 nm (and hence a cross section of 1 × 1 nm$^2$) and (B) diameter of 2.3 nm and height of 3.4 nm (i.e., cross section of 2.3 × 3.4 nm$^2$). In (A), the optimal fit is achieved with $F_2$ and $R_2$ for $p$ equal to 0.8, while in (B), the best fit is observed for $F_5$ and $R_2$ for $p$ equal to 0.35. In both cases, $R_2$ provides a better fit than $F_k$.[27]

The data for $M_1$ and $R_2$ determined from the simulated data for different target geometries and a set of values for $p_s$ were fitted to $\sigma_{5\%}$ data extracted from the PIDE database [124,141] for V79 cells. The model parameter $p_s$ that provides the best fit (least $\chi^2$) was determined for targets of different sizes using a fixed scaling factor between probabilities and cross sections. The optimal $p_s$ parameter varied from 0.8 for targets with a diameter of 1 nm and 0.35 for targets with a diameter of 2.3 nm. $R_2$ curves based on these optimum parameter values can be seen in Fig. 12 to better fit the experimental $\sigma_{5\%}$ data at large $M_1$ values than the fits to $F_2$ and $F_5$ for targets with diameter and height of 1 nm and targets with diameter of 2.3 nm and height of 3.4 nm, respectively.

It should be noted that the presentation of data in [70] and in Fig. 12 uses linear scales, which emphasizes the high LET regions. Plotting the data on logarithmic axes would make the comparison easier to assess. Additionally, it is unclear how the $\chi^2$ values were determined, and the difference in $\chi^2$ between the data fits using $R_2$ and those using $F_2$ or $F_5$ is not reported. This makes the significance of the apparent improvement unclear and hinders a full appreciation of the benefits of introducing an additional adjustable parameter. However, Supplementary Fig. 10 shows that using the probability $R_2$ appears to maintain proportionality between the parameter and the predicted biological effect over the entire range of values. In other words, it solves the problem that predictions based on a parameter $F_k$ face, namely that the biological effect may still vary significantly when $F_k$ approaches its upper limit of unity (Fig. 5).

[27] Fig. 12 reproduced under the CC BY 4.0 license (https://creativecommons.org/licenses/by/4.0/) from International Journal of Molecular Sciences 25, M. Mietelska, M. Pietrzak, A. Bancer, A. Ruciński, Z. Szefliński, B. Brzozowska, Ionization Detail Parameters for DNA Damage Evaluation in Charged Particle Radiotherapy: Simulation Study Based on Cell Survival Database, 5094. Copyright Mietelska et al. (2024).

### 3.10 The nanodosimetric quantity-weighted dose of Yang et al.

Yang et al. [69] compared dose-weighting approaches in carbon-ion therapy using RBE weighting factors derived from microdosimetry and nanodosimetry. They introduced a radiation quantity named nanodosimetric quantity-weighted dose ($D^Q_{a-b}$), as defined by Eq. (53).

$$D^Q_{a-b} = k \times \frac{F^c_{a-b,Q}}{F^c_{a-b,X}} \times D \tag{53}$$

Here, $F^c_{a-b}$ are the conditional probabilities for ICS between $a$ and $b$, as defined in Eq. (3). The superscripts $Q$ and $X$ refer to the radiation quality of the ion beams and of X-rays used as reference radiation, respectively. $D$ is the absorbed dose, and $k$ is an adjustable, cell-line-specific model parameter.

They assessed whether the experimental survival data obtained by Furusawa et al. [143] at different LET values collapsed onto a common survival curve when plotted as a function of the nanodosimetric quantity-weighted dose. Furusawa et al. [143] performed irradiations of HSG and V79 cells at different depths in water with a 135 MeV/u primary carbon ion beam. In the simulations, monoenergetic photons with an energy of 200 keV were used as the reference radiation instead of the 200 kVp X-ray spectrum used in the experiments. This substitution was justified by the argument that photon energy has little influence on nanodosimetric quantities.

In their analysis, Yang et al. [69], tested the cumulative IC probabilities with $a = 3$ and $b \in \{8,10,12\}$ in Eq. (3), as well as the probabilities $F^c_2$, $F^c_3$, and $F^c_4$, for which the upper bound $b$ in Eq. (3) is formally infinity. The experimental cell survival fractions plotted as a function of $D^Q_{3-10}$ appeared to follow the same curve for all considered values of LET. The experimental RBE plotted as a function of the ratio of the $F^c_{a-b}$ probabilities in Eq. (53) showed the highest coefficient of determination with the $F^c_{3-10}$ probability. The cell-specific scaling factor $k$ was found to be in the order of magnitude of unity, differing by about 25 % between the two cell lines.

## 4 Discussion

The models presented in the previous section are summarized in Table 6. The first column distinguishes their different development steps where appropriate. The second column shows the nanodosimetric information used, such as whether the ICS distribution (D) is measured or simulated, the irradiation geometry (G) to which the ICS distribution pertains, and target (T) geometry and material (with water as default). The third column shows the biological aspects: the endpoint (E) considered, the hypotheses (H) made regarding the link between nanodosimetric parameters and radiobiological quantities, and the model's free parameters (P), if applicable. The last column lists the data used for benchmarking (B) or for determining parameters by data fitting (F), as well as known limitations (L).

Table 6: Compilation of the key features of the models linking nanodosimetry and biological effects. If not specified otherwise, the target is in water and surrounded by water. Its dimensions are either the diameter $d_s$ of a target sphere or the diameter $d_c$ and height $h_c$ of a target cylinder.

| **Model** | **Nanodosimetry**<br>D: origin of data for ICS distributions<br>G: irradiation geometry<br>T: target dimensions, material | **Biology**<br>E: targeted endpoint<br>H: hypotheses<br>P: free parameters (if applicable) | **Calibration / validation**<br>B: data used as benchmark<br>F: data for fitting parameters<br>L: limitations (if applicable) |
|---|---|---|---|
| Two-compartment model [36] (section 3.1) | D: derived from microdosimetric simulations in water<br>G: pencil beam<br>T: $d_c$ = 2 nm, $h_c$ = 16 nm | E: cell survival<br>H: $\alpha$ depends on both $q_1$ and $q_2$, $\beta$ only on $q_1$, $q_1$, $q_2$: values of $F_{2-5}$ and $F_{6-10}$ relative to a reference radiation quality | B: literature data of V79 cell survival [115] |
| Grosswendt's approach [77–79] (section 3.2) | D: simulations in water<br>G: circular with diameter 2 nm width [78], 2.3 nm, height 3.4 nm [77,79]<br>T: $d_c$ = 2 nm, $h_c$ = 2 nm [78], $d_c$ = 2.3 nm, $h_c$ = 3.4 nm [77,79] | E: DNA damage (SSBs, DSBs)<br>H: $p_{SSB} \propto p(1)$, $p_{DSB} \propto F^*(1)$, $p_{DSB}$<br>P: proportionality constants for yield of SSB and DSBs and their ratio | B: Simmons and Watt [118], PARTRAC [116,120,122], Nikjoo et al. [123], DNA damage assays [117]<br>L: $F^*(1) \ll 1$ [126] |
| Combinatorial model [45,49] (section 3.3) | D: experiment, scaled to liquid water<br>G: 56 nm height and width<br>T: $d_c$ = 4.5 nm, $h_c$ = 8 nm [45] $d_c$ = 4.3 nm, $h_c$ = 6 nm [49] | E: DNA damage (SSBs, DSBs)<br>H: ICS distribution translates to DNA damage via bi-/trinomial conditional probability<br>P: probability of strand break $p_s$ and (only [49]) base damage | B: PARTRAC [116,120], DNA damage assays [150–154]<br>F: plasmid DNA assays [129] |
| Nanodosimetry-based radiation quality factor [46] (section 3.4) | D: simulated Ion Counter measurements<br>G: circular beam with diameter of 3 cm<br>T: ion collection efficiency map in cube of 5 cm side filled with propane | E: complex DNA DSBs<br>H: radiation risk factor is given by yield of complex DSBs relative to reference radiation<br>P: probability of strand break $p_s$ | B: ICRP Report 60 [155]<br>F: taken from [45] |
| Track-event theory [62] (section 3.5) | D: simulations<br>G: not specified<br>T: $d_s$ = 2.0 nm | E: RBE for cell survival<br>H: two or more DSBs are lethal, single DSB non-lethal, $p_{DSB} = F_2$, Poisson distribution of numbers of OTEs and TTEs<br>P: diameter of clustering volume $d_{CV}$ | F: experimental RBE data [156–158]<br>L: unclear how nanodosimetric parameters link to dose [68] |

| Model | Nanodosimetry<br>D: origin of data for ICS distributions<br>G: irradiation geometry<br>T: target dimensions, material | Biology<br>E: targeted endpoint<br>H: hypotheses<br>P: free parameters (if applicable) | Calibration / validation<br>B: data used as benchmark<br>F: data for fitting parameters<br>L: limitations (if applicable) |
|---|---|---|---|
| Nanodosimetric radiation action model [66] (section 3.5.6) | D: simulations<br>G: not specified<br>T: 2.5 nm | E: cell survival<br>H: two or more DSBs are lethal CL, single DSB non-lethal SL, $p_{DSB} = F_2$, Poisson distribution of numbers of CLs and SLs<br>P: diameter of clustering volume $d_{CV}$, repair probabilities $R_{CL}$ and $R_{SL}$ | F: experimental survival data [159–165]<br>L: ICs and clusters of ICs are not Poisson distributed [68] |
| Conte et al. [56] (section 3.6) | D: experiments and simulations<br>G: experiments: beam centered on target volume; simulations: pencil beam<br>T: experiments: sensitive detector volume; simulations: $d_c = h_c$ = 1.5 nm | E: inactivation cross sections $\sigma$<br>H: $M_1 \propto$ LET, $\sigma \propto F_k$<br>P: proportionality constants for $M_1$ and $F_k$ | F: PIDE database [124] |
| Conte et al. [60] (section 3.6) | D: experiments and simulations<br>G: experiments: beam centered on target volume; simulations: pencil beam<br>T: experiments: sensitive detector volume; simulations: $d_c = h_c$ = 0.5, 1.0, 1.5 nm | E: inactivation cross sections $\sigma$; cell survival<br>H: $M_1 \propto$ LET, $\sigma \propto F_k$<br>P: proportionality constants for $M_1$ and $F_k$, target size $d_c$ | B: experimental survival data [166]<br>F: PIDE database [124] |
| Conte et al. [89] (section 3.6) | D: simulations<br>G: pencil beam<br>T: simulations: $d_s$ = 1, 2, 5, 10 nm | E: inactivation cross sections $\sigma$<br>H: $\sigma \propto$ Linear combination of $F_2$ and $F_3$<br>P: prop. constants and $d_s$ values for $F_2$ and $F_3$ | F: PIDE database [124] |
| Conte et al. [91] (section 3.6) | D: simulations<br>G: pencil beam<br>T: simulations: $d_s$ = 1.0 nm | E: inactivation cross sections $\sigma$ separate for aerobic or hypoxic cells<br>H: $\sigma \propto$ Linear comb. of $F_2$ and $F_3$ or $F_3$ and $F_4$<br>P: prop. constants for $F_2$ and $F_3$ or $F_3$ and $F_4$ | F: PIDE database [124] |
| Logistic nanodosimetry model [65] (section 3.7) | D: simulations, data from [58]<br>G: complex impact parameter distribution for 1800 randomly positioned and oriented targets in a larger cylinder<br>T: $d_c$ = 2.3 nm, $h_c$ = 3.4 nm | E: RBE for cell survival<br>H: $\nu \geq 2$ is a DSB, $\nu = 1$ is SSB, probability of lethal DSB is a logistic function of dose-averaged $M_2^c$; $\beta \propto$ probability $p_{sl}$ of lethal combination of two non-lethal DSB or SSB<br>P: three parameters of logistic function plus probability $p_{sl}$. | F: experimental RBE data [143] |

| **Model** | **Nanodosimetry**<br>D: origin of data for ICS distributions<br>G: irradiation geometry<br>T: target dimensions, material | **Biology**<br>E: targeted endpoint<br>H: hypotheses<br>P: free parameters (if applicable) | **Calibration / validation**<br>B: data used as benchmark<br>F: data for fitting parameters<br>L: limitations (if applicable) |
|---|---|---|---|
| Modified logistic nanodosimetry model [71] (section 3.7) | D: simulations<br>G: complex impact parameter distribution; phase space shrinking to the caps of a cylinder containing 1800 randomly positioned and oriented targets<br>T: $d_c$ = 2.3 nm, $h_c$ = 3.4 nm | E: RBE for cell survival<br>H: $p_{DSB}$ is obtained by binomial weighting from $p^c(\nu)$ for $1 \leq \nu \leq 10$, probability of lethal DSB is a logistic function of $M_1^c$; $\beta \propto p_{sl}$ of lethal combination of two non-lethal DSB<br>P: three parameters of logistic function plus probability $p_{sl}$ | F: experimental RBE data [143,167] |
| Cluster dose [37,72,73,112,144] (section 3.8) | D: simulations<br>G: complex impact parameter distribution for 1800 targets randomly positioned and oriented in a larger cylinder [58] or close-packed in a cube of 100 nm side [112]<br>T: $d_c$ = 2.3 nm, $h_c$ = 3.4 nm | E: RBE for cell survival<br>H: survival curves as a function of cluster dose are independent of ion type<br>P: minimum number of ionizations in IC | F: experimental survival data [148,149,168] |
| Binomial weighting model [70] (section 3.9) | D: simulations<br>G: pencil beam<br>T: $d_c = h_c$= 0.8, 1.0, 1.3, 1.5, 1.8, 2.0, 2.3 and 2.5 nm;<br>$h_c = 1.48 \times d_c$ with $d_c$= 1.0, 2.0, 2.3 nm | E: DSBs and cell inactivation cross section $\sigma$<br>H: $p_{DSB}$ is proportional to binomially weighted ICS distribution. Probability of cell inactivation is proportional to $p_{DSB}$.<br>P: Proportionality factor for $\sigma$, $d_c$, $h_c$, probability $p_s$ of ionization producing strand break | F: PIDE database [124,141] |
| Nanodosimetric quantity-weighted dose [69] (section 3.10) | D: simulations<br>G: complex impact parameter distribution; phase space shrinking to the caps of a cylinder containing 1800 randomly positioned and oriented targets<br>T: $d_c$ = 2.3 nm, $h_c$ = 3.4 nm | E: RBE for cell survival<br>H: RBE is proportional to the ratios of $F_k$ or $F_{a-b}$ of ions to reference radiation<br>P: proportionality factor, minimum and (if applicable) maximum ICS | F: experimental survival data [143,169] |

The models can be roughly categorized into three classes according to their most evident model feature: (a) models that derive weighting factors for biological effectiveness from the probability distribution of ionization cluster formation in a single target; (b) models that use these probability distributions and also account for the synergistic effects of pairs of ionization clusters formed in different targets; (c) models that use an extension of the conceptual framework of nanodosimetry pertinent to 'macroscopic' situations involving many nanometric targets. The first two classes of models essentially aim to correct the absorbed dose by a factor in order to obtain a biological relevant dose quantity that relates directly to the yield of DNA damage or cell survival. The third class of models is translational and aims to construct radiation quantities from the fluence distribution of different particles in mixed radiation fields, such as those encountered in ion-beam radiotherapy. Two different routes have been followed to achieve this. On the one hand, nanodosimetric quantities are averaged over many targets and the components of a mixed radiation field to obtain a nanodosimetric weighting factor for absorbed dose, as in the models of the first two classes. On the other hand, the concept of cluster dose [37] constructs radiation quantities based on particle fluence and macroscopic averages of the absolute frequency of ionization clusters occurrence. Cluster dose cannot be factorized into absorbed dose and a correction factor.

### *4.1 Commonalities among the models*

The models presented in the previous section are based on different approaches. From a formal perspective, they all derive one or more nanodosimetric parameters, $q_n$, as physical descriptors of radiation quality from the ICS distribution. These are subsequently combined, often nonlinearly, with model-specific biological parameters to obtain radiation effectiveness. All models implicitly assume that the parameters $q_n$ can be obtained from ICS distributions by applying the corresponding weighting factors, $w_{n,\nu}$, to the ICS probabilities $p(\nu)$ (Eq. (54)) or weighting factors $w^c_{n,\nu}$ to the conditional ICS probabilities $p^c(\nu)$.

$$q_n(Q,\theta_s,\theta_b) = \sum_{\nu=1}^{\infty} w_{n,\nu}(\theta_s,\theta_b,\theta_m)p(\nu|Q,\theta_s,\theta_b) \tag{54}$$

Some approaches use instead the conceptually equivalent method of applying weighting factors $W_{n,k}$ to the complementary cumulative ICS probabilities $F^*(k)$ (Eq. (55)) or weighting factors $W^c_{n,k}$ to the conditional complementary cumulative ICS probabilities $F^{c*}(k)$.

$$q_n(Q,\theta_s,\theta_b) = \sum_{k=0}^{\infty} W_{n,k}(\theta_s,\theta_b,\theta_m)F^*(k|Q,\theta_s,\theta_b) \tag{55}$$

The weighting factors $w_{n,\nu}$ and $W_{n,k}$ are related by Eq. (56).

$$w_{n,\nu}(\theta_s,\theta_b,\theta_m) = \sum_{k=\nu-1}^{\infty} W_{n,k}(\theta_s,\theta_b,\theta_m) \tag{56}$$

Although Eqs. (54) and (55) are redundant, both are included here since one provides handier notation in some cases and the other in others, depending on the model (Table 7).

The parameters $q_n$ depend on the radiation quality $Q$, a set of parameters $\theta_s$ characterizing the target volume (size), and a set of parameters $\theta_b$ characterizing the irradiation geometry. The ICS probabilities $p(\nu)$ and $F^*(k)$ also depend on radiation quality and the two sets of parameters $\theta_s$ and $\theta_b$, while the weighting factors $w_{n,\nu}$ and $W_{n,k}$ are assumed to be independent of the radiation quality but may depend on an additional set of model-specific parameters $\theta_m$.

Examples of choices of such weighting factors are listed in Table 7. For instance, in the model of Grosswendt [77–79] the weighting factors related to DSB formation are simply $w_{n,\nu} = 0$ for $\nu = 1$ and $w_{n,\nu} = 1$ for $\nu \geq 2$. In the approach of Mietelska et al. [70], the weighting factors $w_{n,\nu}$ are cumulative probabilities of a binomial distribution with success parameter $p_s$ (Eq. (51)). This success parameter corresponds to the model-specific parameter $\theta_m$.

Table 7: Examples of weighting factors used in different models for parameters determined according to Eqs. (54) or (55). ($p_{SSB}$ and $p_{DSB}$ are predicted probability of induction of a single and a double strand break, respectively; $p_{cDSB}$ and $p_{sDSB}$ are predicted probability of a complex and simple double strand break, respectively; $\sigma_\alpha$ and $\sigma_{5\%}$ are inactivation cross sections at low dose and at 5 % survival rate, respectively; $g_w$ is cluster dose, $q_1$ and $q_2$ are radiation quality factors, $D^Q_{3-10}$ is the nanodosimetry-weighted dose, and $\delta_{i,j}$ is the Kronecker symbol, i.e., unity if the two indices agree and else zero.)

| author(s) | quantity | weighting factors in Eqs. (54) or (55) | conditions |
|---|---|---|---|
| Casiraghi [54] | $p_{sDSB}$ | $W_{1,k} = \delta_{k,1} - \delta_{k,3}$ | - |
| | $p_{cDSB}$ | $W_{2,k} = \delta_{k,3} - \delta_{k,10}$ | - |
| Conte [91] ‡ | $\sigma_\alpha$ | $W_{1,k} = c_1\delta_{k,1} + c_2\delta_{k,2}$ | aerobic |
| | | $W_{1,k} = c_3\delta_{k,2} + c_4\delta_{k,3}$ | hypoxic |
| | $\sigma_{5\%}$ | $W_{2,k} = c_5\delta_{k,1} + c_6\delta_{k,2}$ | aerobic |
| | | $W_{2,k} = \delta_{k,2}$ | hypoxic |
| Grosswendt [77] | $p_{SSB}$ | $w_{1,\nu} = \delta_{\nu,1}$ | - |
| | $p_{DSB}$ | $w_{2,\nu} = 1 - \delta_{\nu,1} - \delta_{\nu,2}$ | - |
| Garty [49] | $p_{DSB}$ | $w_{1,\nu} = 1 - 2\left(1 - \frac{p_s}{2}\right)^\nu + (1 - p_s)^\nu$ | - |
| | $p_{cDSB}$ | $w_{2,\nu} = w_{1,\nu} - \binom{\nu}{2}\frac{p_s{}^2}{2}(1 - p_s)^{\nu-2}$ | - |
| Mietelska [70] | $\sigma_{5\%}$ | $w_{1,\nu} = 1 - (1 - p_s)^\nu - p_s\nu(1 - p_s)^{\nu-1}$ | $\nu \geq 2$ |
| Ortiz [72] | $g_w$ | $W_{1,k} = \delta_{k,4}$ | aerobic |
| | | $W_{1,k} = \delta_{k,6}$ | hypoxic |
| Schneider [66] | $p_{DSB}$ | $W_{2,k} = \delta_{k,1}$ | - |
| Schulte [36] | $q_1$ | $W_{1,k} = \delta_{k,1} - \delta_{k,5}$ | - |
| | $q_2$ | $W_{2,k} = \delta_{k,5} - \delta_{k,10}$ | - |
| Yang [69] | $D^Q_{3-10}$ | $W_{1,k} = \delta_{k,2} - \delta_{k,10}$ | - |

‡ The coefficients $c_1, \dots, c_6$ are between 0 and 1 and depend on the cell type.

Although the parameters $q_n$ are weighted linear combinations of nanodosimetric probabilities, the actual model is often non-linear in these parameters, for instance, in Eq. (40).

Another categorization of the presented models is that some are to some extent mechanistic[28] while others are based on empirical correlations. The mechanistic models often come with strong heuristic hypotheses such as the assumption made in [77,66,78,79,132,62,134] that an IC of size two or greater corresponds to a DSB. This assumption implies that the model only considers direct radiation damage and considers DNA damage due to reaction of radicals formed by water radiolysis to be negligible or irrelevant.

## 4.2 Mechanistic models

As demonstrated in Fig. 5, mechanistic models are capable of reproducing experimental data, even when they contain model assumptions such as that the probabilities of an IC of two or larger and of formation of a DSB being equal. These assumption have been argued to lack plausibility [68,83,126]. For some models, the agreement between the model prediction and radiobiological data is related to adjusting one or more free parameters external to nanodosimetry. In the work of Garty et al. [45,49], e.g., this free parameter is the conversion probability between an ionization in the target volume and the occurrence of a strand break or a base damage. In the TET and the model of Schneider et al. [66], the probability of repair of a DSB is another example of such an additional model parameter.

Other mechanistic nanodosimetric models contain implicit (fixed) parameters which have been intentionally or unintentionally chosen to yield agreement of model predictions with radiobiological reference data. Fig. 5 illustrates one example of this. The nanodosimetric parameters used in the model prediction were derived from a simulation using a beam size twice the size of the target volume in both transversal dimensions. If a pencil beam passing the target centrally were used instead, the model would not reproduce the radiobiological data[29]. This indicates that the choice of irradiation conditions and beam size is an implicit parameter of the model that was used to produce the predictions shown in Fig. 5.

It is worth emphasizing that among the mechanistic models, only that of Garty et al. [45,49] genuinely includes indirect radiation effects. Arguments about radical recombination have also been used in [36] and [54]. However, the target cylinder diameter considered is 2 nm, which is the lateral diameter of DNA double strands. Therefore, unlike the model of Garty et al., these approaches do not consider radicals formed outside the volume occupied by DNA and reacting with the DNA molecule after diffusion.

## 4.3 Empirical-correlation models

Empirical correlation models contain adjustable parameters that can be used to fit the model to radiobiological assays. A question that can be answered in this way is the following: For which site diameter $\theta_s$, and which set of weighting factors $w_{n,\nu}$, applied to ICS distributions obtained in spherical targets passed centrally by the primary particle, is there the best correlation to experimental data for inactivation cross sections [56,60,89,91]? Another question is: For which set of weighting factors $w_{n,\nu}$, applied to ICS distributions obtained in cylindrical targets with a diameter of 2.3 nm and a height of 3.4 nm with the irradiation geometry considered in [37], is there the best correlation to cell survival data [72]?

[28] They are mechanistic in the sense that they link ionization clusters with some biological effect.
[29] Private communication by Pavel Kundrát, email on April 13, 2026.

However, the fact that some of the empirical-correlation models use ICS probabilities in targets corresponding to short DNA segments (e.g. in the cluster-dose approach) does not mean that these models also neglect indirect effects. This is because no assumption is made as to where the ICs are formed. The ICs can be formed in DNA or in the surrounding medium. Therefore, the concern that assuming radiation effects in a single target to determine the fate of a cell appears implausible [3] is not valid for the model used in [56,60,81]. This is because the critique implicitly assumes that the models are mechanistic, while they are not.

Similar to the conventional LQ model of cell survival, the empirical correlation models linking nanodosimetry and biological radiation effects do not explain why the observed dependencies are found. Their contribution to radiation physics is that they take into account the microscopic stochasticity of radiation action, enabling new possibilities for optimization of ion beam treatment planning [54,144], for example.

It is not contradictory that different approaches identify different minimal ICSs as being best correlated with biological results. For example, Conte et al. [91] found the best correlation between nanodosimetry and biological effects using the nanodosimetric probabilities $F^*(1)$ and $F^*(2)$ for aerobic cells and $F^*(2)$ and $F^*(3)$ for hypoxic cells. In contrast, Ortiz et al. [72] reported the best association between cell survival and probabilities $F^*(4)$ for aerobic cells and $F^*(6)$ for hypoxic cells.

In fact, these two approaches differ greatly in the nanodosimetric quantity considered even if conceptually the same quantity is used. In the work of Conte et al. [56,60,81,89,91], a single nanometric target traversed centrally by an ideal pencil beam is considered. In the cluster dose approach, the nanodosimetric quantity is averaged over all possible nanometric target volumes of a given size and shape within a range of impact parameters. Additionally, the first approach varied the target size to identify the best correlation, while the target geometry was fixed in the second approach. Both approaches (and variants thereof, such as that of Mietelska et al. [70]) allow identification of nanodosimetry-based parameters that provide the best fit to radiobiological data among the investigated nanodosimetric parameters. Since the models do not include a mechanistic interpretation and apparently achieve the same fit quality and predictive performance, there is no evident criterion at present to select one as being better than the others. Further investigations will be needed to establish their robustness and generalizability. At present, all empirical-correlation nanodosimetry-based models seem to achieve the primary aim of identifying physical descriptors that provide a more consistent relationship with biological effects across different radiation qualities.

## *4.4 Open issues*

### *4.4.1 Understanding the performance of empirical correlation models*

From a heuristic point of view, it would be more important to understand why different empirical correlation approaches succeed in reproducing experimental data despite their different fundamentals. The question that arises is whether this is due to inherent relations between nanodosimetric quantities at different site sizes and irradiation conditions This would require a comprehensive characterization of this relationship. One preliminary step could be exploring which ICS yields best agreement with experiments under the condition that the cluster dose on the *x*-axis of Fig. 10 is determined using the nanodosimetric probabilities determined for the target sizes and irradiation conditions used by Conte et al. [91]. More generally, it would be interesting to investigate the full variability of nanodosimetric quantities as a function of all influencing parameters. Ideally, this would produce functional dependencies similar to those reported in section 2.2.3.

For instance, the influence of the impact parameter range has only been studied for a few possible choices [58,64]. Similarly, most studies have used cylinders of different dimensions and aspect ratios (e.g. [77,56,70,54,78,45,46]) or spheres (e.g. [89,91,103]) as target volumes. However, less common geometries, such as cylinder shell segments [64,100,111] or (almost spherical) dodecahedrons [68], were also employed.

The variation of ICS distributions with the shape of the target volume has only been investigated for a few select cases so far: spheres and cylinders of the same diameter and height equal as the sphere diameter [25], and cylinder shell segments with different aspect ratios [64]. A systematic assessment of the sensitivity of ICS distributions to the choice of target volume shape and size, has not yet been conducted. Such a sensitivity analysis would be the prerequisite for determining the contribution from target geometry to uncertainty and could potentially allow for simpler, more computationally efficient sampling strategies.

The computationally efficient associated volume clustering approach from microdosimetry [170] has already been used in a few nanodosimetric studies [76,103,102]. When used with spherical targets, this scoring approach offers the benefit of mathematical rigor since the associated volume of a track (defined by the union of spheres around each ionization within a track) can be decomposed into sub-volumes of points having a specified number of ionizations within the target radius [76]. Extending associated volume clustering to cylindrical target volumes remains a challenge for two reasons related to the loss of spherical symmetry [76]. First, there is no longer a one-to-one correspondence between target center and the ICS. Second, the associated volume comprises "diffuse" parts in which points have less than 100 % probability of scoring at least one ionization. Additionally, there is computational overhead from the need to sample the cylinder orientation.

If we knew the sensitivity of ICS distributions and their derived parameters to target size and shape and impact parameter range, we could base the choice of these parameters and sampling strategies on the expected uncertainty resulting from such a choice compared to more correct or relevant approaches. For instance, a sphere with diameter of 3 nm has the same volume as the same volume as a cylinder with diameter of 2.3 nm and height of 3.4 nm. If we knew that the ICS distributions are expected to be the same within a given acceptable uncertainty, we could use track-centered associated volume clustering as in [76] instead of making complicated choices of irradiation geometries and target arrangements as described in [112].

#### *4.4.2 Uncertainties*

A major caveat of all models presented is that they rely on simulations.[30] Several studies have revealed large disparities of results obtained using different track structure codes for electron transport [173–176]. A recent systematic analysis indicated that the results for the probability $F_2$ in spherical targets with a diameter of 8 nm produced by 50 eV electrons had a relative standard deviation of 34 % among different codes. Using the same total cross-sections in all codes reduced the standard deviation at 50 eV to about 7 %. However, disparities at larger electron energies were only slightly reduced by this substitution of cross-section data [175]. Furthermore, the common cross-section datasets were only established ad-hoc to study the impact of total cross-sections on nanodosimetric results. There was no claim that this dataset (and the results produced with them) were most realistic. One might expect the discrepancies between codes to be smaller for ion beam simulations, as a significant portion of ionizations originate from ion interactions. For instance, Rabus et al. reported a difference of about 10 % between the number of targets with ionizations in proton beams with energies between 1 MeV and 100 MeV for different options of Geant4-DNA [76]. Nevertheless, a thorough investigation of the uncertainty coming from different codes for ion beam applications is lacking.

---

[30] This also applies to models formulated on experimental data where simulations are needed for extension to radiation qualities at which measurements have not yet been performed or for the characterization of detector performance.

### *4.4.3 Refining the scaling relation*

The scaling relation underlying nanodosimetry (see Appendix A) was initially put forth to address the equivalence of ICS distributions in gas-phase experiments and corresponding targets in liquid water. Its theoretical basis is the assumption that the ICS distribution is determined by the number of interactions of the primary particle in the target volume, as well as the analogy between the compound Poisson process leading to ICS distributions and the microdosimetric multi-event distribution [77,90]. However, Supplement 1 of [76] argues that this analogy may be limited to cases in which the ICs are predominantly produced through primary particle interactions, such as with the small target sizes used in [91]. Even when a large range of impact parameters is covered in scoring and the majority of ICs are produced by secondary electrons, the rationale behind the scaling relation is questionable. This is because it is not trivial that the scaling factor for target sizes between different media, derived from ionization cross sections of ions, should also apply to their secondary electrons.

In mixed radiation fields, the scaling factor for target sizes in different media may differ between different particle types and energies. The cluster dose concept of Faddegon et al. avoids this complication by only considering density scaling [37]. However, Schwarze et al. pointed out that this leads to a cluster dose independent of the material. They also proposed a first approach to adapting the scaling relation to situations of complex radiation fields in different media that considered an electron interaction cross section ratio [147]. One could argue that the relevant information is the yield of ICs in the cell nucleus, where water equivalence could be assumed. This approach would circumvent the issue of scaling between materials, though it requires a theoretical foundation. Notably, it had been demonstrated that ICS probabilities in water and DNA differ in the range of several 10% [125,172].

Although it would be conceptually simple to systematically study the relationship between IC distributions in different materials for a wide range of impact parameters using track structure simulations with different target sizes, it would be potentially tedious. Analyzing such a large dataset would certainly benefit from machine learning. The goal would be to establish an uncertainty associated with the scaling relation for complex radiation fields when considering only density or density and different ionization cross sections across materials.

#### 4.4.4 Identification of the optimum ID parameter

It should be noted that the simplified definition of cluster dose suggested here is contingent on the linear relationship between the ID parameter and the linear yield of ICs. Non-linear dependence is possible within the abstract concept proposed by Faddegon et al. [37]. However, since linear IC yields are dimensioned quantities, a hypothetical non-linear operator would require a reference length scale. Otherwise, the cluster dose would depend on the chosen unit of length. If a non-linear relation eventually proves necessary to characterize radiation quality, the fundamental assumption that nanodosimetric probability distributions are a characteristic of radiation quality would become obsolete. This is because, with a non-linear relation between ID parameter and linear cluster yield, the biological relevance of clusters of a given size $\nu$ would depend on the particular ICS distribution.

### 4.5 Future challenges

#### 4.5.1 Understanding radiobiological effectiveness from nanodosimetry

The development of nanodosimetry and radiation effect models using nanodosimetric track structure features was initially driven by the urge to better understand the fundamental principles behind biological radiation effects. Sophisticated track structure codes can simulate radiation effects across all stages, from physical interactions and radiolysis to chemical reactions and biological consequences. However, these intricate simulations tend to have a black-box character.

Mechanistic models that predict radiation effects from nanodosimetric quantities aim to address this issue. These models try to use only a minimum set of quantities and free parameters. As previously mentioned, a common approach was to fix certain implicit parameters, such as the target size, based on heuristic considerations. In Garty et al.'s work, the contribution of radical species contributing to DNA damage was somewhat mixed with the direct effects in that a joint parameter describing the probability of an ionization to result in a DNA strand break was used. The rationale was to minimize the number of model parameters.

As was shown with the TET and derived models, the applicability of some of the model assumptions can be tested using simulations [68]. Similarly, machine learning approaches have been used in conjunction with simulations to inform optimal parameter choices [171]. However, these examples are only the first steps for the further development of mechanistic models based on nanodosimetry. Machine learning approaches could also be used to determine weighting approaches that are more general than those shown in Eqs. (54) and (55).

For instance, weighting-function estimates could be determined considering the relevance of different target sizes in order to account for radical contributions and their diffusion. For example, consider a hypothetical extension of the approach of Garty et al. that distinguishes between ionizations of the DNA molecule and ionizations in its vicinity. The latter leads to formation of radical species that have a probability of reaching the DNA segment and producing additional DNA damage. This process could be taken into account by a corresponding probability, i.e. an additional model parameter. This model parameter could be determined by fitting data from radiobiological DNA-damage assays, but it would depend on the chosen size of the volume around DNA, which is a further implicit parameter.

Alternatively, we could use adapted simulation tools for the chemistry phase to determine how this probability varies with volume size. We could also use these tools to determine this probability for ICs in a volume near but not intersecting the DNA that produce only radicals or for ICs that only partly overlap with DNA segments. This data could then be used to train machine learning models to predict the weighting function of ICS formed in a set of concentric targets with different sizes, for example. It may be that this weighting function depends only on the set of target sizes, not on how the ICs were produced. In other words, it might be more or less independent of radiation quality. This hypothetical weighting function for target sizes would refine the model of Garty without introducing additional free parameters.

### *4.5.2 Correlation models with pairs of targets*

Radiobiological studies suggest that biological effectiveness depends on processes occurring on several length scales [177,178], as previously proposed in microdosimetry [179–181]. Thus, the interaction between correlated ICs produced by the same particle track, as considered in the TET-based models, may be another factor to be considered in nanodosimetry-based radiation effect models. Ngcezu et al. [68] employed a method akin to hierarchical clustering to score ICs in distinct targets. They only applied their approach to the model geometry used by Schneider et al. [66], but extending it to other geometries appears straightforward. Thus, databases for generalized nanodosimetric concepts could be created, pertaining to quantities such as the number of clusters of ICs formed per primary particle path length. However, fluence-weighting will be more intricate since clusters of ICs may contain ICs formed by different particles present in the same primary particle track, as some TET models suggest. Because this is a type of hierarchical clustering problem, machine learning-based approaches could be useful for generating the corresponding databases.

### *4.5.3 Compact nanodosimeters*

Conventional nanodosimeters are comparatively complex instruments. In a proof-of-principle experiment, Hilgers et al. demonstrated that it is possible to perform measurements with a nanodosimeter in the mixed radiation field of a therapeutic carbon ion beam [109]. The potential future use of nanodosimeters in clinical practice to validate simulation-based nanodosimetric quantities used in treatment planning necessitates novel, more compact detector designs. The feasibility of such compact nanodosimeters has already been studied [21]. However, the most recent developments in this regard indicate a major technological challenge in distinguishing individual ions in an IC during the measurement process [22,88]. In these detectors, an electron avalanche is produced in a gas electron multiplier setup, and the duration of the avalanche interferes with the arrival time distribution of the ions to be detected. Addressing this challenge may require reconsidering nanodosimetric measurement concepts. For instance, one could use information about the temporal pattern of the recorded signal. This could entail novel weighting functions for the different ICSs that have not yet been considered in the proposed models. However, as discussed in Section 4.1, the generic features of the models enable adaption to these potential new nanodosimetric measurement quantities.

## 5 Conclusions and outlook

The reviewed biological effect models based on nanodosimetry vary greatly in their assumptions and approaches. While using harmonized notation helps avoid misinterpretations, it cannot eliminate fundamental differences between the models beyond the targeted biological endpoint. One major factor leading to these differences is the range of impact parameters considered when determining the probability distribution of ionization cluster size. The cluster dose concept framework in the linearized formulation used here may provide a basis for a future unified model connecting nanodosimetry and biological effectiveness of radiation. However, this approach implies a transition from a target-centered to a track-centered view of IC formation and extending the conceptual framework of nanodosimetry. As discussed, this allows for the definition of nanodosimetric analogs for conventional dosimetric quantities, such as LET and absorbed dose: linear yield and number density of ionization clusters.

Important gaps remain along the way to unified nanodosimetry-based biological effect models. One such gap is the need to elucidate latent features of nanodosimetry that enable models based on different nanodosimetric parameters to achieve a comparable degree of association with radiobiological outcomes. Another task to be accomplished with respect to applications in treatment planning is establishing uncertainty budgets for nanodosimetric quantities and model predictions. Another question to be addressed in this context is how to properly apply the scaling relation between different materials in mixed radiation fields and a multitude of targets. Which ionization detail parameter correlates best with biological effects is also an unanswered question.

Beyond the issues related to empirical-correlation models, refining mechanistic nanodosimetric radiation effect models so that their implicit model parameters can be determined rather than heuristically assumed remains a challenge for future work. This includes determining the extent to which models that consider synergistic effects of multiple ICs can be included in macroscopic aggregation frameworks more plausibly than in attempts proposed to date. Finally, potentially including nanodosimetric quantities in treatment planning will require their measurement by compact nanodosimetric detectors, which still need to be developed.

## Appendix A The scaling relation for nanodosimetric target size

The conditions for the equivalence of nanodosimetric ICS distribution for different materials, target sizes, and densities were first studied by Grosswendt and Pszona [84] for 4.6 MeV alpha particles in nitrogen and water and by Grosswendt [15] considering tracks of 5 MeV alpha particles and electrons with energies below 10 keV in propane and water. The approach was formalized in further work by Grosswendt [43] and extended to also consider the non-ideal detection efficiency of nanodosimetric counters by Grosswendt et al. [77,85]. It is hypothesized that the ICS distribution is a unique function of the mean ICS $M_1$. The equivalence of the ICS distributions is thus obtained if at least the same $M_1$ is produced in targets of different materials and densities. The relationship between the corresponding target sizes can then be expressed by Eq. (57).[31]

[31] Note that, in nanodosimetric literature, the target size $d$ is often represented by $D$ and the impact parameter $b$ is often denoted by $d$.

$$(d_s\rho)^{(2)} = (d_s\rho)^{(1)} \times \frac{(\lambda_i\rho)^{(2)}}{(\lambda_i\rho)^{(1)}} \times \frac{m_1^{(1)}\left((b\rho)^{(1)}\right)}{m_1^{(2)}((d\rho)^{(2)})} \times \frac{\eta^{(1)}}{\eta^{(2)}} \tag{57}$$

The superscripts denote the material. $d_s$ denotes the diameter of the target volume, $\rho$ represents the mass density, $\lambda_i$ is the mean free path for ionization by the primary particle. The quantity $m_1$ is the mean number of ionizations produced in the target per ionizing interaction of the primary particle in it. This quantity depends on the impact parameter $b$ of the primary particle with respect to the center of the target volume (i.e., the distance between the two lines in Fig. 1). The quantity $\eta$ is the average detection efficiency for the registration of ions produced in the target volume of a nanometric detector.

It is important to note that the product $\lambda_i\rho$ is independent of density, since it is given by the equation $\lambda_i\rho = \frac{M_m}{\sigma_i}$, where $M_m$ is the molecular mass and $\sigma_i$ is the ionization cross section. Consequently, Eq. (57) is an expression for density scaling of the linear dimensions $d_s$ and $b$, corrected for the different cross-sections for ionizations by the primary particle and secondary electrons.

The scaling relation is generally used to establish the water-equivalent diameter of the target volume of a nanodosimeter [86]. In this context, the simplifying assumption is frequently made that the gain factor $m_1$ from the interactions by secondary electrons is almost the same for different materials. Similarly, as long as the quantity of interest is the frequency distribution of ICs produced by the radiation in a material and not that of the ions detected in a concrete nanodosimeter, the last factor in Eq. (57) is also often neglected.

It is worth noting that the derivation of Eq. (57) is based on a theoretical description of the synergistic generation of ionizations in a target volume by the primary particle and its secondary electrons in a compound Poisson process [77,90]. In analogy to the approach used in microdosimetry for the construction of multi-event distributions from single-event distributions, the ICS distribution is conceived as the weighted superposition of the contributions of multiple ionizing interactions of the primary particle along a "relevant" segment of the primary particle path. The weighting factors are the probabilities of a given number of ionizations of the primary particle to occur, which are assumed to follow a Poisson distribution. The contributions are hypothesized to be convolutions of the probability distributions of the number of ionizations in the target for a single ionization of the primary particle along the path segment. As pointed out in Supplement 1 of [76], this implies the assumption that the probability of obtaining a specific number of ionizations resulting from a single ionization of the primary particle is independent of the location along the relevant path segment in which this interaction occurs.

**Acknowledgement**

The author thanks the reviewer for their feedback, which helped improve the manuscript.

**Author contributions:**

**HR** Writing – original draft, Writing – review & editing

**Data availability:**

No new data was produced.

**Declaration of competing interests:**

The author declares no known competing interests.

**Declaration of AI-assisted technologies in the manuscript preparation process:**

During the preparation of this work, the tool PTB-Translate (a derivate from deepl.com) was used to obtain suggestions for improving the quality of the language and readability. Content produced by this tool was reviewed and used to edit the manuscript if deemed appropriate. The author takes full responsibility for the content of the publication.

# 6 Supplementary Figures

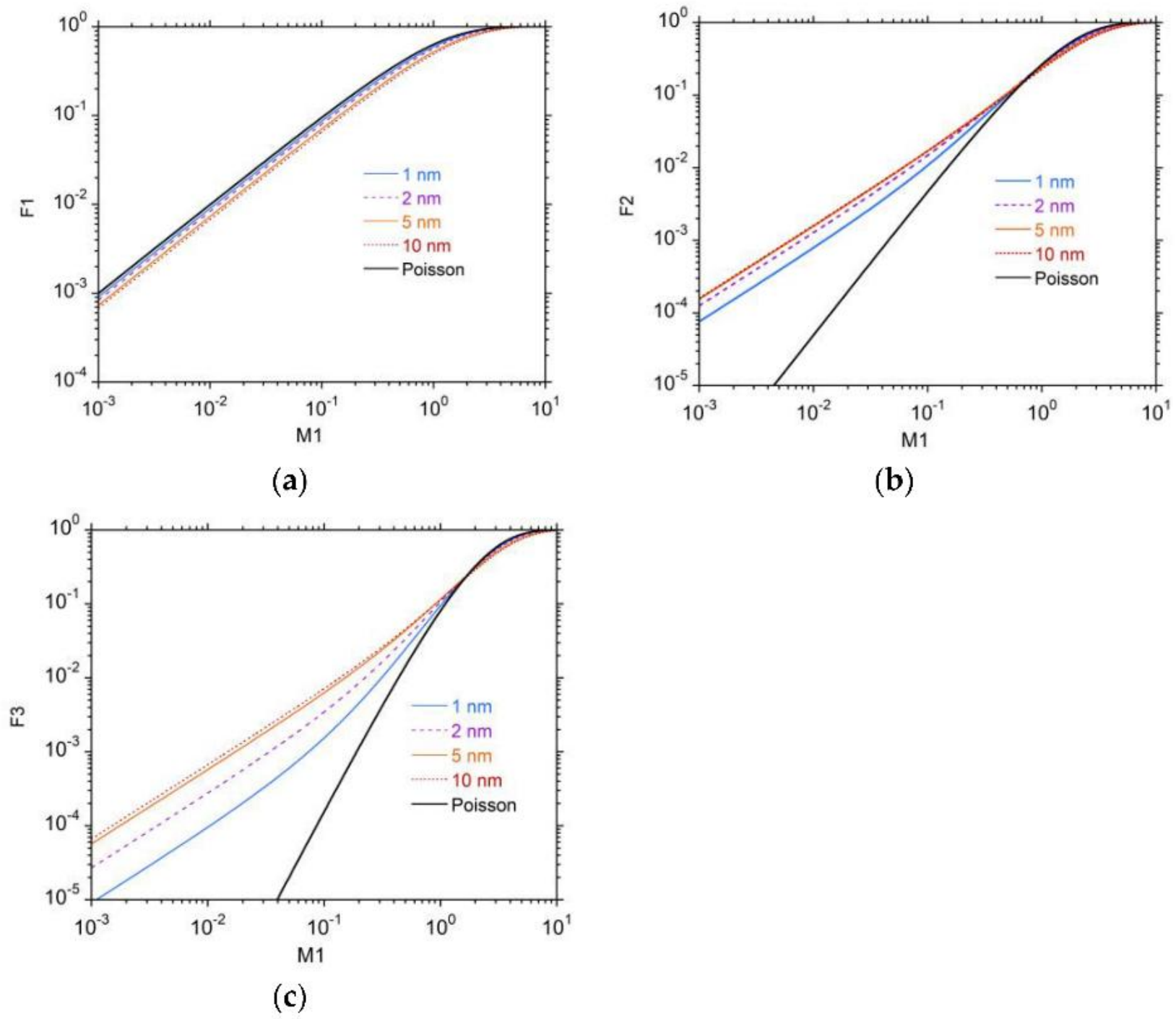

**Supplementary Fig. 1**: Dependence on the cumulative probabilities of ICS of at least one ($F_1$), at least two ($F_2$), and at least three ($F_3$) on the mean ICS $M_1$ for spherical targets of different diameter (see legend) in water. The curves were derived by fitting Eq. (8) in the main text to simulation results for proton, helium and carbon ions. The black lines show the curves of a Poisson distribution for the number of ionizations with the same mean.[32]

[32] Reprinted from Int. J. Mol. Sci., 24, V. Conte et al., Track Structure of Light Ions: The Link to Radiobiology, 5826 (2023) under the CC BY-4.0 license (http://creativecommons.org/licenses/by/4.0).

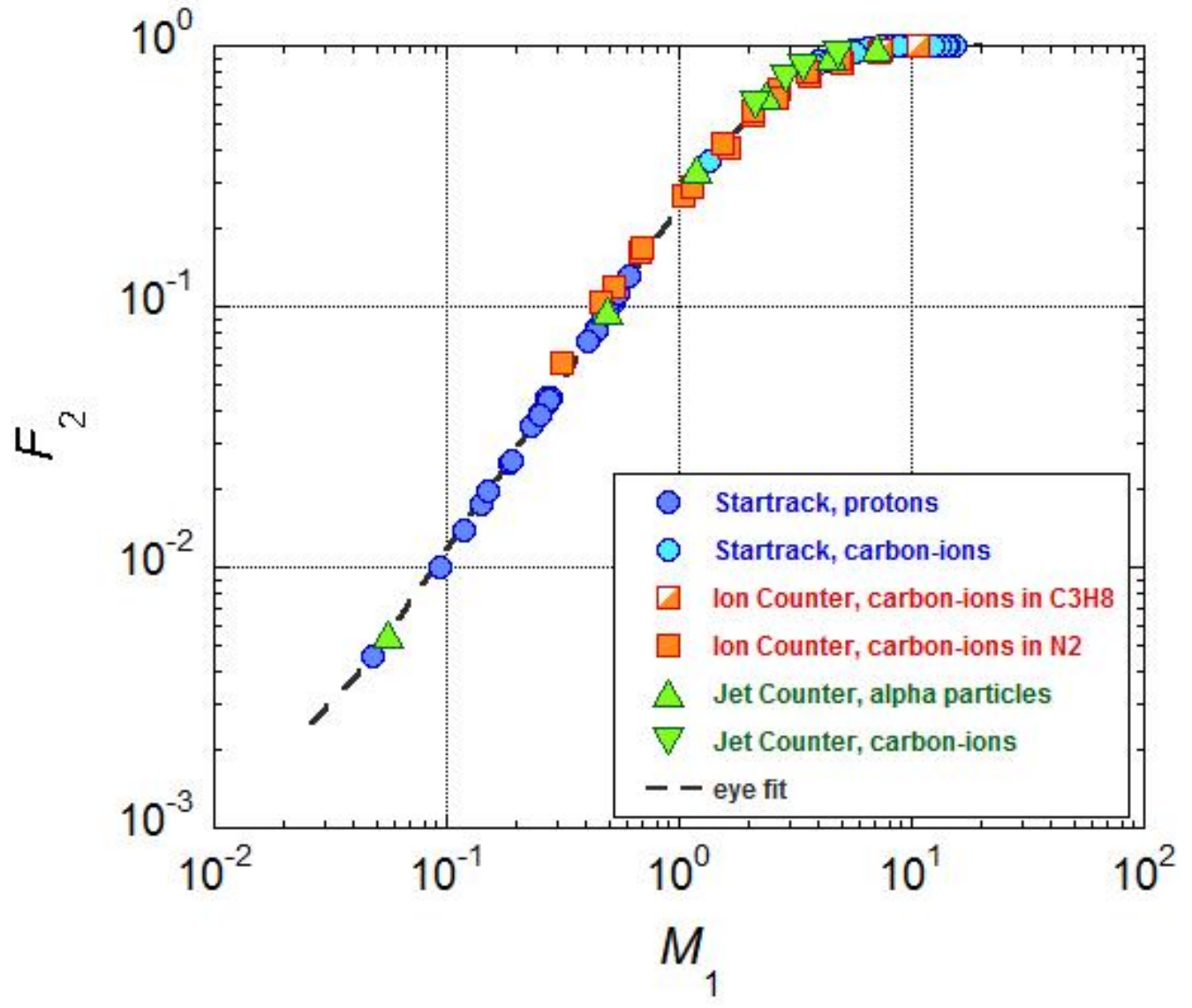

**Supplementary Fig. 2**: Complementary cumulative frequency $F_1^*$ (denoted as $F_2$ on the *y*-axis) of ionization clusters containing two or more ionizations in the target volume of a nanodosimeter as a function of the corresponding mean number of ionizations, $M_1$. The data corresponds to measurements with three different nanodosimeters and different ions as primary particles (see legend). Different data points for the same ion and instrument correspond to different energies of the ion. The Startrack nanodosimeter was always operated with propane ($C_3H_8$), the Jet Counter always with nitrogen ($N_2$). The simulated target size in liquid water varied between about 2 nm for the Ion Counter and 20 nm for the Startrack nanodosimeter.[33]

[33] Reprinted from Radiotherapy and Oncology 119 (Suppl. 1), H. Rabus and V. Conte, Nanodosimetry: from radiation physics to radiation biology, S182-S183, Copyright (2016), with permission from Elsevier.

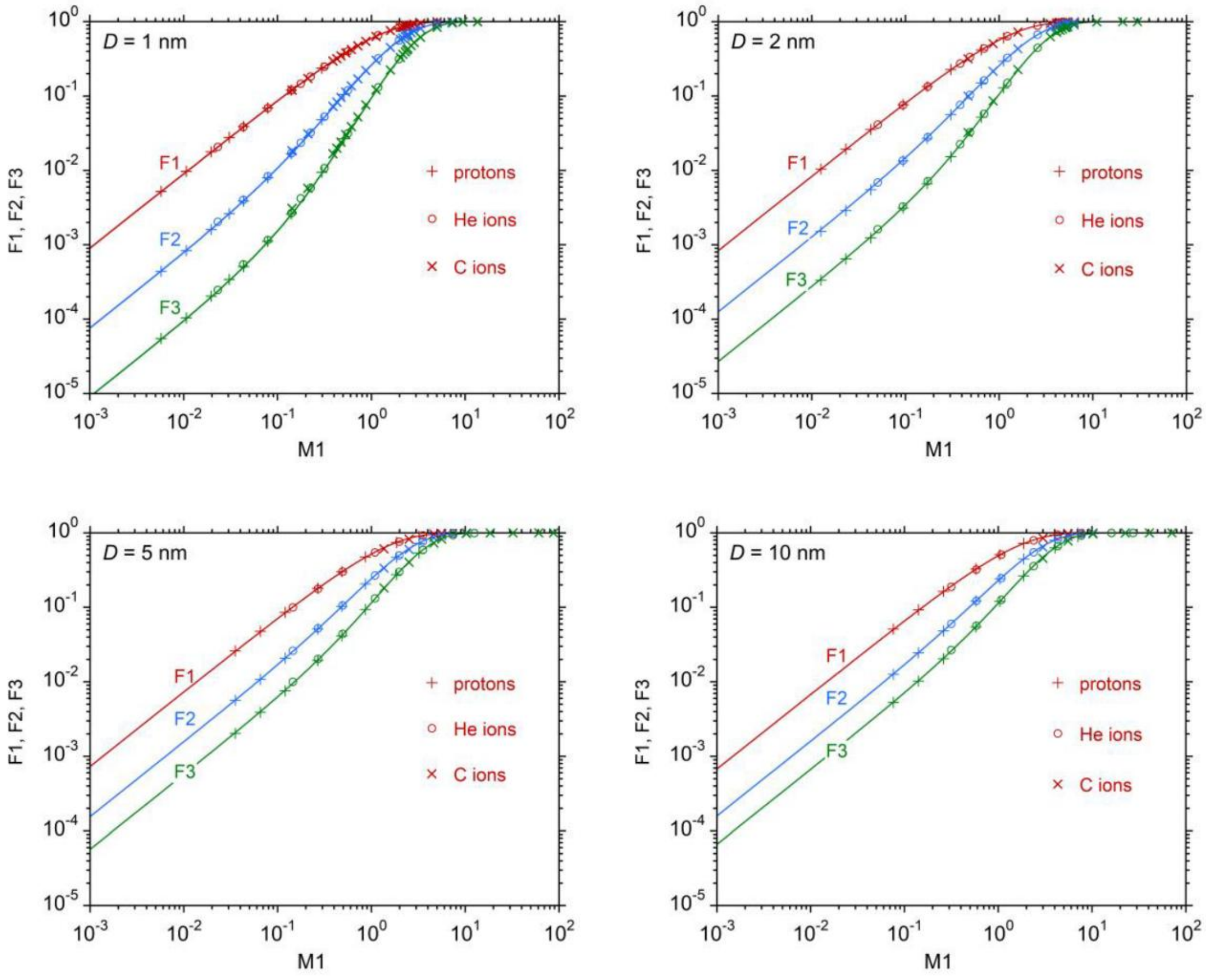

**Supplementary Fig. 3**: Variation of the cumulative probabilities $F_1$, $F_2$, and $F_3$ with mean ICS $M_1$ for spherical targets of different diameter $d$ (denoted by $D$ in the figure labels). The symbols indicate simulation results for protons (plus symbols), helium (circles) and carbon ions (crosses) at different kinetic energies in water. The solid lines are the fit curves according to Eq. (6), using the polynomials in the first column of Table 3.[34]

[34] Reprinted from Int. J. Mol. Sci., 24, V. Conte et al., Track Structure of Light Ions: The Link to Radiobiology, 5826 (2023) under the CC BY-4.0 license (http://creativecommons.org/licenses/by/4.0).

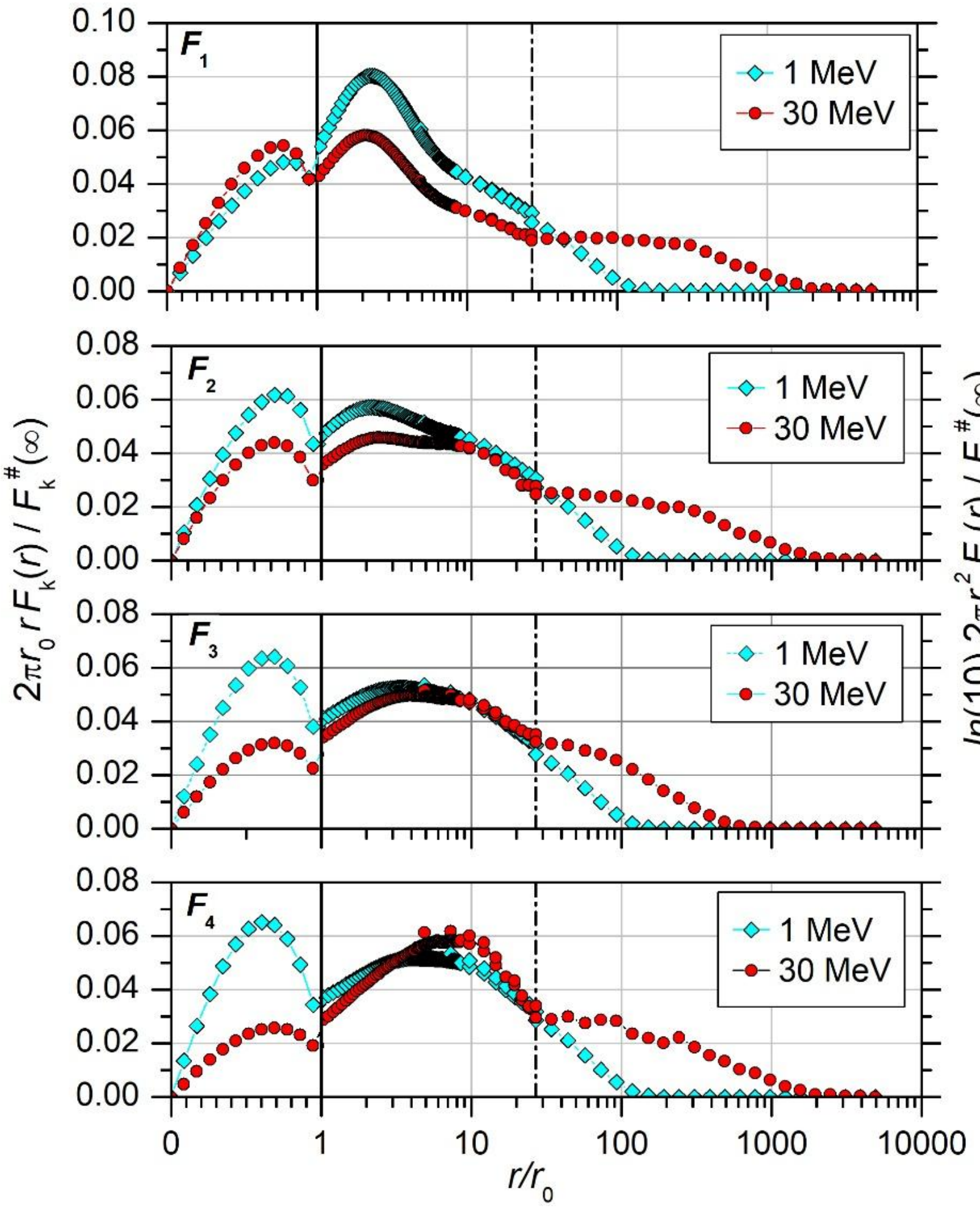

**Supplementary Fig. 4**: Dependence on the cumulative probabilities of ICS of at least one ($F_1$), at least two ($F_2$), at least three ($F_3$), and at least four ($F_4$) ionizations produced by protons with energy of 1 MeV (diamonds) and 30 MeV (circles) in a target volume in liquid water. The data on the left of the dash-dotted vertical line were obtained by scoring in cylinders with radius $r_0$ of 1.15 nm and height of 3.4 nm. On the right of this line, the data were obtained with cylinder shell sectors of the same volume as these cylinders. The *x*-axis shows the radial distance from the proton trajectory normalized to $r_0$. Note that on the left of the vertical solid line the *x*-axis is linear and logarithmic on the right. The *y*-axes show the relative contribution to the area integral from the linear or exponential interval corresponding to a datapoint.[35]

[35] Reprinted from Radiation Physics and Chemistry 176, Thomas Braunroth et al., Three-dimensional nanodosimetric characterisation of proton track structure, 109066 (2020) under the CC BY-NC-ND license (http://creativecommons.org/licenses/by-nc-nd/4.0/).

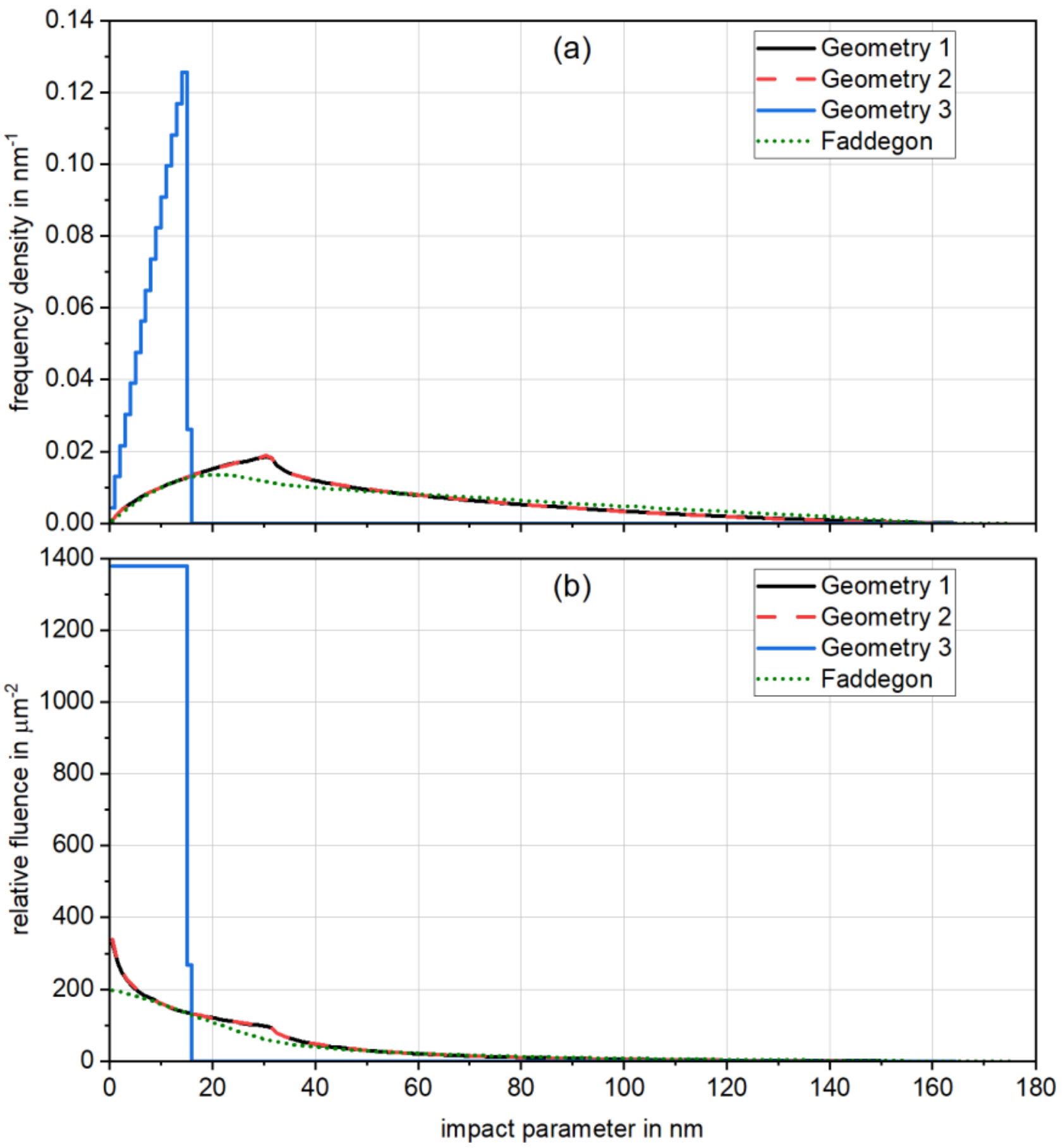

**Supplementary Fig. 5**: Estimated impact parameter distributions corresponding to the different source geometries in the simulations of Ramos-Méndez et al. [58] and Faddegon et al. [37] presented as (a) frequency density per impact parameter interval and (b) relative fluence (frequency per annulus corresponding to an impact parameter bin). Geometry 1 (solid black line) is from a source located at the surface of the scoring region with beams emitted into the scoring region with a direction randomly sampled within the corresponding hemisphere. The scoring region is cylindrical and has 161 nm length and 30.4 nm diameter. With geometry 2 (red dashed line) the source points are located on a sphere surface surrounding the scoring volume, and the emitted particles are directed towards a point randomly chosen in the scoring volume. The sphere radius for the results shown here was 300 nm. The black and red curves essentially overlap. The corresponding data were obtained by starting 1000 beams from a point randomly distributed on the source. For each beam, the histogram of impact parameters was determined for $10^6$ points randomly distributed inside the scoring region. The histograms of all beams were averaged. Geometry 3 is for a beam entering the scoring volume along its axis.

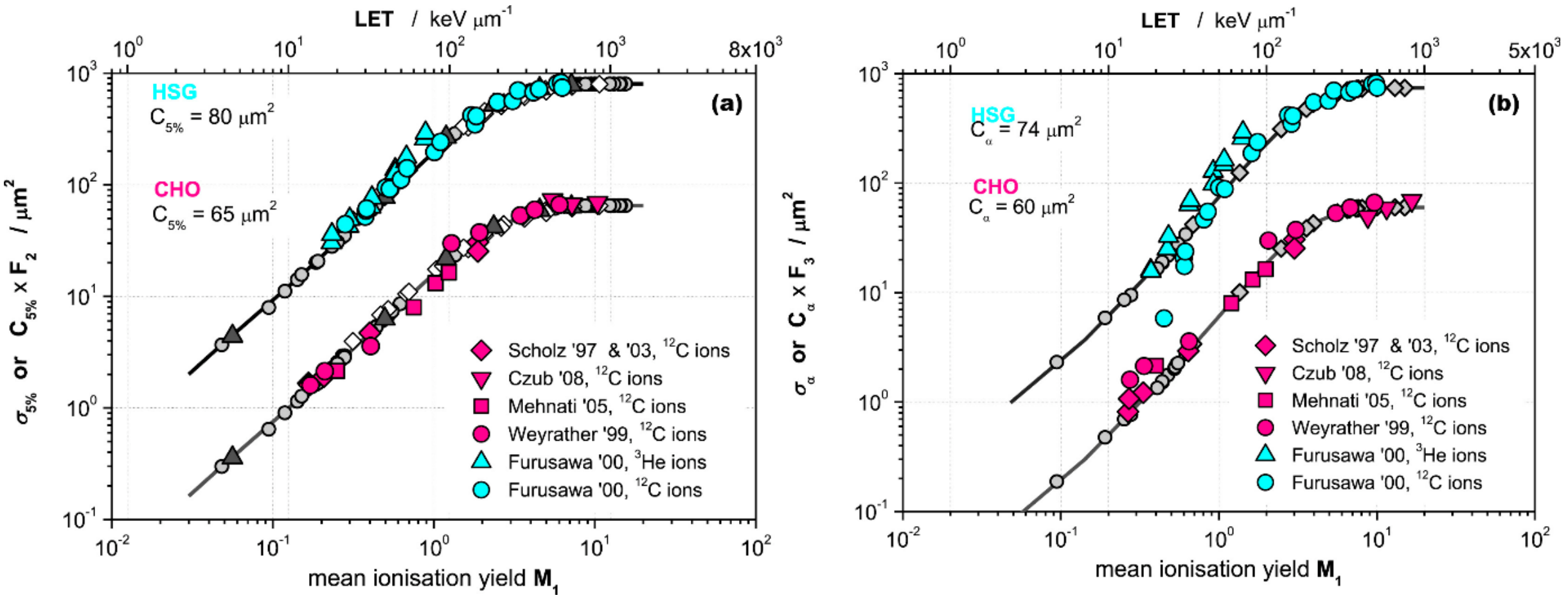

**Supplementary Fig. 6**: Comparison of the inactivation cross sections for HSG and CHO cells at (a) 5 % survival rate ($\sigma_{5\%}$) and (b) low doses ($\sigma_{\alpha}$) as a function of *LET* (upper x-axis) with the scaled nanodosimetric probability $F_2$ and $F_3$ ($F^*(1)$ and $F^*(2)$ in the notation used in the main text) as a function of the mean ionization cluster size $M_1$ (lower x-axis). To facilitate distinguishing the data of the two cell lines, the HSG data (and corresponding $C_{5\%} \times F_2$ and $C_{\alpha} \times F_2$) have been multiplied by 10. Colored symbols represent biological data; grey symbols and solid lines represent the experimental data.[36]

[36] Reprinted from Radiation Measurements 106 (1), V. Conte et al., Track structure characterization and its link to radiobiology, 506-511, Copyright (2017), with permission from Elsevier.

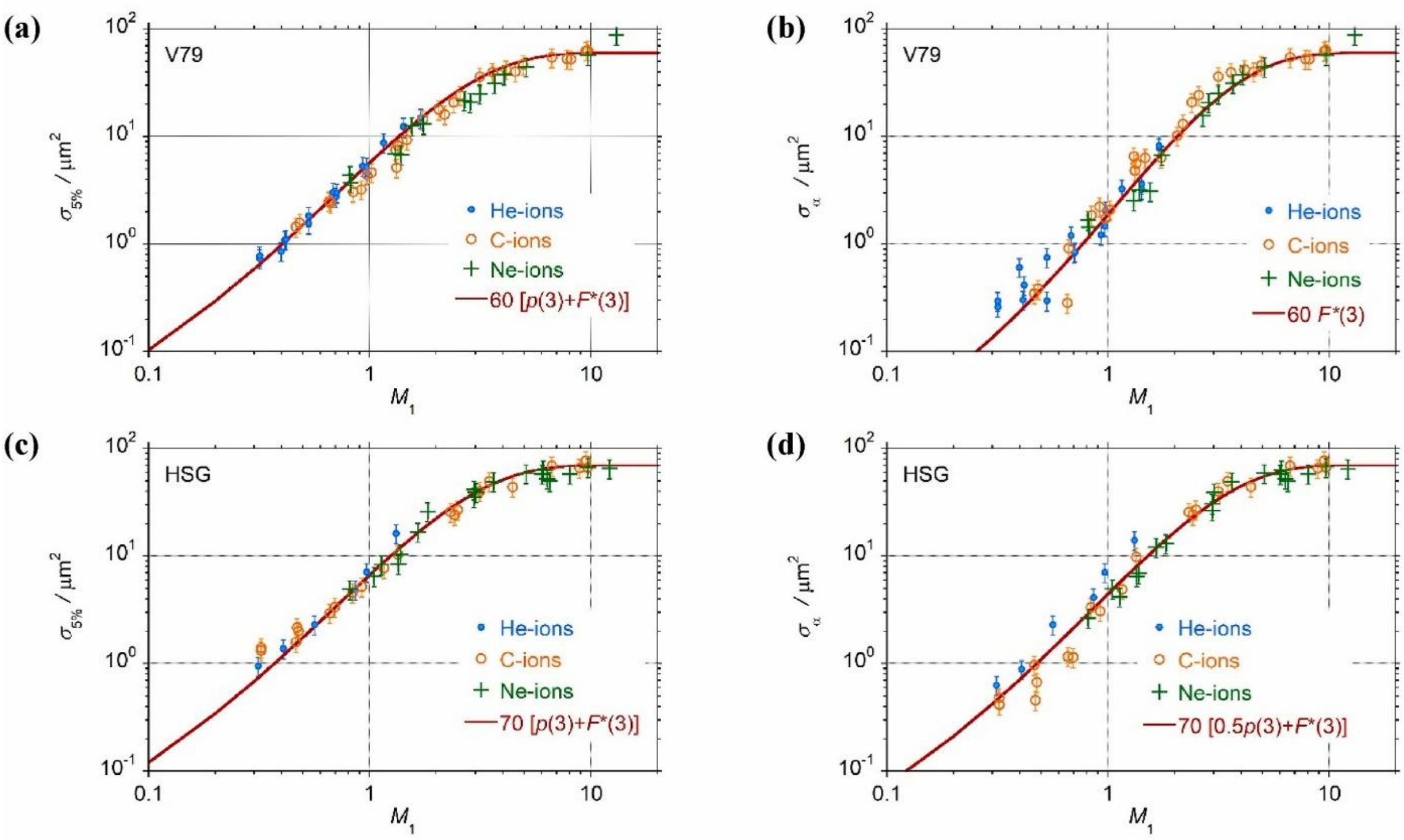

**Supplementary Fig. 7:** Inactivation cross sections for **hypoxic** V79 (panels (a) and (b)) and HSG cells (panels (c) and (d)) at 5% survival (panels (a) and (c)) and at low doses (panels (b) and (d)). Experimental data are drawn with 20% uncertainty bars. The red lines represent the nanodosimetric quantities specified in the legends. Note that $p(3) + F^*(3) = F^*(2)$.[37]

[37] Reprinted from Radiation Physics and Chemistry 222, V. Conte et al., Nanodosimetry applied to aerobic and hypoxic cells, 111835, Copyright (2024), with permission from Elsevier.

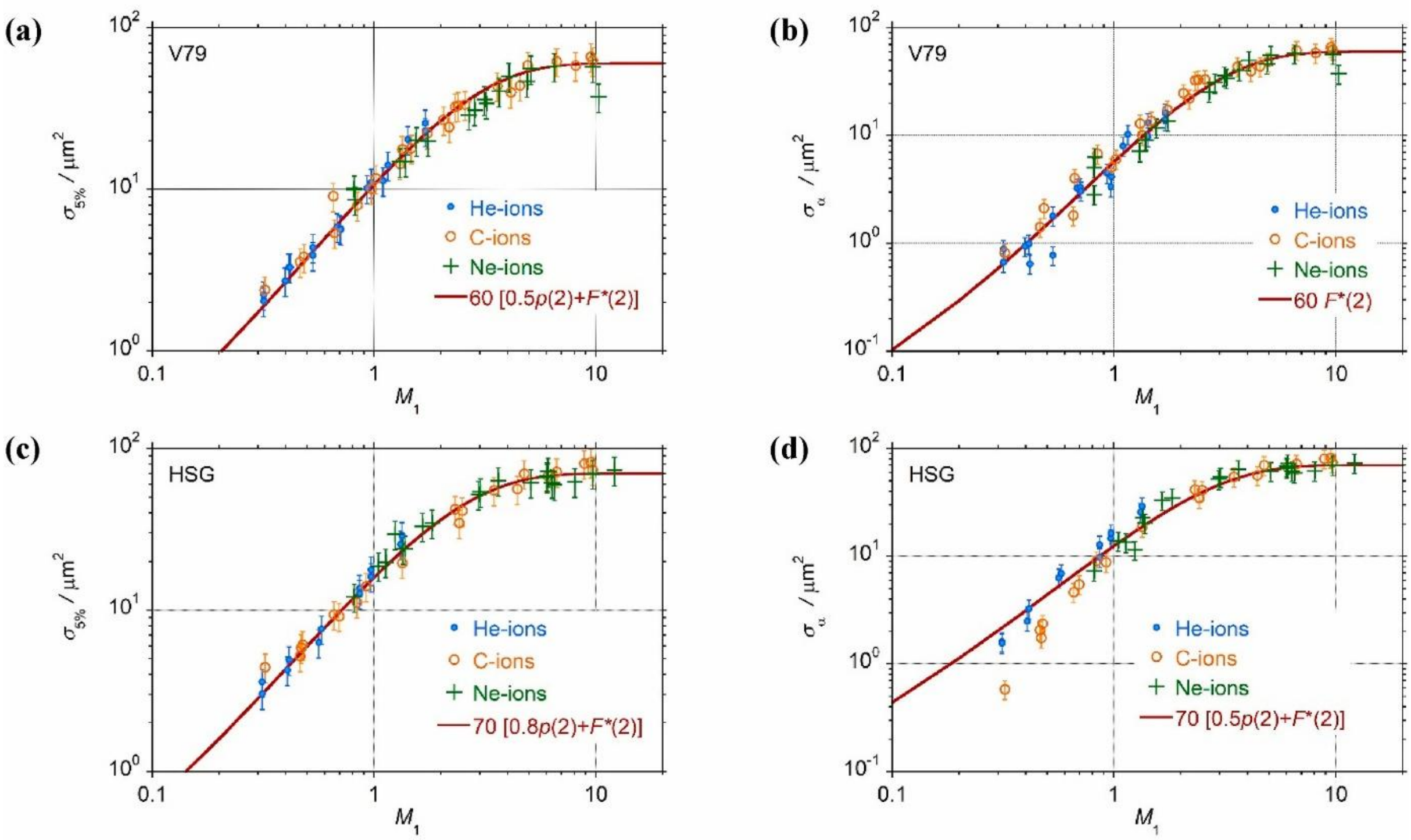

**Supplementary Fig. 8:** Inactivation cross sections for **aerobic** V79 (panels (a) and (b)) and HSG cells (panels (c) and (d)) at 5% survival (panels (a) and (c)) and at low doses (panels (b) and (d)). Experimental data are drawn with 20% uncertainty bars. The red lines represent the nanodosimetric quantities specified in the legends. Note that $p(3) + F^{*}(3) = F^{*}(2)$.[38]

[38] Reprinted from Radiation Physics and Chemistry 222, V. Conte et al., Nanodosimetry applied to aerobic and hypoxic cells, 111835, Copyright (2024), with permission from Elsevier.)

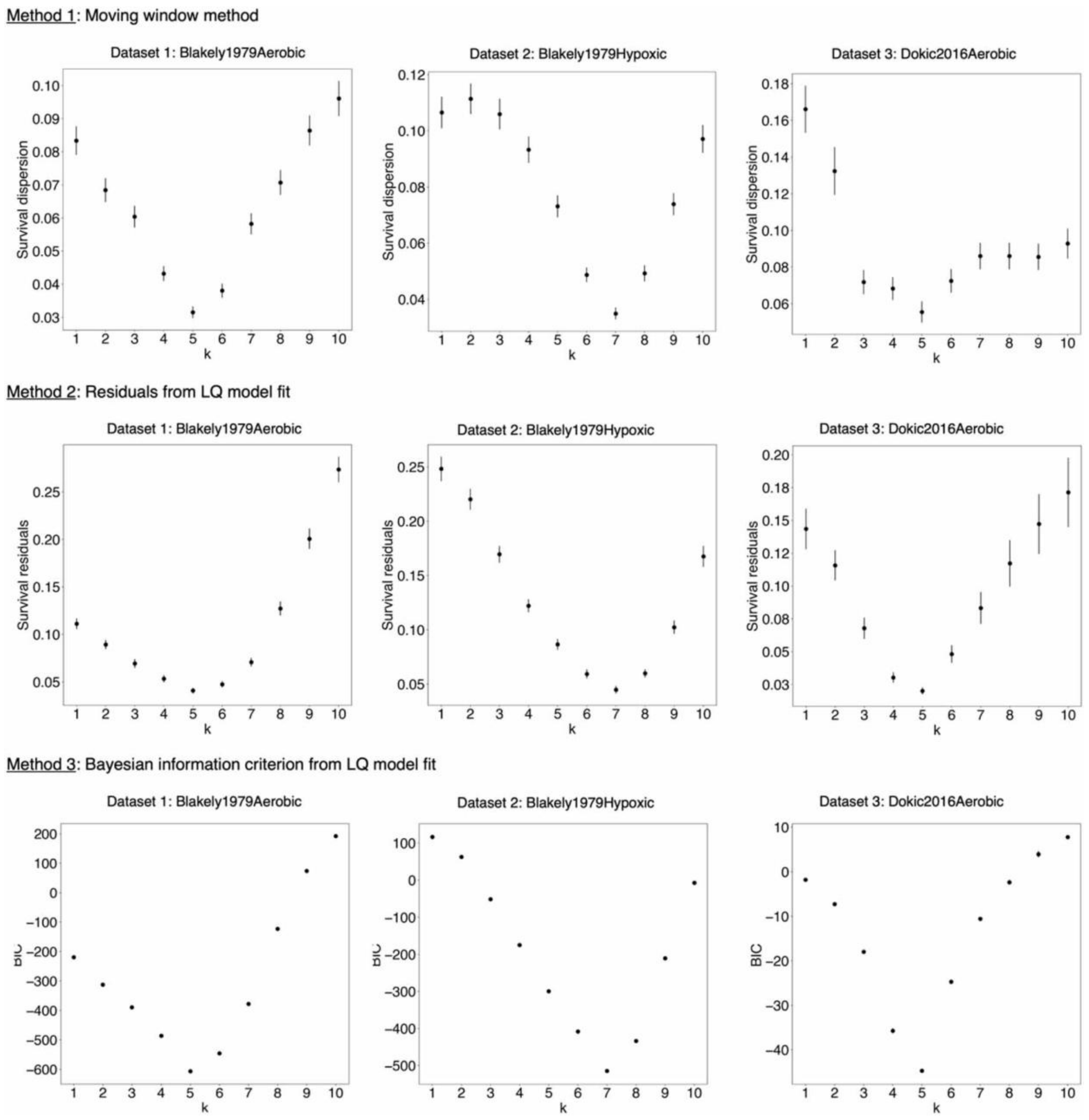

**Supplementary Fig. 9**: Variation of the three statistical metrics for the deviation of the survival data with all ion types from the best fit to the linear-quadratic model (as a function of the cluster dose). The *x*-axis of the plots is the minimum number of ions in an IC. The rows correspond to the three different metrics. The left and middle panels of each row are results for survival data under heavy ion irradiation from [148] under aerobic and hypoxic conditions, respectively. The right panel is for data of aerobic A549 cells [149][39]

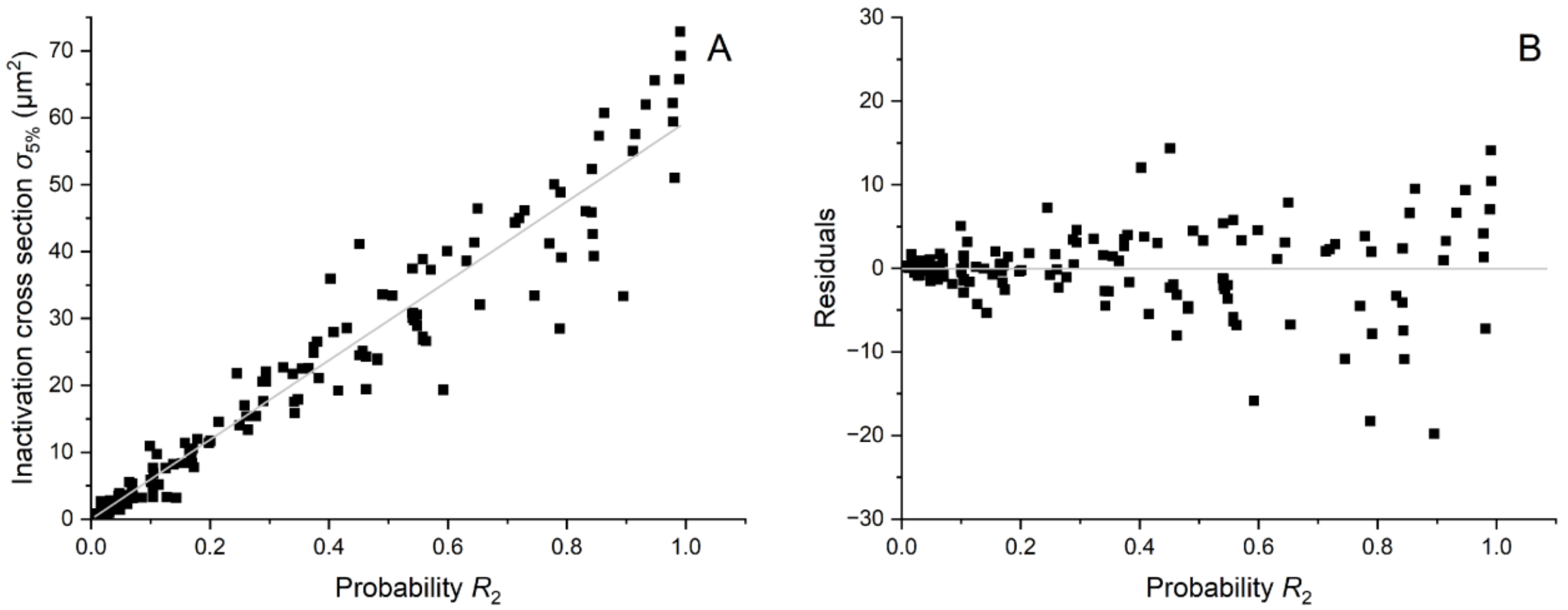

**Supplementary Fig. 10**: (A) Comparison of the probability $R_2$ and the inactivation cross sections at 5 % survival, $\sigma_{5\%}$, and (B) residuals from a linear regression with a line passing the origin. The data corresponds to a target cylinder with diameter of 2.3 nm and height of 3.4 nm in water, and each data point represents a different radiation quality (type of ion and energy).[40]

[40] Supplementary Fig. 10 reproduced from International Journal of Molecular Sciences 25, M. Mietelska, M. Pietrzak, A. Bancer, A. Ruciński, Z. Szefliński, B. Brzozowska, Ionization Detail Parameters for DNA Damage Evaluation in Charged Particle Radiotherapy: Simulation Study Based on Cell Survival Database, 5094, under the CC BY 4.0 license (https://creativecommons.org/licenses/by/4.0/). Copyright Mietelska et al. (2024).

Supplementary Table 2: Synopsis of notations used in different publications for nanodosimetric probabilities and moments. Entries in black correspond to the unconditional probability distribution, and entries in blue to the probability distribution conditional on $\nu \geq 1$, i.e., to targets which receive at least one ionization. Entries in red correspond to the probability distribution conditional on $\nu \geq 2$, i.e., to targets which receive at least two ionizations.

| **Publication(s)** | **Condition** | Probability of $\boldsymbol{\nu}$ ionizations | Probability of $\boldsymbol{\nu \geq k}$ ionizations | Probability of $\boldsymbol{a \leq \nu \leq b}$ ioniz. | $\boldsymbol{n}$**-th moment**[*] | **remarks**[†] |
|---|---|---|---|---|---|---|
| **This work**[‡] | -<br>$\nu \geq 1$<br>$\nu \geq 2$ | $p(\nu)$<br>$p^c(\nu)$<br>$p^{c_2}(\nu)$ | $F^*(k-1) \equiv F_k$<br>$F^{c*}(k-1) \equiv F_k^c$<br>$F^{c_2*}(k-1) \equiv F_k^{c_2}$ | $F_{a-b}$<br>$F_{a-b}^c$<br>$F_{a-b}^{c_2}$ | $M_n$<br>$M_n^c$<br>$M_n^{c_2}$ | Additional parameters to be added in parentheses; e.g., with target size $d$ and radiation quality $Q$: $p(\nu\|d,Q)$, $M_n(d,Q)$ |
| **ICRU Report 98**<br>(Braby *et al* 2023) | -<br>$\nu \geq 1$ | $p(\nu)$<br>$p^c(\nu)$ | $F^*(k-1)$<br>- | -<br>- | $M_n$<br>- | |
| (Alexander *et al* 2015a, 2015b) | $\nu \geq 1$ | $P(\nu\|Q)$ | $F_k(Q)$, $F_k(E)$ | - | $M_n(Q)$, $M_n(E)$ | $E$: primary particle energy |
| (Braunroth *et al* 2020) | - | $P_\nu(r,z\|E_0)$ | $F_k(r,z\|E_0)$ | - | $M_n(r,z\|E_0)$ | $r$: impact parameter, $z$: position of target along primary particle path, $E_0$: primary particle energy |
| (Bueno *et al* 2015) | $\nu \geq 2$ | $p_\nu(Q,V)$ | $f_k(Q,V)$ | - | $m_1(Q,V)$ | $V$: scoring volume in simulation |
| (Bug *et al* 2014) | - | $P(\nu\|Q)$ | $F(k\|Q)$ | | $M_n$ | |
| (Casiraghi and Schulte 2015) | -<br>$\nu \geq 1$ | $P(\nu\|Q)$<br>$P^*(\nu\|Q)$ | -<br>- | $Y_{SC}$[§], $Y_{LC}$[**]<br>- | $M_n$<br>$M_n^*$ | |
| (Conte *et al* 2015) | - | $P_\nu(Q,d)$ | $F_k(Q,d)$ | - | $M_n(Q,d)$ | $d$: impact parameter |
| (Conte *et al* 2017, 2018) | | $P_\nu(Q,V)$ | $F_k(Q,V)$ | - | $M_n(Q,V)$ | $V$: target volume of nanodosimeter |
| (Conte *et al* 2020) | - | $P_\nu(Q)$, $P_\nu(d\rho)$ | $F_k$ | - | $M_n$ | $d$: impact parameter, $\rho$: mass density |
| (Conte *et al* 2023) | | $P_\nu(Q,D)$ | $F_k(Q,D)$ | - | $M_n(Q,D)$ | $D$: diameter of the target |
| (Conte *et al* 2024) | - | $p(\nu)$ | $F^*(k-1)$ | - | $M_n(Q)$ | |

[*] Many publications only use the first moment $M_1$. It is assumed here that analog notation would have been used for higher moments as well.
[†] Note that the symbol $Q$ always indicates radiation quality.
[‡] Note that “≡” used in the third column means that the two symbols designate the same quantity.
[§] for $a = 2$ and $b = 3$
[**] for $a = 4$ and $b = 10$

Supplementary Table to "Prediction of biological radiation effects based on ionization clusters (nanodosimetry)"

| **Publication(s)** | **Condition** | Probability of<br>$\boldsymbol{\nu}$ ionizations | Probability of<br>$\boldsymbol{\nu \geq k}$ ionizations | Probability of<br>$\boldsymbol{a \leq \nu \leq b}$ ioniz. | $\boldsymbol{n}$**-th moment**[*] | **remarks**[†] |
|---|---|---|---|---|---|---|
| **This work**[‡] | -<br>$\nu \geq 1$<br>$\nu \geq 2$ | $p(\nu)$<br>$p^c(\nu)$<br>$p^{c2}(\nu)$ | $F^*(k-1) \equiv F_k$<br>$F^{c*}(k-1) \equiv F_k^c$<br>$F^{c_2*}(k-1) \equiv F_k^{c_2}$ | $F_{a-b}$<br>$F_{a-b}^c$<br>$F_{a-b}^{c_2}$ | $M_n$<br>$M_n^c$<br>$M_n^{c_2}$ | Additional parameters to be added in parentheses; e.g., with target size $d$ and radiation quality $Q$: $p(\nu\|d,Q)$, $M_n(d,Q)$ |
| **ICRU Report 98**<br>(Braby *et al* 2023) | -<br>$\nu \geq 1$ | $p(\nu)$<br>$p^c(\nu)$ | $F^*(k-1)$<br>- | -<br>- | $M_n$<br>- | |
| (Dai *et al* 2020) | $\nu \geq 1$<br>$\nu \geq 2$ | $P(\nu\|Q)$<br>- | $F_k(Q)$<br>$F_k^{c2}(Q)$ | - | $M_n(Q)$<br>$M_n^{c2}(Q)$ | |
| (Garty *et al* 2006) [††] | - | $f(\nu)$ | - | - | - | $f$ is a measured frequency distribution |
| (Garty *et al* 2010) [††] | -<br>$\nu \geq 1$ | $f(\nu)$<br>$\varphi(\nu)$ | -<br>- | -<br>- | -<br>- | $f$ and $\varphi$ refer to the measured frequency distribution of ICs |
| (Grosswendt and Pszona 2002a, 2002b, Grosswendt 2002a, 2002b) | - | $P(T,\nu)$ | - | - | $M_n(T)$ | $T$: primary particle energy |
| (Grosswendt *et al* 2004) | -<br>$\nu \geq 1$ | $P_\nu(T;d)$, $P_\nu(T)$<br>$P_\nu^*(T;d)$ | -<br>- | -<br>- | $M_n(T,d)$<br>- | $T$: primary particle energy, $d$: impact parameter |
| (Grosswendt 2004, 2005) | - | $P_\nu(Q;d)$ | $F_k(T)$ | | $M_n(Q,d)$ | $d$: impact parameter |
| (Grosswendt 2006a) | - | $P_\nu(T;G)$,<br>$P_\nu(Q;G)$, | | | $M_n(T,G)$,<br>$M_n(Q,G)$ | $T$: primary particle energy, $G$: irradiation geometry |
| (Grosswendt 2006b) | - | $P_\nu(Q;d)$,<br>$P_\nu(T;G)$ | $F_k(T,G)$ | | $M_n(Q,d)$,<br>$M_n(T,G)$ | $d$: impact parameter, $T$: primary particle energy, $G$: irradiation geometry |
| (Grosswendt *et al* 2007) | - | $P_\nu(T)$ | $F_k(T)$ | | $M_n(T)$ | $T$: primary particle energy |
| (Mietelska *et al* 2024) | - | $P_\nu$ | $F_k$ | | $M_n$ | |
| (Nettelbeck and Rabus 2011) | - | $P(\nu\|Q)$ | $F_k(Q)$ | - | $M_n(T)$ | $T$: primary particle energy |
| (Ngcezu and Rabus 2021) | - | - | $F_k$ | - | - | |
| (Rabus and Nettelbeck 2011) | - | $P_{\text{ion}}(\nu\|Q)$ | $F_k(Q)$ | | $M_n(Q)$ | |

[††] In these publications, $n_{ion}$ was used as symbol for the ionization cluster size.

| **Publication(s)** | **Condition** | Probability of $\boldsymbol{\nu}$ ionizations | Probability of $\boldsymbol{\nu \geq k}$ ionizations | Probability of $\boldsymbol{a \leq \nu \leq b}$ ioniz. | $\boldsymbol{n}$**-th moment*** | **remarks†** |
|---|---|---|---|---|---|---|
| **This work‡** | -<br>$\nu \geq 1$<br>$\nu \geq 2$ | $p(\nu)$<br>$p^c(\nu)$<br>$p^{c2}(\nu)$ | $F^*(k-1) \equiv F_k$<br>$F^{c*}(k-1) \equiv F_k^c$<br>$F^{c_2*}(k-1) \equiv F_k^{c_2}$ | $F_{a-b}$<br>$F_{a-b}^c$<br>$F_{a-b}^{c_2}$ | $M_n$<br>$M_n^c$<br>$M_n^{c_2}$ | Additional parameters to be added in parentheses; e.g. with target size $d$ and radiation quality $Q$: $p(\nu\|d,Q)$, $M_n(d,Q)$ |
| **ICRU Report 98** (Braby *et al* 2023) | -<br>$\nu \geq 1$ | $p(\nu)$<br>$p^c(\nu)$ | $F^*(k-1)$<br>- | -<br>- | $M_n$<br>- | |
| (Rabus *et al* 2020) | - | $P_\nu(r,x\|x_0,E_0)$ | $F_k(r,z\|x_0,E_0)$ | - | $M_n(r,z\|x_0,E_0)$ | $r$: impact parameter, $x$: position of target along primary particle path, $x_0$: start position of primary particle, $E_0$: primary particle energy |
| (Rabus and Thomas 2025) | -<br>$\nu \geq 1$ | $P_\nu(d_s,r_p)$<br>$p_1(\nu\|Q;d_s,A_p)$ | -<br>- | -<br>- | -<br>- | $d_s$: diameter of site, $r_p$: maximum impact parameter, $A_p$: area of circle with radius $r_p$ |
| (Ramos-Méndez *et al* 2018) | -<br>$\nu \geq 2$ | $P(\nu\|Q)$<br>- | $F_k(Q)$<br>$F_k^{c2}(Q)$ | -<br>- | $M_n(Q)$<br>$M_n^{c2}(Q)$ | |
| (Schneider *et al* 2016) | - | $p_\nu$ | $F_k$ | - | - | |
| (Schneider *et al* 2019, 2020) | - | - | $F_k$ | - | - | |
| (Schulte *et al* 2001) ‡‡ | - | $P(\nu)$ | - | - | - | |
| (Schulte *et al* 2008) §§ | - | $f_{ND}(\nu)$ | - | - | - | $f_{ND}$ is a measured frequency distribution |
| (Schwarze *et al* 2025a, 2025b) | - | $P(\nu)$ | $F_k$ | - | $M_n$ | |
| (Selva *et al* 2019) | - | - | $F_k(Q)$, $F_k(Q,d)$ | - | - | $d$: diameter of the target volume |
| (Yang *et al* 2024) | $\nu \geq 1$ | $P(\nu)$ | $F_k$ | $F_{a-b}$ | $M_n$ | |
| (Yang *et al* 2025) | $\nu \geq 1$ | $P(\nu\|Q)$*** | - | - | $M_n^{C_1}$ | |

‡‡ The symbol used in this publication for the ionization cluster size is $j$.

§§ The symbol used in this publication for the ionization cluster is $n_{ion}$.

*** Yang et al. (Yang *et al* 2025) state that $P(\nu|Q)$ “represents the probability of particles with radiation quality $Q$ producing $\nu$ ionizations within the target volume at nanometer scale”, which could be interpreted such that $P(\nu)$ is an unconditional probability. However, since the work of Yang et al. (Yang *et al* 2025) is a development of the work of Dai et al., the author infers that $P(\nu)$ is meant to be a conditional probability as in (Dai *et al* 2020).

Supplementary Table to “Prediction of biological radiation effects based on ionization clusters (nanodosimetry)”